%% file: SchemeDep.tex
\documentclass[11pt,a4paper]{article}

\usepackage{amsmath,epsfig,a4}
\usepackage{slashed}
\usepackage{jheppub}
\usepackage{color}

\allowdisplaybreaks
  \newcommand{\ccaption}[2]{
    \begin{center}
    \parbox{0.95\textwidth}{
      \caption[#1]{\small{#2}}
      }
    \end{center}
    }
\newcommand{\Eqn}[1]{Eq.~(\ref{#1})}
\newcommand{\Eqns}[2]{Eqs.~(\ref{#1}) and (\ref{#2})}

\newcommand{\sla}[1]{#1\hspace{-2.2mm}\slash}

\newcommand{\RS}{\overline{\text{MS}}}
\newcommand{\Neps}{N_\epsilon}
\newcommand{\alphas}{\alpha_s}
\newcommand{\alphae}{\alpha_e}
\newcommand{\alphafour}{\alpha_{4\epsilon}}

\newcommand{\sss}[1]{{\rm \scriptscriptstyle #1}}
\newcommand{\betaMS}[1]{\bar\beta^s_{#1}}
\newcommand{\betaeMS}[1]{\bar\beta^e_{#1}}

\def\ghat{{\hat{g}}}
\def\gtilde{{\tilde{g}}}
\def\gbar{{\bar{g}}}
\def\HV{{\scshape hv}}
\def\FDH{{\scshape fdh}}
\def\DRED{{\scshape dred}}
\def\CDR{{\scshape cdr}}

\def\RS{{\scshape rs}}

\def\mDRED{{\rm\scriptscriptstyle DRED}}
\def\mCDR{{\rm\scriptscriptstyle CDR}}
\def\mHV{{\rm \scriptscriptstyle HV}}
\def\mFDH{{\rm \scriptscriptstyle FDH}}
\def\mRS{{\rm \scriptscriptstyle RS}}

\begin{document}
\thispagestyle{empty}
\begin{flushright}
PSI-PR-15-06\\
ZU-TH 14/15\\
\today\\
\end{flushright}
\vspace{3em}
\begin{center}
{\Large\bf SCET approach to regularization-scheme dependence of QCD amplitudes}
\\
\vspace{3em}
{\sc A.~Broggio$^a$, Ch.~Gnendiger$^b$, A.~Signer$^{a,c}$, 
D.~St\"ockinger$^b$, A.~Visconti$^{a,c}$
}\\[2em]
{\sl ${}^a$ Paul Scherrer Institut,\\
CH-5232 Villigen PSI, Switzerland \\
\vspace{0.3cm}
${}^b$ Institut f\"ur Kern- und Teilchenphysik,\\
TU Dresden, D-01062 Dresden, Germany\\
\vspace{0.3cm}
${}^c$ Physik-Institut, Universit\"at Z\"urich, \\
Winterthurerstrasse 190,
CH-8057 Z\"urich, Switzerland}
\setcounter{footnote}{0}
\end{center}
\vspace{2ex}
\begin{abstract}
{} We investigate the regularization-scheme dependence of scattering
amplitudes in massless QCD and find that the four-dimensional helicity
scheme (FDH) and dimensional reduction (DRED) are consistent at least
up to NNLO in the perturbative expansion if renormalization is done
appropriately. Scheme dependence is shown to be deeply linked to the
structure of UV and IR singularities. We use jet and soft functions
defined in soft-collinear effective theory (SCET) to efficiently
extract the relevant anomalous dimensions in the different
schemes. This result allows us to construct transition rules for
scattering amplitudes between different schemes (CDR, HV, FDH, DRED)
up to NNLO in massless QCD. We also show by explicit calculation that
the hard, soft and jet functions in SCET are regularization-scheme
independent.
\end{abstract}

\vspace{0.5cm}
\centerline

\newpage
\setcounter{page}{1}

 \noindent\hrulefill
 \tableofcontents
 \noindent\hrulefill

\newpage
\input{01_Introduction}
\input{02_Schemes}
\input{03_SCET}

\input{04_Alternative}
\input{05_Examples}

\input{06_Conclusion}

\bigskip

\section*{Acknowledgments} 
We are grateful to Thomas Becher for useful discussions and for 
providing details on the \CDR\ result of the quark jet function and
to Pier Francesco Monni, Gionata Luisoni, Lorenzo Tancredi and Paolo
Torrielli for useful discussions. We acknowledge financial support
from the DFG grant STO/876/3-1. A.~Visconti is supported by the Swiss
National Science Foundation (SNF) under contract 200021-144252.

\bigskip

\appendix
\input{10_Appendix}
\input{11_AppendixB}


\bibliography{bibliography}{}
\bibliographystyle{JHEP}
\end{document}

%% file: 01_Introduction.tex
\section{Introduction}
\label{sec:introduction}

Higher-order calculations in QCD result in loop integrals that are
often ultraviolet (UV) and/or infrared (IR) divergent. The standard
method to deal with these singularities is dimensional regularization,
where space-time is shifted from 4 to $D\equiv 4-2\epsilon$
dimensions. The UV and IR singularities then manifest themselves as
poles $1/\epsilon^k$.

There are several variants of dimensional regularization.  The most
common scheme is conventional dimensional regularization (\CDR),
where all vector bosons are treated as $D$-dimensional. From a conceptual
point of view this is the simplest possibility and guarantees a
consistent treatment. However, \CDR\, has some disadvantages. Apart
from breaking supersymmetry, it is also not directly compatible with
the helicity method and other computational techniques that rely on 4
dimensions and, hence, leads to more tedious expressions in
intermediate steps of a calculation. Therefore, it is often
advantageous to use other schemes, such as the 't~Hooft-Veltman
scheme (\HV)~\cite{'tHooft:1972fi}, dimensional reduction
(\DRED)~\cite{Siegel:1979wq} or the four-dimensional helicity scheme
(\FDH)~\cite{BernZviKosower:1992}. 

The result for a physical quantity such as a cross section is of
course finite and must not depend on the regularization scheme that
has been used. However, in practise such a result is obtained as a sum
of several contributions, which usually are separately divergent.
Therefore, these partial results can depend on the regularization
scheme. It is often advantageous to use regularization schemes that
are adapted to the technique used for the computation of a particular
contribution. In order to be able to consistently combine the various
partial results it is then imperative to have full control over the
scheme dependence.

The key observation is that the scheme dependence is actually
intimately linked to the structure of UV and IR singularities.
The singularity structure in \FDH\ and \DRED\ is best understood if
the (quasi) 4-dimensional gluons $g$ are split into
$D$-dimensional gluons $\ghat$ and $\Neps = 2 \epsilon$ scalars
$\gtilde$. From a conceptual point of view these so-called
$\epsilon$-scalars $\gtilde$ can be treated as independent fields
with an initially arbitrary multiplicity $\Neps$. The identification
$\Neps = 2 \epsilon$ is to be made only at the end of a calculation.
The decomposition of $g$ into $\ghat$ and $\gtilde$ has to be made in
\DRED\ as well as in \FDH. This seems to be a disadvantage of these schemes.
However, it is useful to gain insight and to derive the scheme
dependence, and for practical purposes, such an explicit separation is
often not required.

The contributions of the $\epsilon$-scalars are UV and IR divergent,
resulting in terms of the form $(\Neps)^i/\epsilon^k$. It is precisely
these terms that -- after setting $\Neps = 2 \epsilon$ -- induce
the scheme dependence in partial results. For a physical cross section the
poles in $\epsilon$ have to cancel, including poles of the form
$\Neps/\epsilon$. This entails that the scheme dependence for a
(finite) physical result can be at most ${\cal O}(\Neps\, \epsilon^0)$
and, hence, will vanish in the limit $\epsilon\to 0$. At
next-to-leading order (NLO) this has been explicitly
demonstrated~\cite{Signer:2008va}. However, virtual corrections
generally are UV and IR divergent and, therefore, scheme dependent. To
find this scheme dependence the structure of UV and IR singularities
has to be understood for a gauge theory with gluons and
$\epsilon$-scalars.

Regarding the UV singularities, the main point is that treating the
$\epsilon$-scalars as independent fields induces additional
couplings. The independence of these couplings and their UV
renormalization was already required in the equivalence proof of
\DRED\, and \CDR~\cite{Capper:1979ns, Jack:1994bn, Jack:1993ws} and in explicit
multi-loop calculations in \DRED~\cite{Harlander:2006rj,
  Harlander:2006xq, Harlander:2007ws}. It has to be stressed that also
in \FDH\ the couplings have to be treated as
independent~\cite{Signer:2008va, Kilgore:2011ta}.

The development regarding the scheme dependence related to the IR
divergent part beyond NLO is more recent. The structure of the IR
singularities for massless gauge amplitudes has a remarkably simple
form~\cite{Gardi:2009qi, Gardi:2009zv, Becher:2009cu,
  Becher:2009qa}. It can be expressed in terms of the cusp anomalous
dimension $\gamma_{\text{cusp}}$ and the anomalous dimensions of the
quark and gluon, $\gamma_{q}$ and $\gamma_{g}$, respectively.
These anomalous dimensions have been extracted from explicit results
of form factors computed in \CDR\ and are consistent with other processes.

It seems natural to assume that this structure can be extended to
other schemes by applying the split of $g$ into $\ghat$ and
$\gtilde$. This results in modified (i.e. scheme dependent) anomalous
dimensions. At NLO, this leads to results that are consistent with the
well-known scheme dependence of NLO amplitudes~\cite{Kunszt:1994}.
Based on this assumption, $\gamma_{\text{cusp}}$, $\gamma_{q}$ and
$\gamma_{g}$ have been extracted in the \FDH\ scheme at
NNLO~\cite{Kilgore:2012tb, Gnendiger:2014nxa}, by comparing the
generalized IR structure to explicit results of two-loop amplitudes
for the $\gamma^*\to q \bar{q}$ and $H\to g g$ form factors and the
process $q\bar{q}\to g \gamma$.  Considering all these processes
together yields an over-constrained system for the extraction of
$\gamma_{\text{cusp}}$, $\gamma_{q}$ and $\gamma_{g}$ in the
\FDH\ scheme. The fact that there is a solution to this system
suggests that \FDH\ is a well defined scheme beyond NLO.

The main results of this paper are the following: First, we will
provide further evidence that with a proper definition \FDH\ can be
used for loop calculations beyond NLO. To this end we show that the
anomalous dimensions $\gamma_{\text{cusp}}$, $\gamma_{q}$ and
$\gamma_{g}$ can be computed directly in soft-collinear effective
theory (SCET) \cite{Bauer:2000ew,Bauer:2000yr,Bauer:2001ct,Bauer:2001yt,
Beneke:2002ph,Beneke:2002ni,Hill:2002vw,Becher:2014oda,Lee:2014xit}
by relating them to the jet- and soft
functions. We repeat the original calculation of the quark-jet
function~\cite{Becher:2006qw} and gluon-jet
function~\cite{Becher:2010pd} in the \FDH\ scheme and also determine
the soft function in \FDH. This gives us an independent determination
of $\gamma_{\text{cusp}}$, $\gamma_{q}$ and $\gamma_{g}$ in the
\FDH\ scheme and the results we find are in agreement with previous
findings. Note that the \FDH\ as we use it \cite{Signer:2008va,
  Kilgore:2012tb} is slightly different from
previous implementations~\cite{Bern:2002zk}.

Second, we extend the scheme dependence study to \DRED. While the
anomalous dimensions in \DRED\ are the same as in \FDH\ we also need
to consider amplitudes with external $\epsilon$-scalars. Determination
of the IR structure of these amplitudes requires the knowledge of
$\gamma_{\epsilon}$, the anomalous dimension of the $\epsilon$-scalar
$\gtilde$.  We compute $\gamma_{\epsilon}$ in SCET via the calculation
of the $\gtilde$-jet and soft functions and give the generalization of
the IR structure to amplitudes with external $\gtilde$. Furthermore,
we verify that this result for $\gamma_{\epsilon}$ is in agreement
with the result extracted from an explicit computation of $H\to
\gtilde \gtilde$ at NNLO~\cite{Broggio:2015ata}. We thus obtain a
complete understanding of the relations between NNLO amplitudes with
gluons and massless quarks computed in \CDR, \HV, \FDH, and \DRED.

Finally, we gain insights into how the regularization-scheme
dependence cancels for fully differential cross sections at
NNLO. While a complete study of this issue is beyond the scope of this
work, our calculations in SCET show that the jet- and soft functions
are separately scheme independent. The same is true for the hard
function. Hence, if the cross section is written as a convolution of
hard-, soft-, and jet functions it is manifestly regularization-scheme
independent. Recently there has been a lot of activity in performing
fully differential NNLO calculations using the SCET framework. This
development started with the computation of top-quark
decay~\cite{Gao:2012ja} and has then been extended to more generic
cases~\cite{Boughezal:2015dva, Boughezal:2015aha, Gaunt:2015pea}.  The
results of our work show how to apply a particular regularization
scheme for the calculation of either the hard-, soft- or jet function.
For each of these building blocks separately, the most convenient
regularization scheme can be used. This opens up possibilities for
further technical advances.

The paper is organized as follows: In Section~\ref{sec:schemes} we
briefly review the various regularization schemes and discuss how they
affect the IR structure of scattering amplitudes.
Section~\ref{sec:scet} is devoted to the computation of the anomalous
dimensions that are required for the IR structure. These computations
are done in SCET. An alternative determination of the anomalous
dimension of the $\epsilon$-scalar is presented in
Section~\ref{sec:alt}, where we extract $\gamma_\epsilon$ from the
gluon form factor computed in \DRED.  In Section~\ref{sec:examples} we
use these results to obtain explicit transition rules for two-loop
amplitudes between \HV\ and \FDH, as well as between \FDH\ and
\DRED. The transition rules are then checked with explicit
examples. Our conclusions including a discussion on the scheme
independence of cross sections at NNLO are presented in
Section~\ref{sec:conclusion}. Finally, we give some explicit results
of the SCET computations in Appendix~\ref{app:A} and list the required
anomalous dimensions and $\beta$ functions in all schemes in
Appendix~\ref{app:B}.

%% file: 02_Schemes.tex
\section{Schemes and structure of IR singularities}
\label{sec:schemes}

\subsection{Regularization schemes}

Dimensional reduction has been shown to be mathematically consistent
\cite{Stockinger:2005gx} and equivalent to dimensional regularization
\cite{Jack:1993ws,Jack:1994bn} on the level of IR finite Green
functions. In the way we define it, the 
\FDH\ scheme has the same properties.
The consistent implementation of the considered regularization schemes
requires the introduction of three vector spaces. Apart from the strictly
4-dimensional space (4S) with metric $\gbar^{\mu\nu}$ two
infinite-dimensional spaces have to be introduced, the quasi 4-dimensional
space Q4S~\cite{Avdeev:1981vf, Avdeev:1982xy, Stockinger:2005gx} with
metric $g^{\mu\nu}$ satisfying $g^{\mu}_{\ \mu} = 4$ and quasi
$D$-dimensional space Q$D$S with metric $\ghat^{\mu\nu}$ satisfying
$\ghat^{\mu}_{\ \mu} = D$. The structure Q4S $\supset$ Q$D$S $\supset$
4S is reflected in the properties of the various metric tensors:
$g^{\mu\nu}\ghat_{\nu\rho}=\ghat^\mu_{\ \rho}$ and
$\ghat^{\mu\nu}\gbar_{\nu\rho}=\gbar^\mu_{\ \rho}$.

For a detailed discussion and a precise definition of the four considered
regularization schemes (\RS) we refer to Ref~\cite{Signer:2008va}.
Here we only repeat the most important aspects to facilitate the
following discussion. The various \RS\ differ in the way ``internal
gluons'' (part of a one-particle irreducible loop diagram or
unresolved final state gluon) and ``external gluons'' (all remaining
gluons) are treated. This is summarized in Table~\ref{tab:RSs} taken
from Ref.~\cite{Signer:2008va}. Since external gluons are treated as
stricly 4-dimensional in \FDH\ and \HV\ these schemes are best adapted
to be used in connection with the helicity method.

\begin{table}
\begin{center}
\begin{tabular}{l|cccc}
&\CDR&\HV&\DRED&\FDH\\
\hline
internal gluon&$\ghat^{\mu\nu}$&$\ghat^{\mu\nu}$&
$g^{\mu\nu}$&$g^{\mu\nu}$\\
 external gluon&$\ghat^{\mu\nu}$&$\gbar^{\mu\nu}$&
$g^{\mu\nu}$&$\gbar^{\mu\nu}$
\end{tabular}
\end{center}
\ccaption{}{
Treatment of internal and external gluons in the four different \RS,
i.e.\ prescription for which metric tensor is to be used in propagator
numerators and polarization sums.\label{tab:RSs}}
\end{table}

The cleanest way to understand the scheme differences is to
consistently apply the split of the (quasi) 4-dimensional gluon into a
$D$-dimensional gluon and an $\epsilon$-scalar. This is done at the
level of the Lagrangian writing the field of the 4-dimensional gluon
field of \FDH\ and \DRED\ as $A^\mu = \hat{A}^\mu + \tilde{A}^\mu$,
where $\hat{A}^\mu$ and $\tilde{A}^\mu$ are the $D$-dimensional gauge
field and the $\epsilon$-scalar field, 
respectively~\cite{Capper:1979ns}.  We will denote the associated
'particles' as $\ghat$ and $\gtilde$, respectively.  The
$\epsilon$-scalars have an initially independent multiplicity
$N_\epsilon$ and the metric $\gtilde^{\mu\nu}$ associated with
$\gtilde$ satisfies the orthogonality relation
$\ghat^{\mu\nu}\gtilde_{\nu\rho}=0$ and $\gtilde^{\mu\nu}
\gtilde_{\mu\nu} = \Neps$. Scheme differences have their origin in UV
and IR divergent contributions due to these $\epsilon$-scalars. These
contributions are of the form $(N_\epsilon)^i/\epsilon^k$ and after
setting $N_\epsilon \to 2\epsilon$ result in the scheme
differences. This connection to UV and IR singular terms allows for a
completely systematic treatment of the \RS\ dependence.

Regarding UV renormalization, \FDH\ and \DRED\ behave in the same
way. The possible split of internal gluons into gauge fields and
$\epsilon$-scalars implies that in principle five different couplings
need to be distinguished (see in particular
\cite{Jack:1993ws,Harlander:2006rj,Kilgore:2011ta}): the gauge
coupling $\alpha_s$,
the $\gtilde q \bar{q}$ coupling $\alpha_e$, and 
three different independent quartic $\gtilde$-couplings
$\alpha_{4\epsilon,i}$ with $i=1,2,3$. In general, we write the
perturbative expansion of a \RS-dependent quantity $X^{\mRS}(\{\alpha\})$ as 
\begin{align} \label{eq:generalex}
X^{\mRS}(\{\alpha\}) = \sum^{\infty}_{m,n,k,l,j}
\left(\frac{\alpha_s}{4 \pi}\right)^{m}
\left(\frac{\alpha_e}{4 \pi}\right)^{n}\,
\left(\frac{\alpha_{4\epsilon,1}}{4 \pi}\right)^{k}\,
\left(\frac{\alpha_{4\epsilon,2}}{4 \pi}\right)^{l}\,
\left(\frac{\alpha_{4\epsilon,3}}{4 \pi}\right)^{j}\,
X^{\mRS}_{mnklj} \, .
\end{align} 
Accordingly, the $\beta$ functions for $\alpha_s$ and
$\alpha_e$ in full generality are written as
\begin{subequations}
\label{eq:alphaRGE}
\begin{align}
 \mu^2\frac{d}{d\mu^2}\frac{\alpha_s}{4\pi}&=
  -\epsilon\frac{\alpha_s}{4\pi}\ -
 \sum_{\Sigma\ge 2}
 \left(\frac{\alpha_s}{4\pi}\right)^m
 \left(\frac{\alpha_e }{4\pi}\right)^n
\left(\frac{\alpha_{4\epsilon,1}}{4 \pi}\right)^{k}\,
\left(\frac{\alpha_{4\epsilon,2}}{4 \pi}\right)^{l}\,
\left(\frac{\alpha_{4\epsilon,3}}{4 \pi}\right)^{j}\,
\beta^{s\, \mRS}_{m n k l j} ,
 \label{eq:alphasRGE}\\
 \mu^2\frac{d}{d\mu^2}\frac{\alpha_e}{4\pi}&=
  -\epsilon\frac{\alpha_e}{4\pi}\ -
 \sum_{\Sigma\ge 2}
 \left(\frac{\alpha_s}{4\pi}\right)^m
 \left(\frac{\alpha_e }{4\pi}\right)^n
 \left(\frac{\alpha_{4\epsilon,1}}{4 \pi}\right)^{k}\,
 \left(\frac{\alpha_{4\epsilon,2}}{4 \pi}\right)^{l}\,
 \left(\frac{\alpha_{4\epsilon,3}}{4 \pi}\right)^{j}\,
\beta^{e\, \mRS}_{m n k l j}
 \label{eq:alphaeRGE}
\end{align}
\end{subequations}
with analogous expansions for the $\beta$ functions for
$\alpha_{4\epsilon,i}$. In the sums, $\Sigma\ge2$ is an abbreviation
for $m+n+k+l+j\ge2$. The later results of the present paper will show
that the $\beta$ functions of the $\alpha_{4\epsilon,i}$ are not
needed and that we do not need to distinguish between them; hence we
will often denote them generically by
$\alpha_{4\epsilon}$.\footnote{We remark that in practice the
  couplings can often be identified; only the bare couplings and the
  associated renormalization constants and $\beta$ functions must be
  kept different. Section \ref{sec:examples} will provide further
  discussion and examples.}  Note that in \Eqn{eq:alphaRGE} all
quantities are finite and the scheme dependence is
$\mathcal{O}(\Neps)$. Thus, after setting $\Neps\to 2 \epsilon$ and
then $\epsilon\to 0$, the scheme dependence disappears and we refrain
from using an \RS\ label on the l.h.s. of \Eqn{eq:alphaRGE}. In
particular we write $\alpha_{s}$ and $\alpha_{e}$ without an \RS\ label.

According to Table~\ref{tab:RSs}, in \DRED\ external gluons are
(quasi) 4-dimensional.  The decomposition of these external gluons
into $\ghat$ and $\gtilde$ also allows to avoid all problems related
to factorization theorems~\cite{Signer:2005} in \DRED\ regularized
QCD. However, this split results in a larger number of 'independent'
diagrams.  Applying the decomposition of $g$ into $\ghat$ and
$\gtilde$ then implies that in \DRED\ amplitudes with external
$\epsilon$-scalars have to be considered. This is not the case in the
other schemes. As this leads to additional complications, we will
first restrict our discussion of the scheme dependence to the schemes
\CDR, \HV\ and \FDH\ in Section~\ref{sec:chv}. Then we will consider
\DRED\ in a second step in Section~\ref{sec:dred}.

\subsection{IR structure in  CDR, HV and FDH}
\label{sec:chv}

After UV renormalization, on-shell scattering amplitudes in massless
QCD still contain IR poles $1/\epsilon^k$. In the framework of
\CDR\ it has been shown that these singularities can be subtracted in
the $\overline{\mathrm{MS}}$ scheme, using the procedure described in
\cite{Becher:2009qa,Becher:2009cu,
Magnea:2012pk,Gardi:2009qi,Gardi:2009zv,DelDuca:2011ae,Bret:2011xm},
via a multiplicative renormalization factor $\mathbf{Z}$ which is a matrix
in colour space. This can be generalized not only to the \HV\ but also to
the \FDH\ and \DRED\ schemes~\cite{Kilgore:2012tb, Gnendiger:2014nxa}.

For the following discussion we find it more convenient to work with
amplitudes squared. More precisely, we consider
\begin{align}\label{def:M}
\mathcal{M}^{\mRS*}(\epsilon,\Neps,\{p\}) \equiv 
  2\, {\rm Re}\, \langle
  \mathcal{A}_0^{\mRS*}(\epsilon,\Neps,\{p\})|
  \mathcal{A}^{\mRS*}(\epsilon,\Neps,\{p\})\rangle \, ,
\end{align}
where $|\mathcal{A}^{\mRS*}(\epsilon,\Neps,\{p\})\rangle$ is a UV
renormalized, on-shell $n$-parton scattering amplitude containing IR
poles and $\langle \mathcal{A}_0^{\mRS*}(\epsilon,\Neps,\{p\})|$ is
the corresponding tree-level amplitude\footnote{Strictly speaking, the
  tree-level amplitudes in the \RS*-schemes do not depend on
  $\Neps$. Nevertheless, we keep the dependence on $\Neps$ in the
  notation to simplify the generalization to \DRED\ in Section~\ref{sec:dred}.} Both the
$\epsilon$- and the $\Neps$-dependence differ in the four
regularization schemes. For the moment we restrict ourselves to \CDR,
\HV, \FDH, as indicated by the label \RS*. Then the regularized
external gluons behave completely as gauge fields and do not have to
be split into gauge fields and $\epsilon$-scalars. The set $\{p\}$
denotes the set of partons of the process under consideration and
contains only quarks or gluons.

The regularization-scheme dependence of $ \mathcal{M}^{\mRS*}$ is
related to the IR poles and can be absorbed by a scheme-dependent
factor $(\mathbf{Z}^{\mRS*})^{-1}$.  We can define IR subtracted
finite squared amplitudes as
\begin{align}\label{eq:zcdr}
\mathcal{M}_{\mathrm{sub}}^{\mRS*}(\epsilon,\Neps,\{p\},\mu)
= 2\, {\rm Re} \langle \mathcal{A}_0^{\mRS*}(\epsilon,\Neps,\{p\})| 
\big(\mathbf{Z}^{\mRS*}(\epsilon,\Neps,\{p\},\mu)\big)^{-1} 
|\mathcal{A}^{\mRS*}(\epsilon,\Neps,\{p\})\rangle\, ,
\end{align}
where $\mu$ represents the factorization scale. The expression on the
l.h.s of \Eqn{eq:zcdr}, $\mathcal{M}_{\mathrm{sub}}^{\mRS*}$, denotes the
finite remainder of the amplitude where the poles have been subtracted
in a minimal way.  $\mathcal{M}_{\mathrm{sub}}^{\mRS*}$ still depends on
$\epsilon$ (and $N_\epsilon$) but does not contain poles
$1/\epsilon^k$ any longer. Hence, the limit $\epsilon\to 0$ can be
taken and then we obtain a scheme independent finite matrix element
squared
\begin{align}\label{eq:fin}
\mathcal{M}_{\mathrm{fin}}(\{p\},\mu) = \lim_{(N)_\epsilon\to 0}
\mathcal{M}_{\mathrm{sub}}^{\mRS*}(\epsilon,\Neps,\{p\},\mu)\, .
\end{align}
The limit $(N)_\epsilon\to 0$ indicates that first we set
$N_\epsilon\to 2\epsilon$ and then $\epsilon\to 0$. To put it
differently, after setting $N_\epsilon \to 2 \epsilon$, the scheme
dependence of $\mathcal{M}_{\mathrm{sub}}^{\mRS*}$ is only in the
terms ${\cal O}(\epsilon)$.

The starting point for a typical NNLO calculation is the computation
of the two-loop virtual corrections in a particular regularization
scheme. This corresponds to $\mathcal{M}^{\mRS*}$ as defined in
\Eqn{def:M}. To understand the IR divergence structure and obtain
transition rules between schemes we want to exploit the relation of
the scheme-dependent $\mathcal{M}^{\mRS*}$ to the scheme-independent
$\mathcal{M}_{\mathrm{fin}}$. The key quantity for this is the scheme
dependent factor $\mathbf{Z}^{\mRS*}$ to which we turn now.

The all-order amplitude
$|\mathcal{A}^{\mRS*}(\epsilon,\Neps,\{p\})\rangle$ in \Eqn{eq:zcdr}
is independent of the factorization scale $\mu$. It follows that the
IR subtracted amplitude squared satisfies a renormalization group
equation (RGE)
\begin{align}\label{eq:rge}
\frac{d}{d \ln \mu} 
\mathcal{M}_{\mathrm{sub}}^{\mRS*}(\epsilon,\Neps,\{p\},\mu) 
 = \mathbf{\Gamma}^{\mRS*} (\Neps,\{p\},\mu)\,
 \mathcal{M}_{\mathrm{sub}}^{\mRS*}(\epsilon,\Neps,\{p\},\mu) \, ,
\end{align}
where the anomalous dimension
$\mathbf{\Gamma}^{\mRS*}(\Neps,\{p\},\mu)$ is related to the
$\mathbf{Z}^{\mRS*}$ factor through
\begin{align} \label{rel:ZGamma}
\mathbf{\Gamma}^{\mRS*} (\Neps,\{p\},\mu) = 
- \big(\mathbf{Z}^{\mRS*}(\epsilon,\Neps,\{p\},\mu) \big)^{-1}
 \frac{d}{d \ln \mu} \mathbf{Z}^{\mRS*}(\epsilon,\Neps,\{p\},\mu) \, .
\end{align}
This equation can be formally solved to obtain a path-ordered
exponential with respect to colour matrices
\begin{align}\label{eq:zsol}
\mathbf{Z}^{\mRS*}(\epsilon,\Neps,\{p\},\mu) 
= \mathcal{P} \exp \int_{\mu}^{\infty}  \frac{d \mu^\prime}{\mu^\prime}
 \mathbf{\Gamma}^{\mRS*} (\Neps,\{p\},\mu^\prime)\, .
\end{align}

In \cite{Gardi:2009qi, Gardi:2009zv, Becher:2009cu, Becher:2009qa} it
has been shown that in \CDR\ the general structure of the anomalous
dimension operator $\mathbf{\Gamma}$, which controls the IR
divergences of QCD scattering amplitudes, is exactly known up to
two-loop level and only involves colour dipoles. In those papers it
was also conjectured, by using soft-collinear factorization
constraints and symmetry arguments, that this simple structure is more
general and it is valid to all orders in  perturbation
theory. Generalizing this from \CDR\ to other schemes and suppressing the
dependence on $\Neps$, we write according to Refs.~%
\cite{Kilgore:2012tb,Gnendiger:2014nxa}
\begin{align}\label{eq:conj}
\mathbf{\Gamma}^{\mRS*} (\{p\},\mu)  
= \sum_{(i,j)} \frac{\mathbf{T}_i \cdot \mathbf{T}_j}{2}\, 
\gamma_{\textrm{cusp}}^{\mRS*}\, \ln{\frac{\mu^2}{-s_{ij}}}
 + \sum_{i=1}^n \gamma_i^{\mRS*} \, ,
\end{align}
where $s_{ij}=\pm 2 p_i\cdot p_j + i0$, the sign ``+'' is chosen when
both momenta $p_i$ and $p_j$ are incoming or outgoing and the sign
``$-$'' when one momentum is incoming and the other one outgoing. The
first sum in \Eqn{eq:conj} runs over all pairs $i\neq j$ of distinct
parton indices $i,j\in \{1,2,\ldots,n\}$, where $n$ is the number of
external partons. The universal quantity
$\gamma_{\textrm{cusp}}^{\mRS*}$ that appears as coefficient of the
two-particle correlation term,
$\mathbf{T}_i\cdot\mathbf{T}_j\equiv\mathbf{T}^c_i\mathbf{T}^c_j$, is
called ``cusp" anomalous dimension. The quantity $\gamma_i^{\mRS*}$ is
a single-particle term which depends on the type of the external
particle, $\gamma_{q}^{\mRS*}\equiv \gamma_{\bar{q}}^{\mRS*}$ in the
case of a (anti)quark and $\gamma_{g}^{\mRS*}$ in the case of a gluon.
The explicit form of the colour generator associated to the $i$-th
parton, $\mathbf{T}^a_i$, is as follows: For final-state quarks or
initial-state antiquarks, the colour matrices $\textbf{T}$ are defined
by $\left(\mathbf{T}^c\right)_{ba}=t^c_{ba}$, where $t^c$ is a SU($N$)
generator. For final-state antiquarks or initial state quarks one has
instead $\left(\mathbf{T}^c\right)_{ba}=-t^c_{ab}$, while for gluons
$\left(\mathbf{T}^c\right)_{ba}=i f^{abc}$.

As a consequence the IR structure can be described by a set of three
constants, which depend on the scheme
\begin{align} \label{eq:listgammasRS}
\mRS*\in\{\mCDR,\mHV,\mFDH\}&:\quad
\gamma^{\mRS*}_{\textrm{cusp}},
\gamma^{\mRS*}_{q},
\gamma^{\mRS*}_{g}.
\end{align}
Thanks to the simple structure of the anomalous dimension
matrix $\mathbf{\Gamma}$, one can find an explicit solution for the
perturbative expansion of $\mathbf{Z}$.  It is also possible to drop
the path-ordering symbol in \Eqn{eq:zsol} since the colour structure
of $\mathbf{\Gamma}$ is independent of $\mu$.  The following notation
is often introduced
\begin{align}
\Gamma^{\prime \mRS*} (\{p\}) \equiv 
\frac{\partial}{\partial \ln \mu} 
\mathbf{\Gamma}^{\mRS*}(\{p\},\mu)
 = -\gamma_{\textrm {cusp}}^{\mRS*}\,\sum_i C_i\, ,
\end{align}
where the last equality follows from colour conservation,
$C_i=C_{\bar{q}}=C_q=C_F$ for (anti)quarks and $C_i=C_g = C_A$ for
gluons.  

All scheme-dependent quantities introduced so far potentially depend on all couplings $\{\alpha(\mu)\}
\equiv \{\alpha_s(\mu), \alpha_e(\mu), \alpha_{4\epsilon,i}(\mu) \}$.
Thus, in general the perturbative expansion is of the form of
\Eqn{eq:generalex}. 

Solving the differential equation \Eqn{rel:ZGamma} one obtains a
perturbative expression for $\ln\mathbf{Z}^{\mRS*}$ which also depends
on the $\beta$ functions. Suppressing the arguments, in particular the
dependence on the process $\{p\}$, it can be written up to NNLO as
\begin{align}
\label{eq:lnZfullRS}
 \ln\mathbf{Z}^{\mRS*}&=
 \left(\frac{\vec{\alpha}}{4\pi}\right)\cdot
\left(\frac{\vec{\Gamma}^{\prime\, \mRS*}_{1}}{4\epsilon^2}
      +\frac{\vec{\mathbf{\Gamma}}^{\mRS*}_{1}}{2\epsilon}\right)\notag\\
 &+
 \sum_{\Sigma=2}  
\left(\frac{\alpha_s}{4\pi}\right)^m
 \left(\frac{\alpha_e }{4\pi}\right)^n
 \left(\frac{\alpha_{4\epsilon,1}}{4 \pi}\right)^{k}\,
 \left(\frac{\alpha_{4\epsilon,2}}{4 \pi}\right)^{l}\,
 \left(\frac{\alpha_{4\epsilon,3}}{4 \pi}\right)^{j}
 \notag \\
& \quad \Bigg( 
-\frac{3\vec{\beta}^{\mRS*}_{mnklj} \cdot \vec{\Gamma}^{\prime\, \mRS*}_{1}}{16\epsilon^3}
- \frac{\vec{\beta}^{\mRS*}_{mnklj} \cdot \vec{\mathbf{\Gamma}}^{\mRS*}_{1}}{4\epsilon^2}
+\frac{\Gamma^{\prime\, \mRS*}_{mnklj}}{16\epsilon^2}
+\frac{\mathbf{\Gamma}^{\mRS*}_{mnklj}}{4\epsilon}
\Bigg)
+ \mathcal{O}(\alpha^3) \, .
\end{align}
Here the sum $\Sigma=2$ denotes a sum over all terms satisfying $m+n+k+l+j=2$, and
the following vector notation for terms involving pure
one-loop quantities has been used:
\begin{subequations}
\begin{align}
\label{eq:oneloopnotation}
\vec{\alpha}\cdot
\vec{\mathbf{\Gamma}}^{\mRS*}_{1}
\equiv &\
\alpha_{s} \, \mathbf{\Gamma}^{\mRS*}_{10000}
+ \alpha_e \, \mathbf{\Gamma}^{\mRS*}_{01000}
+ \alpha_{4\epsilon,1} \, \mathbf{\Gamma}^{\mRS*}_{00100}
+ \alpha_{4\epsilon,2} \, \mathbf{\Gamma}^{\mRS*}_{00010}
+ \alpha_{4\epsilon,3} \, \mathbf{\Gamma}^{\mRS*}_{00001}\,,
\\[5pt]
\label{eq:defvectornotation}
\vec{\beta}^{\mRS*}_{mnklj} \cdot \vec{\mathbf{\Gamma}}^{\mRS*}_{1}
\equiv &\
\beta^{s\, \mRS*}_{mnklj} \,  \mathbf{\Gamma}^{\mRS*}_{10000}
+ \beta^{e\, \mRS*}_{mnklj} \,  \mathbf{\Gamma}^{\mRS*}_{01000}
\nonumber\\
&
+ \beta^{4\epsilon,1\, \mRS*}_{mnklj} \,  \mathbf{\Gamma}^{\mRS*}_{00100}
+ \beta^{4\epsilon,2\, \mRS*}_{mnklj} \,  \mathbf{\Gamma}^{\mRS*}_{00010}
+ \beta^{4\epsilon,3\, \mRS*}_{mnklj} \,  \mathbf{\Gamma}^{\mRS*}_{00001} \, ,
\end{align}
\end{subequations}
and analogously for the combinations involving $\vec{\Gamma}^\prime_{1}$.
The dependence of $\mathbf{\Gamma}$ on the individual couplings and the
appearance of the different $\beta$ functions constitutes an important
difference to the \CDR\ case, where only the $\alpha_s$ and $\beta^s$
terms appear. It can be obtained by setting
$\alpha_e,\alpha_{4\epsilon,i} \to 0$ in \Eqn{eq:lnZfullRS} and identifying
$\mathbf{\Gamma}_{m0000} = \mathbf{\Gamma}_m$ etc.

\Eqn{eq:lnZfullRS} shows that the one-loop IR divergences are
described by the one-loop coefficients of $\Gamma^\prime$, which depend
on the process-independent quantity $\gamma_{\textrm {cusp}}^{\mRS*}$,
and of $\mathbf{\Gamma}$.
Both anomalous dimensions depend on the partons involved in the process.
At the two-loop level, the full $1/\epsilon^3$ and parts of the
$1/\epsilon^2$ divergences are predicted by one-loop $\beta$ and $\Gamma$
coefficients. The remaining $1/\epsilon^2$ and the $1/\epsilon$ poles are
described by genuine two-loop anomalous dimensions.

Eq.~\eqref{eq:zcdr} together with Eq.~\eqref{eq:lnZfullRS} allows to describe
the RS dependence of the squared amplitude $\mathcal{M}^{\mRS*}$:
\begin{itemize}
\item \CDR-\HV:
Since internal gluons are treated in the same way in \CDR\ and \HV\ we have
\mbox{$\mathbf{Z}^{\mCDR} = \mathbf{Z}^{\mHV}$} and all the anomalous
dimensions are the same in these two schemes.
The difference in the squared matrix element comes entirely from using
different metric tensors for the polarization sum due to external gluons.
In \CDR, where external gluons are $D$-dimensional, this polarization sum
involves $\ghat^{\mu\nu}$, whereas in \HV\ $\gbar^{\mu\nu}$ is to be used.
\item \HV-\FDH:
Since internal gluons are treated differently in \HV\ and \FDH\ we have
\mbox{$\mathbf{Z}^{\mHV} \neq \mathbf{Z}^{\mFDH}$} and the anomalous
dimensions are not the same in these two schemes. This results in further
scheme differences of the squared  matrix element. However, external gluons
are treated in the same way in \HV\ and \FDH\ and the metric tensors in
polarization sums are the same in the two schemes.
\end{itemize}

\subsection{IR structure in  DRED}
\label{sec:dred}

Understanding the IR structure of
\DRED\ processes with external gluons is more complicated. Each
external quasi-4-dimensional gluon can be split into a $\ghat$ and a
$\gtilde$, and the squared matrix element for a process with
$\#g$ external gluons can be decomposed into $2^{\#g}$ terms. Following
Ref.~\cite{Signer:2008va}, we can write for the amplitude squared for such
a process
\begin{align}\label{dred:dec}
\mathcal{M}^{\mDRED}(\ldots g_{1}\ldots g_{\#g}\ldots) =
\sum_{\breve{g}_{1}\in\{\ghat,\gtilde\}} \ldots
\sum_{\breve{g}_{\#g}\in\{\ghat,\gtilde\}}
\mathcal{M}^{\mDRED}(\ldots \breve{g}_{1}\ldots 
\breve{g}_{\#g}\ldots) \, .
\end{align}
Reinstating all variables explicitly, we write the same relation in a
more compact way as
\begin{align}\label{dred:dec2}
\mathcal{M}^{\mDRED}(\epsilon,\Neps,\{p\},\mu) =
\sum_{\{\breve{p}\}}
\mathcal{M}^{\mDRED}(\epsilon,\Neps,\{\breve{p}\},\mu) \, .
\end{align}
Hence, the partons appearing in the list $\{\breve{p}\}$ on the
r.h.s.\ can be either quarks or $\ghat$, $\gtilde$, but not full
quasi-4-dimensional gluons.  We stress that practical calculations are
not as complicated as implied by \Eqns{dred:dec}{dred:dec2}.
The l.h.s.\ will typically be computed directly as a whole with quasi
4-dimensional gluons, i.e.\ 4-dimensional numerator algebra.
Even the renormalized couplings $\alpha_s$, $\alpha_e$,
$\alpha_{4\epsilon}$ can be identified, see section \ref{sec:examples}
for further discussion. However, from a conceptual point of view each
term in the sum on the r.h.s.\ of \Eqns{dred:dec}{dred:dec2} can be
considered as an independent process and the couplings as independent.
Then, each of these processes behaves as the processes in \CDR, \HV,
\FDH\ discussed in the previous subsection, and it becomes possible to
understand the IR structure and construct IR subtraction terms and
transition rules to other schemes. 

For each process on the r.h.s.\ of \Eqns{dred:dec}{dred:dec2} a corresponding
factor $\mathbf{Z}(\epsilon,\Neps,\{\breve{p}\},\mu)$ and a subtracted squared amplitude
$\mathcal{M}_{\mathrm{sub}}^{\mDRED}(\epsilon,
\Neps,\{\breve{p}\},\mu)$ can be constructed, like for
$\mathcal{M}^{\mRS*}$ in \Eqn{def:M} and \Eqn{eq:zcdr}. 
Overall, one can then define the full subtracted squared amplitude in
\DRED\ as
\begin{align}
\mathcal{M}^{\mDRED}_{\mathrm{sub}}(\epsilon,\Neps,\{p\},\mu) =
\sum_{\{\breve{p}\}}
\mathcal{M}^{\mDRED}_{\mathrm{sub}}(\epsilon,\Neps,\{\breve{p}\},\mu) \, .
\end{align}
It satisfies an equation analogous to \Eqn{eq:rge},
\begin{align}\label{eq:rgedred}
\frac{d}{d \ln \mu} 
\mathcal{M}_{\mathrm{sub}}^{\mDRED}(\epsilon,\Neps,\{p\},\mu) 
 = \sum_{\{\breve{p}\}}
\mathbf{\Gamma}^{\mDRED} (\{\breve{p}\},\mu) 
 \mathcal{M}_{\mathrm{sub}}^{\mDRED}(\epsilon,\Neps,\{\breve{p}\},\mu) \, ,
\end{align}
The $\mathbf{\Gamma}^{\mDRED}$'s for the individual parton sets
$\{\breve{p}\}$ satisfy relations analogous to Eqs.~\eqref{rel:ZGamma},
\eqref{eq:zsol} and \eqref{eq:conj}. Likewise, the
subtraction factors $\mathbf{Z}$ can be written as
\begin{align}
\label{eq:lnZfullDRED}
\nonumber
 \ln\mathbf{Z}^{\mDRED}&=
 \left(\frac{\vec{\alpha}}{4\pi}\right)\cdot
\left(\frac{\vec{\Gamma}^{\prime\, \mDRED}_{1}}{4\epsilon^2}
      +\frac{\vec{\mathbf{\Gamma}}^{\mDRED}_{1}}{2\epsilon}\right)\\
 &+
 \sum_{\Sigma=2}  
\left(\frac{\alpha_s}{4\pi}\right)^m
 \left(\frac{\alpha_e }{4\pi}\right)^n
 \left(\frac{\alpha_{4\epsilon,1}}{4 \pi}\right)^{k}\,
 \left(\frac{\alpha_{4\epsilon,2}}{4 \pi}\right)^{l}\,
 \left(\frac{\alpha_{4\epsilon,3}}{4 \pi}\right)^{j}\\
& \quad \Bigg( 
-\frac{3\vec{\beta}^{\mDRED}_{mnklj} \cdot \vec{\Gamma}^{\prime\, \mDRED}_{1}}{16\epsilon^3}
- \frac{\vec{\beta}^{\mDRED}_{mnklj} \cdot \vec{\mathbf{\Gamma}}^{\mDRED}_{1}}{4\epsilon^2}
+\frac{\Gamma^{\prime\, \mDRED}_{mnklj}}{16\epsilon^2}
+\frac{\mathbf{\Gamma}^{\mDRED}_{mnklj}}{4\epsilon}
\Bigg)
+ \mathcal{O}(\alpha^3) \, .
\nonumber 
\end{align}
Like in the corresponding \Eqn{eq:lnZfullRS} the arguments are
suppressed. An important difference to the \RS* schemes is that in
\DRED\ the individual split processes $\{\breve{p}\}$ have to be used.
This implies that the set of $\gamma$'s needed to describe the IR
structure is different in \DRED\ compared to the other schemes,
\begin{align} \label{eq:listgammasDRED}
\mDRED&:\quad
\gamma^{\mDRED}_{\textrm{cusp}},
\gamma^{\mDRED}_{{q}},
\gamma^{\mDRED}_{\ghat},
\gamma^{\mDRED}_{\gtilde}\, .
\end{align}
This should be compared with \Eqn{eq:listgammasRS}.
There are however several obvious relations, since internal gluons are
treated equally in  \FDH\ and \DRED:
\begin{subequations}
\label{eq:Defgammabar}
\begin{align}
\bar{\gamma}_{\textrm{cusp}} &\equiv \gamma_{\textrm{cusp}}^{\mFDH} =
\gamma_{\textrm{cusp}}^{\mDRED},\\*
\bar{\gamma}_q &\equiv
\gamma_q^{\mFDH} = \gamma_q^{\mDRED},\\*
\bar{\gamma}_g &\equiv
\gamma_g^{\mFDH} = \gamma_{\ghat}^{\mDRED}\, .
\end{align}
\end{subequations}
Thus, the $\epsilon$-scalar anomalous dimension
$\gamma^{\mDRED}_{\gtilde}$ is the only additional ingredient in \DRED.
To highlight this, we introduce the notation $\bar\gamma_{\epsilon}$
for this quantity,
\begin{align}
\label{eq:Defgammaeps}
\bar{\gamma}_\epsilon &\equiv
\gamma_{\gtilde}^{\mDRED}\, .
\end{align}

It is instructive to compare the individual processes with external
$\ghat$ or $\gtilde$ in \DRED\ to a process in \FDH.
The squared amplitude for a process with at least one external
$\gtilde$ has an overall factor $N_\epsilon$ from the
$\epsilon$-scalar polarization sum. As long as we consider the
UV renormalized, but not yet IR subtracted matrix element, we cannot
set  $(N)_\epsilon \to 0$ since there are still IR poles
present. However, once these have been subtracted, the squared matrix
element is free of poles in $\epsilon$ and still contains a factor
$N_\epsilon$. Hence, 
\begin{align} \label{eq:dred1}
\mathcal{M}_{\mathrm{fin}}^{\mDRED}(\ldots \gtilde\ldots) =
\lim_{(N)_\epsilon\to 0}
    \mathcal{M}_{\mathrm{sub}}^{\mDRED}(\ldots \gtilde\ldots) = 0 
\end{align}
and
\begin{align} \label{eq:dred2}
\lim_{(N)_\epsilon\to 0}
\mathcal{M}_{\mathrm{sub}}^{\mDRED}(\ldots g_{1}\ldots g_{\#g}\ldots) =
\lim_{(N)_\epsilon\to 0}
\mathcal{M}_{\mathrm{sub}}^{\mDRED}(\ldots \ghat_{1}\ldots \ghat_{\#g}\ldots) =
\mathcal{M}_{\mathrm{fin}}(\ldots g\ldots) \, ,
\end{align}
i.e. once the amplitudes are properly subtracted and the limit
$(N)_\epsilon\to 0$ is taken, processes with external $\gtilde$ do not
contribute any longer and the finite squared amplitude is equal in all
four regularization schemes.

%% file: 03_SCET.tex
\section{SCET approach to scheme dependence}
\label{sec:scet}

In Section~\ref{sec:schemes} it has been shown that the
regularization-scheme dependence of any massless QCD amplitude can be
absorbed into a re-definition of the factor $\mathbf{Z}$. Hence, it is
important to study the scheme dependence of the anomalous dimension
$\mathbf{\Gamma}$ governing the RG equation for the
$\mathbf{Z}$-factor. We work at NNLO, and at this order the anomalous
dimension has a sum-over-dipoles structure. Thus, we need to compute
the three relevant anomalous dimensions in \Eqn{eq:conj},
$\gamma_{\textrm {cusp}}$, $\gamma_q$ and $\gamma_g$ in the several
schemes considered in this work, particularly in \FDH\ (in \DRED, also
$\gamma_\epsilon$ is needed). In principle $\gamma_q$ and $\gamma_g$
can be directly extracted from the IR divergences of the on-shell
quark and gluon form factors computed in the three schemes.  This
approach~\cite{Kilgore:2012tb, Gnendiger:2014nxa}, which at first
glance seems to be totally straightforward, turned out to hide highly
non-trivial technical complications related to the UV renormalization
procedure in schemes like \FDH\ and \DRED.

Here we show that the same $\gamma$'s can be also extracted by
combining the anomalous dimensions of the quark and gluon jet
functions together with the anomalous dimensions of the corresponding
soft functions (for Drell-Yan or Higgs production) defined through
SCET operators.  The soft and the jet functions can be computed with a
standard diagrammatic procedure, and they are free of the
renormalization difficulties that appear in the form factor
calculations.  This is an easier and more direct way to
perform such a calculation. We have carried out this calculation at NNLO.
In addition,  the computation has also been carried out using the more
traditional method to have an independent check of the results
presented in this work and to show that the scheme dependence of these
anomalous dimensions is universal and does not depend on the
particular process analyzed.

\subsection{Outline of the method}
\label{sec:outlinescet}

In the following we present the procedure for the direct calculation
of the relevant anomalous dimensions in the four schemes via a SCET
approach.  The anomalous dimensions are obtained not from QCD
scattering amplitudes but from soft and jet functions defined in
SCET. Schematically, we get
\begin{subequations}
\begin{align}
\text{soft function} \quad&\Rightarrow\quad
\gamma^{\mRS}_{\text{cusp}},\gamma^{\mRS}_{W_{\{\textrm{DY, H}\}}} \, ,\\
\text{jet function} \quad&\Rightarrow\quad
\gamma^{\mRS}_{\text{cusp}},\gamma^{\mRS}_{J_{\{q,g\}}} \, ,
\end{align}
\end{subequations}
where $\gamma^{\mRS}_{W_{\{\textrm{DY, H}\}}}$ governs the single-logarithmic
evolution of the soft function for the case with an
initial quark and an anti-quark (Drell-Yan) or two initial gluons
(Higgs production), respectively. $\gamma^{\mRS}_{J_{\{q,g\}}}$ is
defined similarly via the jet function. In \DRED, one has to
distinguish the jet functions for $D$-dimensional gluons $\ghat$ and
$\epsilon$-scalars $\gtilde$ and the corresponding
$\gamma^{\mDRED}_{J_{\ghat}}$ and $\gamma^{\mDRED}_{J_{\epsilon}}$.
The present discussion applies to these two cases in an analogous way.

Thus, the cusp anomalous dimension $\gamma_{\textrm{cusp}}^{\mRS}$ and
its scheme dependence can be easily extracted independently either
from the soft or the jet functions.  The situation is slightly more
involved for the quark and the gluon 
anomalous dimensions where we need to exploit some known relations
between anomalous dimensions to determine $\gamma_q^{\mRS}$ and
$\gamma_g^{\mRS}$. In the case of Drell-Yan and Higgs production, these
relations hold as a consequence of the factorization of the cross
section in the threshold region \cite{Becher:2007ty}. In particular
one finds
\begin{align}\label{eq:andim1}
\gamma^{\mRS}_{W_{\{\textrm{DY, H}\}}}
= 2 \gamma^{\mRS}_{\phi_{\{q, g\}}}+2 \gamma^{\mRS}_{\{q,g\}}\, ,
\end{align}
where $\gamma_{\phi_{\{q, g\}}}^{\mRS}$ is one half the
coefficient of the $\delta(1-x)$ term in the Altarelli-Parisi
splitting functions and controls the parton distribution functions
(PDFs) evolution. A similar relation involving the jet anomalous dimension
instead of the soft anomalous dimension is found for DIS \cite{Becher:2006mr}
\begin{align}\label{eq:andis}
\gamma^{\mRS}_{\phi_{\{q, g\}}}
= \gamma^{\mRS}_{J_{\{q,g\}}} - 2 \gamma^{\mRS}_{\{q,g\}}\, .
\end{align}
By combining Eq.~(\ref{eq:andim1}) with Eq.~(\ref{eq:andis}) to eliminate the universal
PDF anomalous dimension one obtains \cite{Becher:2007ty}
\begin{align}\label{eq:andim2}
\gamma^{\mRS}_{\{q,g\}}
= \gamma^{\mRS}_{J_{\{q,g\}}}
  -\frac{\gamma^{\mRS}_{W_{\{\textrm{DY, H}\}}}}{2}\, ,
\end{align}
The validity of
Eq.~(\ref{eq:andis}) is a consequence of the factorization theorem for
deep-inelastic scattering in the threshold region.  The factorization
proof is explicitly derived in \cite{Becher:2006mr} only for the quark
current. Nevertheless by replacing the photon with a Higgs boson and
after integrating out the heavy top loop, the factorization theorem
for a gluon current follows in total analogy to the quark case.
Indeed it can be explicitly checked that this relation holds both for
the quark and gluon cases up to two-loop order by directly
substituting the known expressions for the anomalous dimensions in \CDR.

Before we turn to the evaluation of the various anomalous dimensions
we introduce some notation. As explained in Section~\ref{sec:dred} the
anomalous dimensions in \FDH\ and \DRED\ are equal, except for the
appearance of the additional $\bar\gamma_\epsilon \equiv
\gamma^\mDRED_{\gtilde}$, see
\Eqns{eq:Defgammabar}{eq:Defgammaeps}. Likewise, the anomalous
dimensions in \CDR\ and \HV\ are equal. Thus, we will drop the label
\RS\ whenever possible and denote \FDH/\DRED\ quantities with a bar,
schematically
\begin{align}
\gamma\equiv\gamma^\mCDR&=\gamma^\mHV,&
\bar\gamma\equiv\gamma^\mFDH&=\gamma^\mDRED.
\label{barconvention}
\end{align}
In principle all perturbative expansions are carried out in terms of
the five couplings $\{\alpha\}$, as indicated in \Eqn{eq:generalex}.
However, for the results presented in this paper it is not necessary
to distinguish the various $\alpha_{4\epsilon,i}$. Therefore, a
coefficient in the perturbative expansion of the quantity $X$ will
have at most three labels, $X_{mnk}$, indicating the power of
$\alpha_s$, $\alpha_e$ and $\alpha_{4\epsilon}$, respectively. Very
often, the quantities do not depend on $\alpha_{4\epsilon}$, i.e. the
last of the three indices is zero. In this case we often drop this
label altogether and write the perturbative expansion with two labels
only by setting $X_{mn} = X_{mn0}$.%
\footnote{In the \CDR\ and \HV\ schemes,
all quantities of course only depend on $\alpha_s$. However, our
notation will be adapted for the cases of \FDH\ and \DRED, unless
noted otherwise.}

We mention two special cases. First, the $\beta$
functions are defined with a negative sign,
\begin{align}
\label{not:beta}
\beta^{s\,\mRS} = -
\sum_{mn} \left(\frac{\alpha_s}{4 \pi}\right)^{m}
\left(\frac{\alpha_e}{4 \pi}\right)^{n}
\beta^{s\,\mRS}_{mn} \, ,
\end{align}
so the one-loop renormalization factors of $\alpha_s$
and $\alpha_e$ in the various schemes are given by
\begin{subequations}
\begin{align}
Z^{\mRS}_{\alpha_s} &= 1
 - \beta^{s\,\mRS}_{20} \frac{\alpha_s}{4 \pi \epsilon} +
 \mathcal{O}(\alpha^2)\, \\[3pt]
Z^{\mRS}_{\alpha_e} &= 1
 - \beta^{e\,\mRS}_{11} \frac{\alpha_s}{4 \pi \epsilon} 
 - \beta^{e\,\mRS}_{02} \frac{\alpha_e}{4 \pi \epsilon} +
 \mathcal{O}(\alpha^2)\,
\end{align}
\end{subequations}
where the explicit form of the coefficients of the $\beta$ functions
are listed in Appendix~\ref{app:B}.
Second, we also introduce an abbreviation for the cusp anomalous
dimension multiplied with a colour factor,
\begin{align}
\label{def:Cusp}
\Gamma_{\text{cusp}}^\mRS \equiv C_R \, \gamma_{\text{cusp}}^\mRS
=  \sum_{mn} \left(\frac{\alpha_s}{4 \pi}\right)^{m}
\left(\frac{\alpha_e}{4 \pi}\right)^{n}
\Gamma_{mn}^\mRS\, ,
\end{align}
where the colour factor $C_R$ is either $C_F$ or $C_A$, depending on
the quantity under consideration. For brevity we
omit the superscript $\text{cusp}$ in the expansion
coefficients $\Gamma_{mn}^\mRS$ of $\Gamma_{\text{cusp}}^\mRS$.

\subsection{Computation and scheme dependence of the soft
  functions and  $\gamma_{W}$}
\label{sec:soft}

In this subsection we describe the calculation of the two-loop soft
functions for Drell-Yan and Higgs production in momentum space and the
extraction of the soft anomalous dimensions $\gamma_{W_{\textrm{DY}}}$
and $\gamma_{W_{\textrm{H}}}$ in the different regularization schemes
considered in this work.  In the partonic threshold region, where the
emitted gluons in the final state are soft, the Drell-Yan and Higgs
production hard-scattering kernels factorize into the product of soft
functions and hard functions. The factorization proof can be found in
\cite{Becher:2007ty, Becher:2014oda}. The soft functions describe the
real emission of soft gluons and contain singular distributions of the
gluon energy while the hard functions depend on the virtual
corrections and are regular functions of their variables.  The soft
matrix elements $\hat{W}_{\{\sss{DY,H}\}}(x)$ arise in the cross
section after the decoupling transformation which separates the soft
and collinear sectors in the leading power SCET Lagrangian.

The building blocks for the soft functions are the soft Wilson lines
\begin{align}
{\bf{S}}_i(x)=\mathcal{P} 
\exp\left(i g_s \int^{0}_{-\infty}ds \,
 n_i\cdot A^a_s(x+s n_i)\mathbf{T}^a_{i}\right)\, ,
\end{align}
where $A^a_s(x)$ is a soft gluon field in SCET and $n_i =\{n,\,
\bar{n}\}$ ($n_\mu=(1,0,0,1)$, $\bar{n}_{\mu}=(1,0,0,-1)$ are
light-like reference vectors in the direction of the two incoming
partons). The path-ordering acts on the colour generators
$\mathbf{T}^a_{i}$ in the representation appropriate for the $i$th
field. For the conjugate quark fields one finds
$\mathbf{T}^a_i=-(t^a)^T$ which turns into anti-path-ordering.  The
soft matrix elements $\hat{W}_{\{\sss{DY,H}\}}(x)$ are defined in
terms of a soft operator
\begin{align}
\mathbf{O}_s(x) = \left[{\bf S}_{\bar{n}} {\bf S}_n\right] (x)\, ,
\end{align}
as an expectation value of products of soft Wilson lines forming a closed Wilson loop
\begin{align}\label{eq:softdef}
\hat{W}_{\{\sss{DY,H}\}}(x) = 
\frac{1}{d_R}\mathrm{tr}\langle0|\bar{T}\big(\mathbf{O}^\dagger_s(x)\big) 
T \big(\mathbf{O}_s(0)\big)|0\rangle\, ,
\end{align}
where $d_R=N_c$ for Drell-Yan and $d_R=N_c^2-1$ for Higgs production,
$T$ and $\bar{T}$ are the time-ordering and anti-time-ordering
operators, respectively.

Since the collinear and soft sectors no longer interact, it is worth
noting that $\hat{W}_{\{\sss{DY,H}\}}(x)$ in \Eqn{eq:softdef}
still contains the information about the colour and the direction of
the initial quarks/gluons, but it is insensitive to the spin of the external
particles due to the eikonal approximation.  The soft function is
defined as the Fourier transform of the soft matrix element
$\hat{W}_{\{\sss{DY,H}\}}(x)$ in \Eqn{eq:softdef}:
\begin{align}\label{eq:fouriersoft}
S_{\{\sss{DY,H}\}}(\omega) = \int \frac{d x^0}{4 \pi} \, 
e^{i x^0 \omega/2} \, \hat{W}_{\{\sss{DY,H}\}}(x^0,\vec{x}=0) \, .
\end{align}
The Drell-Yan and Higgs production soft functions are closely related
to each other; up to NNNLO they differ by Casimir scaling replacements
\cite{Li:2014afw}. At NNLO the situation is even simpler and the
following replacement holds~\cite{Ahrens:2008nc}:
\begin{align}\label{eq:dytoh}
S_{\mathrm{H}}(\omega) = 
S_{\mathrm{DY}}(\omega)\big|_{C_F\to C_A} + \mathcal{O}(\alpha^3_s)\, .
\end{align} 
Thus, we directly compute the soft function for Drell-Yan and
obtain the Higgs soft function by using \Eqn{eq:dytoh}. In the
\DRED\ scheme the soft function for external $\epsilon$-scalars is also
needed. Since soft gluon interactions are insensitive to the spinorial
structure of the external particles, it turns out that the soft
function for external $\epsilon$-scalars is the same as the one for
external gluons. Therefore we will not discuss it further.

\begin{figure}
\begin{center}
\begin{align}
\includegraphics[width=5cm]{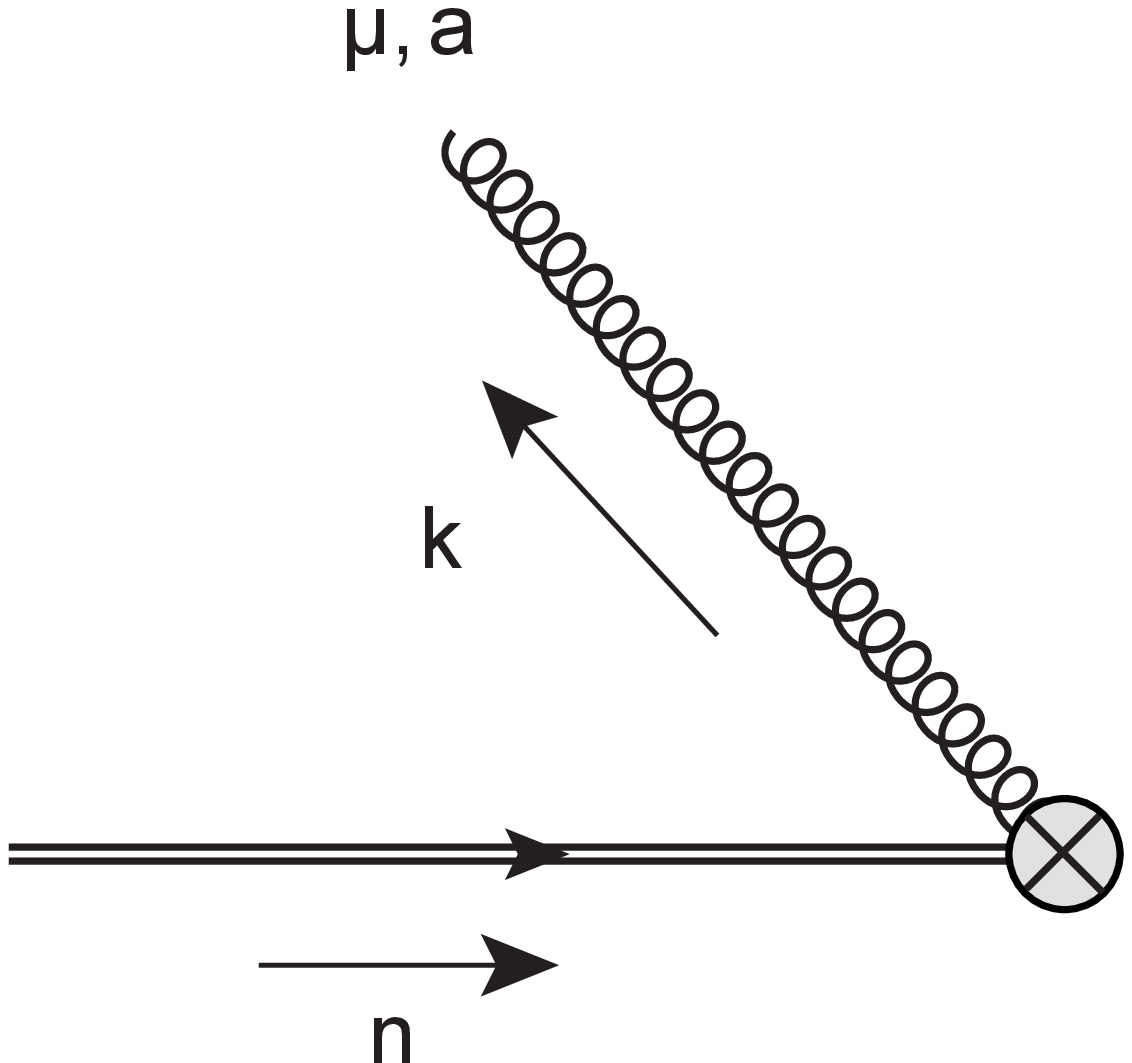}
&
\begin{aligned}
\notag
\quad\rightarrow\quad g_s {\Large \frac{n^{\mu}}{n \cdot k}}\,\mathbf{T}^a
\\[3cm]
\end{aligned}
\\
\includegraphics[width=5cm]{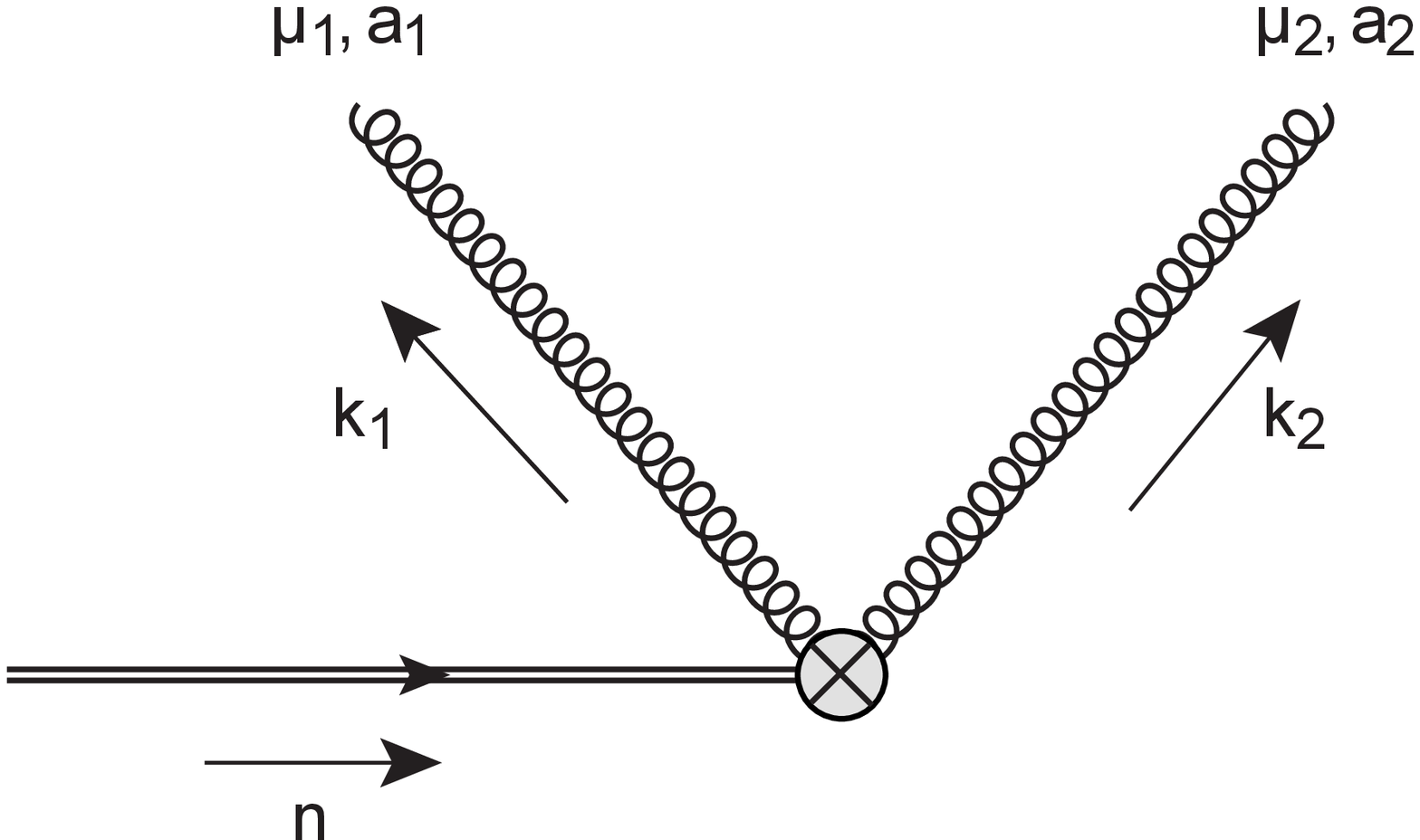}
&
\begin{aligned}
\notag
\quad\rightarrow\quad g_s^2\, n^{\mu_1} n^{\mu_2}\,
{\Large \left[\frac{\mathbf{T}^{a_1} \mathbf{T}^{a_2}}{n\cdot k_2 \,
n\cdot(k_1+k_2)} + \frac{\mathbf{T}^{a_2} \mathbf{T}^{a_1}}{n\cdot k_1 \,
n\cdot(k_1+k_2)}\right]}
\\[3cm]
\end{aligned}
\end{align}
\end{center}
\vspace*{-1cm}
\caption{\label{fig:fw} Feynman rules for the emission of one and two
  gluons from a Wilson line. Figure taken from \cite{Becher:2014oda}.}
\end{figure}

In momentum space it is more convenient to rewrite the soft function
in \Eqn{eq:fouriersoft} as a squared amplitude by inserting a
complete set of states
\begin{align}\label{eq:softsquared}
S(\omega) = \frac{1}{d_R} \sum_{X_{s}}  \, {\rm tr} 
\langle 0 | \bar{T}\big(\mathbf{O}^\dagger_s(0)\big) | X_{s}\rangle 
\langle X_{s} | T \big(\mathbf{O}_s(0)\big) | 0 \rangle 
\delta(\omega - 2 E_{X_s}) \, ,
\end{align}
where $X_{s}$ refers to a final state made of unobserved soft gluons
carrying energy $E_{X_s}$. For simplicity in \Eqn{eq:softsquared} we
drop the subscripts $\{\mathrm{DY,H}\}$.  To perform this calculation,
we need not only the usual QCD Feynman rules but also the
momentum-space Feynman rules for gluons emitted from Wilson lines up
to $\mathcal{O}(\alpha^2_s)$. We report them in Figure~\ref{fig:fw}.

The $\mathcal{O}(\alpha^2_s)$ \cite{Belitsky:1998tc} Drell-Yan soft
functions in the \CDR\ scheme have been originally calculated in
position space directly from the definition in \Eqn{eq:softdef}. An
exclusive soft function for Drell-Yan at $\mathcal{O}(\alpha^2_s)$ has
been computed in \cite{Li:2011zp}.
The state of the art $\mathcal{O}(\alpha^3_s)$ soft functions for
Higgs and Drell-Yan production have been computed very recently in a
series of papers \cite{Li:2013lsa,Li:2014bfa,Li:2014afw}.  We also
mention that related soft functions for thrust distribution and
N-jettiness have been computed at $\mathcal{O}(\alpha^2_s)$ in
\cite{Monni:2011gb,Kelley:2011ng} and \cite{Boughezal:2015eha}
respectively.

In order to study the higher-order corrections of the soft functions
in the  regularization schemes different from \CDR\ we define expansion
coefficients of the perturbative series as
\begin{align}\label{eq:softexpansion}
S_{\mathrm{bare}}^{\mRS} (\omega) = \delta(\omega) 
+ a_s(\omega) \, S^{\mRS}_{10}(\omega) 
+ a_s^2(\omega)\, S^{\mRS}_{20}(\omega)+ \ldots\, ,
\end{align}
where we have introduced the superscript $\mRS$ to indicate the scheme
dependence.  In the above equation we have introduced 
\begin{align}
\label{eq:asdef}
a_s(\omega) \equiv e^{-\epsilon \gamma_E} (4\pi)^\epsilon\, 
\left(\frac{1}{\omega^2}\right)^\epsilon
\frac{\alpha^{\text{bare}}_s}{4\pi} 
= \left(\frac{\mu^2}{\omega^2}\right)^\epsilon
\frac{Z^{\mRS}_{\alpha_s} \alpha_s}{(4\pi)}
\end{align}
and expressed the bare coupling $\alpha^{\mathrm{bare}}_s$ in terms of
the renormalized coupling $\alpha_s\equiv \alpha_s(\mu)$ in the
$\overline{\text{MS}}$ scheme.  Note that $a_s(\omega)$ and
$\alpha^{\text{bare}}_s$ are actually scheme independent, but if
expressed in terms of the $\overline{\text{MS}}$ coupling
$\alpha_s(\mu)$ depend on the scheme-dependent renormalization factor
$Z^{\mRS}_{\alpha_s}$.  The all-order bare soft function in
\Eqn{eq:softexpansion} is independent of the renormalization scale
$\mu$. Up to NNLO the soft function depends only on $\alpha_s$ and not
on $\alpha_e$ or $\alpha_{4\epsilon}$.

At NLO only two diagrams contribute to the soft functions; they
describe the real emission of one soft gluon from the Wilson lines. At
NLO the bare soft function turns out to be scheme independent,
\begin{align}
\bar{S}_{10}(\omega) &= \frac{8}{\omega} 
 C_R \frac{e^{\epsilon \gamma_E}\Gamma (-\epsilon)}{\Gamma (1-2 \epsilon)}\, .
\end{align}
As a result, the soft anomalous dimensions must be scheme independent,
too. This reproduces the well-known fact that $\gamma_{\text{cusp}}$
is scheme independent at NLO, and it implies $\gamma^{W\,\mRS}_{10}=0$
in all \RS. The reason is that for the \FDH\ and \DRED\ schemes there
are no additional diagrams involving $\epsilon$-scalars compared to
\CDR\ and \HV. This is a consequence of the fact that dot products of
a $\epsilon$-scalar field $\tilde{A}$ with the vectors $n$, $\bar{n}$
are vanishing, i.e. $n\cdot \tilde{A}=\bar{n}\cdot \tilde{A}=0$. It
follows that soft $\epsilon$-scalars cannot be emitted from the Wilson
lines.  This explains in a direct way the result \cite{Signer:2008va}
that the scheme dependence of general NLO amplitudes is contained in
the parton anomalous dimensions.

\begin{figure}[t]
\begin{center}
\begin{tabular}{cc}
\includegraphics[width=4cm]{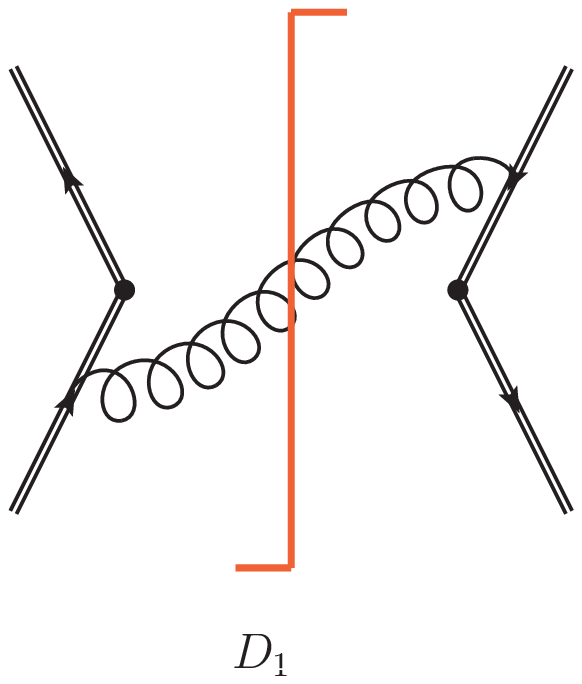} & \includegraphics[width=4cm]{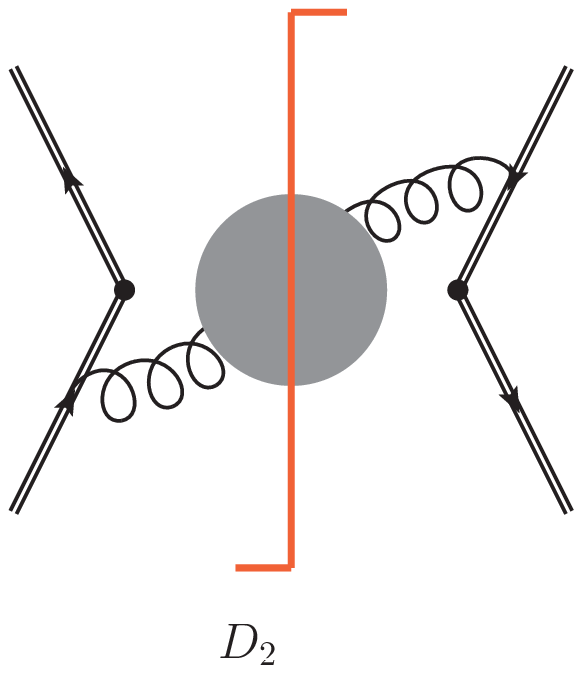} \\
\includegraphics[width=4cm]{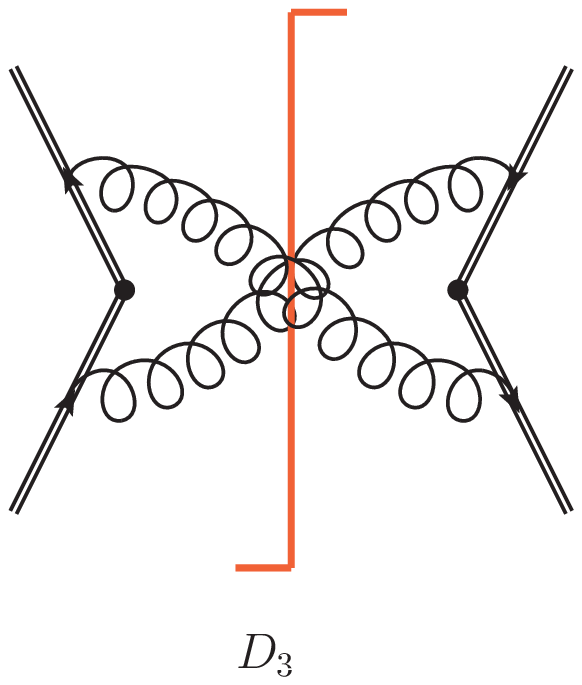} & \includegraphics[width=4cm]{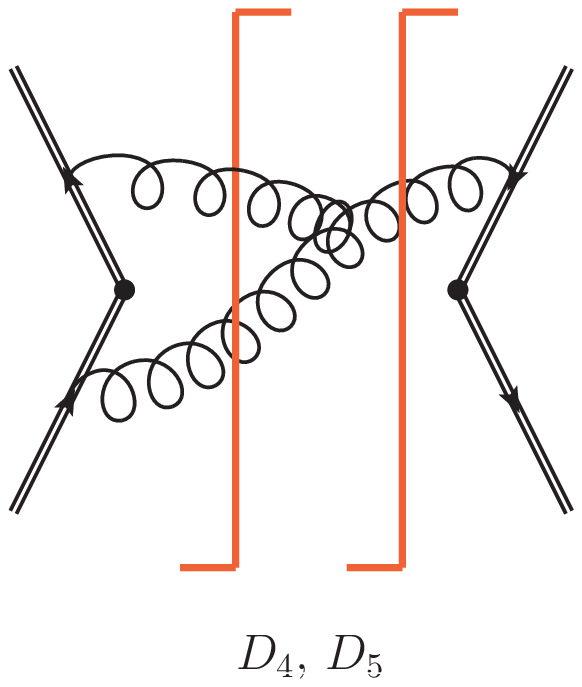}
\end{tabular}
\end{center}

\caption{Selected non-zero Feynman diagrams contributing to the
  one-loop and two-loop soft functions. A complete list of diagrams
  can be found in \cite{Belitsky:1998tc}. Double lines indicate the
  direction of Wilson lines while the red vertical cut indicates
  on-shell partons. The scheme dependence originates from the diagram
  $D_2$. Diagrams $D_2$, $D_3$ and $D_4$ represent double real soft
  emissions while diagram $D_5$ represents a single virtual-real
  emission.  }
\label{fig:diagrams}
\end{figure}

At NNLO the situation is more involved; diagrams with two real soft
emissions and virtual diagrams with one real soft emission are
present.  The soft functions and soft anomalous dimensions at NNLO
have a scheme dependence, which originates from the $\epsilon$-scalar
cut bubble contributing to the second diagram in
Figure~\ref{fig:diagrams}. The grey blob represents the quark, gluon,
ghosts and $\epsilon$-scalar contributions. The latter is present only
in \FDH\ and \DRED.  After calculating the non-vanishing integrals
using the techniques described in
\cite{Becher:2012za,Ferroglia:2012uy} and summing all the
contributions we obtain the NNLO coefficient in \Eqn{eq:softexpansion}
in \FDH/\DRED,
\begin{align}
\bar{S}_{20}(\omega) &=  \frac{1}{\omega}
 C_R \left[C_A\, \bar{S}_A+ N_F T_R\, \bar{S}_f 
  + C_R\, \bar{S}_R\right]\label{eq:soft2lfdh}\, ,
\end{align}
with
\begin{subequations}
\begin{align}
\bar{S}_A &=
\frac{1}{\epsilon^2}\left(-\frac{44}{3}+\frac{2 N_\epsilon}{3}\right) 
+ \frac{1}{\epsilon}\left(\frac{16 N_{\epsilon}}{9}+\frac{4 \pi^2}{3}
-\frac{268}{9}\right) -\frac{7 \pi^2 N_{\epsilon}}{9}
+\frac{104 N_{\epsilon}}{27}\, \nonumber\\
&+56 \zeta_3 + \frac{154 \pi^2}{9}-\frac{1616}{27} 
+ \bigg(-\frac{124 N_{\epsilon} \zeta_3}{9}-\frac{56\pi^2 N_\epsilon}{27}
\nonumber \\
&+\frac{640 N_\epsilon}{81}+\frac{2728 \zeta_3}{9}-\frac{4 \pi ^4}{9}
+\frac{938 \pi^2}{27}-\frac{9712}{81}\bigg) \epsilon 
+ \mathcal{O}(\epsilon^2)\, ,\\
\bar{S}_f &= \frac{16}{3 \epsilon^2}+\frac{80}{9 \epsilon}
-\frac{56 \pi^2}{9}+\frac{448}{27} + 
\left(-\frac{992\zeta_3}{9}+\frac{2624}{81}-\frac{280\pi^2}{27}
\right) \epsilon+ \mathcal{O}(\epsilon^2)\, ,\\
\bar{S}_R &= -\frac{32}{\epsilon^3} 
+\frac{112 \pi^2}{3 \epsilon} + \frac{1984 \zeta_3}{3} 
+\frac{4 \pi^4 \epsilon}{5} + \mathcal{O}(\epsilon^2)\, ,
\end{align}
\end{subequations}
where $C_R=C_F$ for Drell-Yan and $C_R=C_A$ for Higgs production.

We now turn to the determination of the soft and cusp anomalous
dimension from the soft function. In order to do this we need to
discuss the singularities of the soft function that remain after
coupling renormalization. From the point of view of ordinary QCD
computations, these remaining singularities are closely related to IR
singularities. However, from the SCET point of view they simply
correspond to UV singularities and are to be removed by
renormalization within the effective theory. For convenience this is
done in Laplace space by introducing the Laplace transformed soft
function as
\begin{align}
\label{soft:Lap}
s^{\mRS}(\kappa) = \int^\infty_0 d\omega  
\exp\left(-\frac{\omega}{\kappa\, e^{\gamma_E}}\right) S^{\mRS}(\omega)\, ,
\end{align}
where the integral transform can be easily carried out by using the relation
\begin{align}
\int^\infty_0 d\omega \exp\left(-b \omega\right) \omega^{-1-n\epsilon} 
= \Gamma(-n \epsilon) b^{n \epsilon}\, .
\end{align}
The remaining UV divergences of the soft function can be subtracted
multiplicatively, 
\begin{align}
\label{eq:renS}
s_{\mathrm{sub}}^{\mRS}(\kappa,\mu)=
Z^{\mRS}_s(\kappa,\mu)\, s_{\mathrm{bare}}^{\mRS}(\kappa)\, .
\end{align} 

Like in the case of general amplitudes in \Eqn{eq:rge} and
\Eqn{rel:ZGamma}, the RGE
\begin{align}
\label{rge:softZ}
\frac{d}{d\ln\mu}\, s_{\mathrm{sub}}^{\mRS}(\kappa,\mu)
= \frac{d\, Z^{\mRS}_s(\kappa,\mu)}{d\ln\mu} 
\big(Z^{\mRS}_s(\kappa,\mu)\big)^{-1} \, 
s_{\mathrm{sub}}^{\mRS}(\kappa,\mu)\, 
\end{align}
holds, and the corresponding anomalous dimension has a structure
similar to \Eqn{eq:conj},
\begin{align}\label{rge:softG}
\frac{d}{d \ln \mu}  s_{\mathrm{sub}}^{\mRS}(\kappa,\mu) 
= \left[-4 \, \Gamma^{\mRS}_{\mathrm{cusp}} \, L_\kappa
-2 \gamma^{\mRS}_{W}\right] s_{\mathrm{sub}}^{\mRS}(\kappa,\mu)\, ,
\end{align}
which is derived from the RG invariance of the cross sections in the
threshold region in analogy to the \CDR\ case in
Ref.~\cite{Becher:2007ty}.  In \Eqn{rge:softG} we have defined
$L_\kappa \equiv \ln(\kappa/\mu)$ and $C_R=C_F$ for Drell-Yan and
$C_R=C_A$ for Higgs production. Comparison of the previous two
equations yields an expression for the \FDH\ renormalization factor
$\bar{Z}_s(\kappa,\mu) \equiv Z^{\mFDH}_s(\kappa,\mu)$ in terms of the
soft and cusp anomalous dimensions. This expression has the same
structure as \Eqn{eq:lnZfullRS}, but can be written in a simpler form
because up to NNLO the soft function does not depend on $\alpha_e$ and
$\alpha_{4\epsilon}$:
\begin{align}\label{eq:lnz}
\ln \bar{Z}_s &= \left(\frac{\alpha_s}{4 \pi}\right)\left[
-\frac{\bar\Gamma_{10}}{\epsilon^2}
+\frac{1}{\epsilon}\left(2 \bar\Gamma_{10} L_\kappa 
+\bar\gamma^{W}_{10}\right)\right]\,\\
&+ \left(\frac{\alpha_s}{4 \pi}\right)^2 \left[
\frac{3 \bar\beta^{s}_{20} \bar\Gamma_{10}}{4 \epsilon^3}
-\frac{\bar\beta^{s}_{20}}{2 \epsilon^2}
 \left(2 \bar\Gamma_{10} L_\kappa  +\bar\gamma^{W}_{10}\right)
-\frac{\bar\Gamma_{20}}{4 \epsilon^2}
+\frac{1}{2 \epsilon}\left(2 \, \bar\Gamma_{20} L_\kappa 
+ \bar\gamma^{W}_{20}\right)\right]\, \nonumber\\
&  + \mathcal{O}(\alpha^3_s)\, . \nonumber
\end{align}
By requiring that the renormalization factor $\bar{Z}_s$ in
\Eqn{eq:lnz} minimally subtracts all of the divergences of the bare
soft function (in \FDH, treating $N_\epsilon$ as an independent
multiplicity), we extract the expressions for the anomalous dimensions
in the \FDH\ scheme
\begin{subequations}
\label{eq:softgammas}
\begin{align}
\bar\Gamma_{\text{cusp}} &= 
\left(\frac{\alpha_s}{4\pi}\right) C_R \,(4)
\nonumber \\
&+ \Big{(}\frac{\alpha_s}{4\pi}\Big{)}^2\,
C_R \Big[C_A\Big{(}\frac{268}{9} - \frac{4}{3}\pi^2\Big{)} 
- \frac{80}{9} T_R N_F-N_\epsilon\frac{16}{9} C_A \Big]  +
\mathcal{O}(\alpha^3) \, , \\
\bar{\gamma}_{W} & =
\Big{(}\frac{\alpha_s}{4\pi}\Big{)}^2
C_R 
\bigg[C_A \Big{(}-\frac{808}{27} + \frac{11}{9}\pi^2 +28 \zeta_3
  + N_\epsilon\frac{52}{27} - N_\epsilon \frac{\pi ^2}{18} \Big{)} 
+  T_R N_F \Big{(}\frac{224}{27} - \frac{4}{9}\pi^2\Big{)} \bigg] 
\nonumber \\
& + \mathcal{O}(\alpha^3) \, .
\label{eq:gammaW}
\end{align}
\end{subequations}
The fact that $\bar{\Gamma}_{\rm cusp}=C_R\bar\gamma_{\textrm{cusp}}$,
with the known expression of the cusp anomalous dimension in the
\FDH\ scheme, $\bar\gamma_{\textrm{cusp}}$, is a consistency check of
the method. $\bar{\gamma}_{W}$ is a new result.  The corresponding
expressions in \CDR/\HV\ can be obtained by simply using the
appropriate $\beta$ functions and anomalous dimensions in \Eqn{eq:lnz}
and by setting $\Neps=0$ in \Eqn{eq:softgammas}. They are consistent
with the literature~\cite{Becher:2007ty}.

Finally we remark that in analogy to \Eqn{eq:fin} we can obtain a
finite and scheme independent soft function $s_{\mathrm{fin}}$ through
\begin{align}
\label{eq:Sfin}
s_{\mathrm{fin}}(\kappa,\mu) = 
\lim\limits_{(N)_\epsilon\to 0} s_{\mathrm{sub}}^{\mRS}(\kappa,\mu)\, .
\end{align} 
The explicit expression for $s_{\mathrm{fin}}$ is given in
\Eqn{eq:sfinres} of Appendix~\ref{app:A}.

\subsection{Computation and scheme dependence of the quark
  jet function and  $\gamma_{Jq}$}

The quark jet function has been calculated at NNLO in
\CDR~\cite{Becher:2006qw}. Referring to \cite{Becher:2006qw} for more
details,  we describe here the corresponding calculation in
\FDH\ (which is identical to the one in \DRED, but for simplicity we
will only refer to \FDH\ in the present subsection).  The jet function is 
given in terms of the hard-collinear quark propagator
\begin{align}
\frac{\sla{n}}{2}\;\bar{n}\cdot p\,\mathcal{J}^{\mRS}_q(p^{2})&=
\int d^{4}x\, e^{ipx}\langle 0
|T\{\chi_{hc}(x)\bar{\chi}_{hc}(0)\} | 0\rangle 
\nonumber \\
&=\int d^{4}x\, e^{ipx}\langle 0|T\{\frac{\sla{n}\,\sla{\bar{n}}}{4}
W^{\dagger}(x)\psi (x)\bar{\psi}(0)W(0)
\frac{\sla{\bar{n}}\,\sla{n}}{4}\} | 0\rangle\label{jet1}\, ,
\end{align}
with Wilson lines 
\begin{equation}
W(x)=\mathcal{P}\,\exp\Big{(}ig_s\int_{-\infty}^{0}ds\, \bar{n}\,\cdot 
\,A(x+s\bar{n})\Big{)}\, ,\label{jet2}
\end{equation}
where $A^\mu = A^\mu_a\, t^a$.  The field $\chi_{hc}(x)$ is the
gauge-invariant (under both soft and hard-collinear gauge
transformations) effective-theory field for a massless quark after a
decoupling transformation has been applied, which removes the
interactions of soft gluons with hard-collinear fields in the
leading-power SCET Lagrangian.  As shown in \Eqn{jet1}, we can rewrite
the propagator in terms of standard QCD fields.

\begin{figure}[t]
\begin{center}
\includegraphics [width=0.6\textwidth]{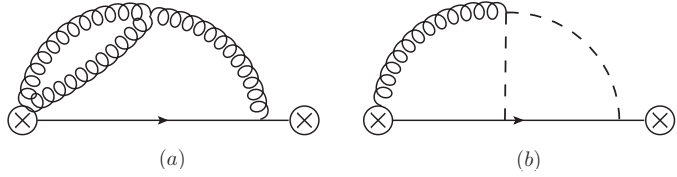}
\caption{Examples of two-loop diagrams contributing to the quark jet
  function. Gluons emitted from the crossed circles originate from the
  Wilson lines. Diagram (a) contributes in \CDR\ and \FDH, whereas
  diagram~(b) with two $\epsilon$-scalars contributes only in \FDH.}
\label{jetfig1}
\end{center}
\end{figure}

The hard-collinear quark propagator $\mathcal{J}_q^{\mRS}$ as defined
in \Eqn{jet1} is scheme dependent. The
fields $\chi_{hc}$ and $\psi$ on the r.h.s.\ of \Eqn{jet1}  are
Heisenberg fields, so applying the
usual perturbative expansion results in loop diagrams
contributing to the propagator. The scheme dependence is related to
UV singularities of such diagrams.  Examples of two-loop diagrams are
shown in Figure~\ref{jetfig1}. In \FDH\ the computation is similar to
the \CDR\ scheme. However there are additional diagrams, which include
the $\epsilon$-scalars and also depend on the coupling
$\alpha_{e}$. An example of a two-loop diagram needed for the jet
function in \FDH\ (and not present in the \CDR\ scheme) is shown in
Figure~\ref{jetfig1} $(b)$. Since $\bar{n}$ is a $D$-dimensional
vector, there are no $\epsilon$-scalars originating from the Wilson
lines.  Indeed, the scalar product in \Eqn{jet2} will vanish in the
case of the $\epsilon$-scalar.

The jet function $J_q^{\mRS}(p^{2})$ is the discontinuity of the propagator, i.e.
\begin{equation}
J_q^{\mRS}(p^{2})=
\frac{1}{\pi}\text{Im}\Big{[}i\,\mathcal{J}_q^{\mRS}(p^{2})\Big{]}\, .
\end{equation}
To highlight the similarities with the discussion in
Section~\ref{sec:schemes} and the soft function it is convenient to
work in Laplace space, so we define $\jmath_q^{\mRS}(Q^2)$, the
Laplace transform of the jet function as
\begin{equation}\label{qjet:Lap}
\jmath_q^{\mRS}(Q^2) \equiv \int_0^\infty d p^2\, 
\exp\left(-\frac{p^2}{Q^2 e^{\gamma_E}}\right) J_q^{\mRS}(p^{2})\, .
\end{equation}
The analogous equation in the case of the soft function is \Eqn{soft:Lap}.

To compute the propagator in the \FDH\ scheme,
$\bar{\mathcal{J}}_q(p^{2})$, the diagrams have been generated with
QGRAF~\cite{Nogueira:1991ex} and the colour algebra has been done with
ColorMath~\cite{Sjodahl:2012nk}. For the reduction of the integrals
Reduze~2~\cite{vonManteuffel:2012np} has been used. The master
integrals needed for the \FDH\ jet function are the same as for the
\CDR\ scheme. After taking the imaginary part and performing the
Laplace transform, the bare quark jet function at NNLO in \FDH\ is
obtained as
\begin{align}\label{eq:Qbare}
\nonumber
\bar{\jmath}_{q\,\text{bare}}(Q^2)&=1
+ a_{s}(Q^2)\, C_F\, \Big{(}\frac{4}{\epsilon^{2}}+\frac{3}{\epsilon}
  +7-\frac{2\,\pi^{2}}{3}+\epsilon\big{(}14-\frac{\pi^{2}}{2}
  -8\zeta_{3}\big{)}\Big{)}
 \\
&+a_{e}(Q^2)\, C_F\,\Neps\, \Big{(}-\frac{1}{2\,\epsilon}-1
  +\epsilon\big(-2+\frac{\pi^{2}}{12}\big)\Big{)}
\nonumber\\
&+  a_{s}^{2}(Q^2)\, \Big(
  C^2_{F}\,\bar{\jmath}^{\, q;F}_{20}
+ C_{F}C_{A}\,\bar{\jmath}^{\, q;A}_{20}
 + C_{F}T_{R}N_F\,\bar{\jmath}^{\, q;f}_{20}\Big{)}
 \phantom{\frac{1}{1}}\nonumber\\
&+ a_{e}^{2}(Q^2)\, \Big{(}
  C^2_{F}\, \bar{\jmath}^{\, q;F}_{02}
+ C_{F}C_{A}\, \bar{\jmath}^{\, q;A}_{02}
  + C_{F}T_{R} N_F\, \bar{\jmath}^{\, q;f}_{02}\Big{)}
  \phantom{\frac{1}{1}}\nonumber\\
&+ a_{s}(Q^2)a_{e}(Q^2)\, \Big{(}
  C^2_{F}\,\bar{\jmath}^{\, q;F}_{11}
+C_{F}C_{A}\,\bar{\jmath}^{\, q;A}_{11}\Big)
+{\cal O}(a^3)\, .\phantom{\frac{1}{1}}
\end{align}
In analogy to \Eqn{eq:asdef} we have defined
\begin{align}
a_s(Q^2) \equiv e^{-\epsilon \gamma_E} (4\pi)^\epsilon\, 
\left(\frac{1}{Q^2}\right)^\epsilon
\frac{\alpha^{\text{bare}}_s}{4\pi}
= \left(\frac{\mu^2}{Q^2}\right)^\epsilon
\frac{\bar{Z}_{\alpha_s} \alpha_s}{(4\pi)}\, .
\end{align}
with an analogous equation for $a_e$. The explicit expression for the
two-loop coefficients are given in Appendix~\ref{app:A}. Note that
$\bar{\jmath}_{q\,\text{bare}}(Q^2)$ is independent of $\mu$.

The renormalization procedure in any regularization scheme can easily
be generalized from the corresponding procedure in
\CDR~\cite{Becher:2006qw}. A renormalization factor
$Z_{J_q}^{\mRS}(Q^2,\mu)$ absorbing the UV divergences of the bare jet
function is introduced such that
\begin{equation}\label{eq:ren}
\jmath_{q\, \text{sub}}^{\mRS}(Q^2,\mu)= 
Z_{J_q}^{\mRS}(Q^2,\mu)\, \jmath_{q\, \text{bare}}^{\mRS}(Q^2)
\end{equation}
is finite. This equation is analogous to \Eqns{eq:zcdr}{eq:renS}.
Requiring minimal subtraction with $N_\epsilon$ as an independent
multiplicity determines the explicit form of $Z_{J_q}^{\mRS}(Q^2,\mu)$
uniquely in terms of the bare quark jet function
$\bar{\jmath}_{q\,\text{bare}}(Q^2)$. In principle, $Z_{J_q}^{\mRS}$
depends on all couplings $\{\alpha\}$. However, in \FDH, up to NNLO
there is no dependence on $\alpha_{4\epsilon}$.

To relate $Z_{J_q}^{\mRS}(Q^2,\mu)$ to the cusp anomalous dimension
$\gamma_{\textrm {cusp}}^{\mRS}$ and the quark jet anomalous dimension
$\gamma_{J_q}^{\mRS}$ we follow the same procedure as for the soft
anomalous dimension. We compare the RGE of the quark jet function in
the form
\begin{align}
\label{rge:qjetZ}
\frac{d}{d\ln\mu}\, \jmath_{q\,\text{sub}}^{\mRS}(Q^2,\mu)
= \frac{d\,  Z^{\mRS}_{J_q}(Q^2,\mu)}{d\ln\mu}
\left(Z^{\mRS}_{J_q}(Q^2,\mu)\right)^{-1} \,  
\jmath_{q\,\text{sub}}^{\mRS}(Q^2,\mu)\, 
\end{align} 
to the RGE  written in terms of
$\Gamma_{\textrm{cusp}}^{\mRS}$ and $\gamma_{J_q}^{\mRS}$,
\begin{align} 
\label{rge:qjetG}
\frac{d}{d \ln \mu}\jmath_{q\,\text{sub}}^{\mRS}(Q^2,\mu)&=
\left[-2\Gamma_{\text{cusp}}^{\mRS}\, L_Q 
 - 2\gamma_{J_q}^{\mRS} \right]
\jmath_{q\,\text{sub}}^{\mRS}(Q^2,\mu) \, .
\end{align}
This relation is analogous to \Eqns{eq:rge}{rge:softG}; we have used
$L_Q\equiv \ln(Q^2/\mu^2)$ and $\Gamma_{\text{cusp}}^{\mRS} = C_F
\gamma_{\text{cusp}}^{\mRS}$. With the help of
\Eqns{rge:qjetZ}{rge:qjetG} we can express $\bar{Z}_{J_q}$ in terms of
the \FDH\ anomalous dimensions. Up to NNLO, the expression for
$\ln\,\bar{Z}_{J_q}$ has the same structure as
\Eqns{eq:lnZfullRS}{eq:lnz}. We write it explicitly, using that up to
NNLO only the two couplings $\alpha_s$ and $\alpha_e$ appear:
\begin{align}
\ln\bar{Z}_{J_q}
&=\frac{\alpha_{s}}{4\pi}\Big{[}-\frac{\bar{\Gamma}_{10}}{\epsilon^{2}}
+\frac{1}{\epsilon}\Big{(}\bar{\Gamma}_{10}\,L_Q+\bar{\gamma}_{10}^{J_q}\Big{)}\Big{]}
+\frac{\alpha_{e}}{4\pi}\Big{[}-\frac{\bar{\Gamma}_{01}}{\epsilon^{2}}
+\frac{1}{\epsilon}\Big{(}\bar{\Gamma}_{01}\,L_Q+\bar{\gamma}_{01}^{J_q}\Big{)}\Big{]}
\nonumber\\
&+\big{(}\frac{\alpha_{s}}{4\pi}\big{)}^{2}
\Big{[}\,\frac{3\,\big{(}\bar\beta^s_{20}\bar{\Gamma}_{10}
+\bar{\beta}^{e}_{20}\bar{\Gamma}_{01}\big{)}}{4\epsilon^{3}}
-\frac{\bar\beta^s_{20}}{2\,\epsilon^{2}}\Big{(}\bar{\Gamma}_{10}\,L_Q
+\bar{\gamma}_{10}^{J_q}\Big{)}-\frac{\bar\beta^e_{20}}{2\,\epsilon^{2}}
\Big{(}\bar{\Gamma}_{01}\,L_Q+\bar{\gamma}_{01}^{J_q}\Big{)}\nonumber\\
&\qquad\qquad-\frac{\bar{\Gamma}_{20}}{4\,\epsilon^{2}}
+\frac{1}{2\,\epsilon}
\Big{(}\bar{\Gamma}_{20}\,L_Q+\bar{\gamma}_{20}^{J_q}\Big{)}\Big{]}
\nonumber\\
&+\big{(}\frac{\alpha_{e}}{4\pi}\big{)}^{2}
\Big{[}\,\frac{3\,\big{(}\bar\beta^s_{02}\bar{\Gamma}_{10}
+\bar{\beta}^{e}_{02}\bar{\Gamma}_{01}\big{)}}{4\epsilon^{3}}
-\frac{\bar\beta^s_{02}}{2\,\epsilon^{2}}
\Big{(}\bar{\Gamma}_{10}\,L_Q+\bar{\gamma}_{10}^{J_q}\Big{)}
-\frac{\bar\beta^e_{02}}{2\,\epsilon^{2}}
\Big{(}\bar{\Gamma}_{01}\,L_Q+\bar{\gamma}_{01}^{J_q}\Big{)}\nonumber\\
&\qquad\qquad-\frac{\bar{\Gamma}_{02}}{4\,\epsilon^{2}}
+\frac{1}{2\,\epsilon}
\Big{(}\bar{\Gamma}_{02}\,L_Q+\bar{\gamma}_{02}^{J_q}\Big{)}\Big{]}
\nonumber\\
&+\left(\frac{\alpha_{s}}{4\pi}\right)\left(\frac{\alpha_{e}}{4\pi}\right)
\Big{[}\,\frac{3\,\big{(}\bar\beta^s_{11}\bar{\Gamma}_{10}
+\bar{\beta}^{e}_{11}\bar{\Gamma}_{01}\big{)}}{4\epsilon^{3}}
-\frac{\bar\beta^s_{11}}{2\,\epsilon^{2}}
\Big{(}\bar{\Gamma}_{10}\,L_Q+\bar{\gamma}_{10}^{J_q}\Big{)}
-\frac{\bar\beta^e_{11}}{2\,\epsilon^{2}}
\Big{(}\bar{\Gamma}_{01}\,L_Q+\gamma_{01}^{J_q}\Big{)}\nonumber\\
&\qquad\qquad-\frac{\bar{\Gamma}_{11}}{4\,\epsilon^{2}}
+\frac{1}{2\,\epsilon}
\Big{(}\bar{\Gamma}_{11}\,L_Q+\bar{\gamma}_{11}^{J_q}\Big{)}\Big{]}
+{\cal O}(\alpha^3)\,.
\label{eq:logZ}
\end{align}
On the one hand this formula gives strong consistency checks. It allows 
for an independent extraction of the cusp anomalous dimension and the
coefficients of the $\beta$ functions of $\alpha_s$ and $\alpha_e$ in
the \FDH\ scheme. These coefficients agree with the well-known results
in the literature \cite{Kilgore:2012tb,Gnendiger:2014nxa}.

On the other hand, comparing \Eqn{eq:logZ}, in particular the
$1/\epsilon$ pole, to the explicit result for the bare quark jet
function allows to read off the anomalous dimension
$\bar\gamma_{J_q}$.  We obtain the following explicit expression in
the \FDH\ scheme:
\begin{align}
\bar\gamma_{J_q} &=  \left(\frac{\alpha_s}{4\pi}\right)\,
\left(-3\,C_{F}\right) 
+ \left(\frac{\alpha_e}{4\pi}\right)\, \frac{N_{\epsilon}}{2}\,C_{F}
\nonumber \\
&+ \left(\frac{\alpha_s}{4\pi}\right)^2\, \Big[
C_{F}^{2}\Big{(}-\frac{3}{2}+2\pi^{2}-24\zeta_{3}\Big{)}
+C_{F}C_{A}\Big{(}-\frac{1769}{54}-\frac{11\pi^{2}}{9}
+40\zeta_{3}\Big{)} \nonumber \\
& \qquad\qquad +\ C_F T_R N_F\Big{(}\frac{242}{27}
+\frac{4\pi^{2}}{9}\Big{)} +\frac{N_{\epsilon}}{2}\,\Big{(}\frac{271}{54}
+\frac{\pi^{2}}{9}\Big{)}C_{F}C_{A}\Big]
\nonumber\\
&+ \left(\frac{\alpha_s}{4\pi}\right)
\left(\frac{\alpha_e}{4\pi}\right) 
\Big[\frac{N_{\epsilon}}{2}\,\Big{(}11C_{F}C_{A}
-4 C_{F}^{2}-\frac{2}{3}C_{F}^{2}\pi^{2}\Big{)} \Big]
\nonumber \\
&+ \left(\frac{\alpha_e}{4\pi}\right)^2\, 
\Big[ 
-\frac{N_{\epsilon}^{2}}{8}C_{F}^{2}
-\frac{3\, N_{\epsilon}}{2}\, C_{F} T_R N_F \Big]
+{\cal O}(\alpha^3)\,.
\label{eq:yJquark}
\end{align} 
Using this expression together with \Eqns{eq:andim2}{eq:gammaW} the
quark anomalous dimension in the \FDH\ scheme, $\bar{\gamma}_q$ can be
found. Thus the computation of the soft and quark jet functions
provides an alternative determination of $\bar{\gamma}_q$.  The result
agrees with previous determinations~\cite{Kilgore:2012tb,
  Gnendiger:2014nxa} and is listed in Appendix~\ref{app:B} for
completeness. Of course, setting $N_\epsilon=0$ only the pure
$\alpha_s$ terms survive and the well known results in the
\CDR/\HV\ scheme are recovered.

This is also true for the quark jet
function as a whole. In analogy to \Eqn{eq:Sfin} we can define
\begin{equation}\label{eq:Jfin}
\jmath_{q\,\text{fin}}(Q^2,\mu) = 
\lim_{(N)_\epsilon\to 0} \jmath_{q\,\text{sub}}^{\mRS}(Q^2,\mu) \, ,
\end{equation}
so the finite quark jet function is scheme independent and can be
obtained using any of the regularization schemes. The explicit result
is given in Appendix~\ref{app:A}.

\subsection{Computation and scheme dependence of the gluon
  jet function and $\gamma_{Jg}$}

The discussion of the previous subsection can be readily adapted to
the gluon case. We closely follow Ref.~\cite{Becher:2010pd}, where
the gluon jet function $J_g(p^2)$ has been calculated at NNLO in
\CDR. The starting point is the gauge-invariant field
$\mathcal{A}^{\mu}$, related to the collinear gluon field
$A^{\mu}_c(x)$ through
\begin{equation}
\mathcal{A}^{\mu}(x)=\mathcal{A}^{a\mu}(x)t_{a}
=W^{\dagger}(x)[i D_c^{\mu}W(x)] \, .
\end{equation}
The treatment of this vector field depends on the regularization
scheme; we will give the details below. In all schemes
the field $\mathcal{A}^{\mu}$ satisfies $\bar{n}\cdot
\mathcal{A} = 0$; hence it can be decomposed as $\mathcal{A}^{\mu}
=\mathcal{A}^{\mu}_\perp + (n\cdot\mathcal{A}) \bar{n}^\mu/2$ and the
leading term is $\mathcal{A}^{\mu}_\perp$. The gluon jet propagator
$\mathcal{J}_{g}(p^{2})$ is then defined as
\begin{align}\label{eq:gpropA} 
\delta^{ab} g_{s}^{2}
\left(-g^{\mu\nu}_\perp\right)
\mathcal{J}^{\mRS}_{g}(p^{2})
=  \int d^{4}x\, e^{ipx} \langle 0\vert T
\{\mathcal{A}_\perp^{a\mu}(x)\mathcal{A}_\perp^{b\nu}(0)\}\vert0\rangle
\, .
\end{align}
For the calculation of $\mathcal{J}^{\mRS}_{g}(p^{2})$ it is actually more
convenient to use an equivalent definition in terms of the
time-ordered product of the full fields 
$\mathcal{A}^{\mu}$,
\begin{align}
&
\delta^{ab}g_{s}^{2}\Big{[}
\Big(-g_{\mu\nu}+\frac{\bar{n}_\mu p_\nu+p_\mu \bar{n}_\nu}{\bar{n}\cdot p}\Big)
\mathcal{J}^{\mRS}_{g}(p^{2})
+\frac{\bar{n}_{\mu}\bar{n}_{\nu}}{(\bar{n}\cdot p)^{2}}
\mathcal{K}^{\mRS}_{g}(p^{2})\Big{]} 
\label{eq:gprop} \\
& \qquad \qquad 
=  \int d^{4}x e^{ipx} \langle 0\vert 
T\{\mathcal{A}_{\mu}^{a}(x)\mathcal{A}_{\nu}^{b}(0)\}\vert0\rangle
\nonumber
\end{align}
and then extract $\mathcal{J}^{\mRS}_{g}(p^{2})$ using a
projection. The gluon jet function $J^{\mRS}_g(p^{2})$ is the
discontinuity of the leading part of the propagator, more precisely
$J^{\mRS}_g(p^{2})= \text{Im}
[i\,\mathcal{J}^{\mRS}_g(p^{2})]/\pi$. The function $K^{\mRS}_g$ is
related to power-suppressed terms and will not be considered any
further in this paper.

As in the case of the quark jet function, after decoupling of the soft
fields, the collinear Lagrangian is equivalent to the QCD Lagrangian.
Exploiting the gauge invariance of $\mathcal{J}^{\mRS}_{g}$ we work in
the light-cone gauge $\bar{n}\cdot A=0$.  This is particularly
convenient as in this gauge $W(x)=1$ and, therefore, no diagrams with
additional emission of gluons from the Wilson lines have to be
considered.  Therefore, for the calculation of
$\mathcal{J}^{\mRS}_{g}$ only standard QCD Feynman rules are
required. Of course, ghost loops are also absent in this gauge.

\begin{figure}[t]
\begin{center}
\includegraphics [width=0.6\textwidth]{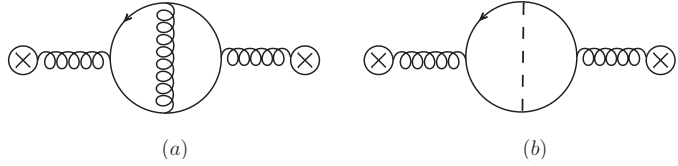}
\caption{ Sample two-loop diagrams contributing to the gluon jet
  function. Diagram (a) is present both in \CDR\ and \FDH, diagram
  (b) including an $\epsilon$-scalar contributes only in \FDH.}
\label{jetgluon1}
\end{center}
\end{figure}

Now we give details on the regularization scheme dependence.  Typical
examples of two-loop diagrams contributing to $\mathcal{J}^{\mRS}_{g}$
are shown in Figure~\ref{jetgluon1}. In \CDR\ all gluons are
$D$-dimensional gluons $\ghat$ and no $\epsilon$-scalar diagrams are
present. Correspondingly, the metric tensor in \Eqn{eq:gprop} is
$\hat{g}^{\mu\nu}$ in \CDR.  In \HV\ and \FDH\ the external gluons are
understood to be strictly 4-dimensional. Thus, the gluons attached to
the Wilson lines in Figure~\ref{jetgluon1} are to be interpreted as
$\bar{g}$, and the metric tensor in \Eqn{eq:gprop} is
$\bar{g}^{\mu\nu}$ in these schemes. Furthermore, in \FDH\ internal
gluons are treated as $g$ and hence are decomposed into $\ghat$ and
$\gtilde$, as indicated in the left and right panel of
Figure~\ref{jetgluon1}. In \DRED\ the definitions of the present
subsection apply to external $D$-dimensional gluons $\ghat$. For
these, the calculation and the result are the same as the
corresponding \FDH\ calculation, see \Eqn{eq:Defgammabar}. Hence for
simplicity we will only refer to \FDH\ in the remainder of the
subsection.

After an explicit calculation of the diagrams in \FDH, taking the
imaginary part and performing the Laplace transform, we obtain for the
bare gluon jet function in \FDH
\begin{align}
\bar{\jmath}_{g\,\text{bare}}(Q^2)&= 1
+a_{s} \Big{(} C_{A}\Big{[}\frac{4}{\epsilon^{2}}+\frac{11}{3\epsilon}
+\frac{67}{9}-\frac{2\pi^{2}}{3}
+\epsilon\,  \big(\frac{404}{27}-\frac{11\pi^{2}}{18}-8\zeta_{3}\big)\Big{]}
\nonumber\\
&\qquad \qquad +N_F T_{R}\Big{[}-\frac{4}{3\epsilon}-\frac{20}{9}+\epsilon\,
  \big(\frac{2\pi^{2}}{9}-\frac{112}{27}\big)\Big{]}\
\nonumber \\
&\qquad \qquad +\frac{N_{\epsilon}}{2}\,C_{A}\Big{[}
-\frac{1}{3\epsilon}-\frac{8}{9}+\epsilon \big{(}\frac{\pi^{2}}{18}
-\frac{52}{27}\big{)}\Big{]}\Big{)}\
\nonumber\\
&+ a_{s}^{2}\, \Big{(}
C_{A}^{2}\, \bar{\jmath}^{\, g;\, AA}_{20}
+C_{A}N_FT_{R}\, \bar{\jmath}^{\, g;\, Af}_{20}
+C_{F}N_FT_{R}\, \bar{\jmath}^{\, g;\, Ff}_{20}
+N_F^{2}T_{R}^{2}\, \bar{\jmath}^{\, g;\, ff}_{20}\Big{)}
\phantom{\frac{1}{1}}\nonumber \\
&+ a_{s} a_{e}\, \Big{(}
C_{A}N_FT_{R}\, \bar{\jmath}^{\, g;\, Af}_{11}
+C_{F}N_FT_{R}\, \bar{\jmath}^{\, g;\, Ff}_{11}\Big)
+{\cal O}(\alpha^3)\,.\phantom{\frac{1}{1}}
\label{eq:jetgbare}
\end{align}
The explicit results of the two-loop coefficients are given in
Appendix~\ref{app:A}.  In the limit $N_\epsilon \to 0$ all terms
proportional to $\alpha_e$ vanish and we obtain the results in \CDR,
in agreement with Ref.~\cite{Becher:2010pd}.

The renormalization procedure is the same as for the quark jet
function. In Laplace space, the renormalized gluon jet function in the
\FDH\ scheme is obtained by multiplying \Eqn{eq:jetgbare} by a factor
$\bar{Z}_{J_g}$. This factor is the same as in \Eqn{eq:logZ} apart
from the replacement $\bar{\gamma}_{ij}^{J_q} \to
\bar{\gamma}_{ij}^{J_g}$ and $\Gamma_{\text{cusp}}^{\mRS} =
C_A \gamma_{\text{cusp}}^{\mRS}$. After renormalization of the coupling, all
divergences of the bare gluon jet function have to be absorbed by
$\bar{Z}_{J_g}(Q^2,\mu)$. This allows to determine the anomalous
dimension of the gluon jet in the \FDH\ scheme as
\begin{align}
\bar\gamma_{J_g} &=  \left(\frac{\alpha_s}{4\pi}\right)\,
\left(-\frac{11}{3}C_{A}+\frac{4}{3}N_FT_{R}+\frac{N_\epsilon}{6}C_{A}
\right) 
\nonumber \\
&+ \left(\frac{\alpha_s}{4\pi}\right)^2\,
\Big[C_{A}^{2}\Big{(}-\frac{1096}{27}+\frac{11\pi^{2}}{9}+16\zeta_{3}\Big{)}
+C_{A}N_F T_{R}\Big{(}\frac{368}{27}-\frac{4\pi^{2}}{9}\Big{)}
+4C_{F}T_{R}N_F\nonumber\\ 
& \qquad\qquad +\frac{N_{\epsilon}}{2}\Big{(}\frac{248}{27}
-\frac{\pi^{2}}{9}\Big{)}C_{A}^{2} \Big] \nonumber \\
&+ \left(\frac{\alpha_s}{4\pi}\right)
\left(\frac{\alpha_e}{4\pi}\right) 
\Big[-N_{\epsilon}\,(2\,C_{F}N_F T_{R}) \Big]
+{\cal O}(\alpha^3)\,.
\label{eq:yJgluon}
\end{align} 
Of course, it is again also possible to extract the cusp anomalous
dimension as well as the $\beta$ functions of $\alpha_s$ and
$\alpha_e$ from $\bar{Z}_{J_g}(Q^2,\mu)$. The fact that we obtain
again the same results for these quantities is a strong consistency
check on the procedure.

From $\bar\gamma_{J_g}$ we can determine $\bar\gamma_{g}$ with the
help of \Eqn{eq:andim2}. The result is in agreement with previous
determinations~\cite{Kilgore:2012tb, Gnendiger:2014nxa} and is listed
in Appendix~\ref{app:B} for
completeness, but the present procedure provides a more direct
alternative determination of $\bar\gamma_{g}$.

Finally, as for the soft and quark jet function, we can obtain a
finite and scheme independent gluon jet function as
\begin{equation}\label{eq:Jgfin}
\jmath_{g\,\text{fin}}(Q^2,\mu) = 
\lim_{(N)_\epsilon\to 0} \jmath_{g\,\text{sub}}^{\mRS}(Q^2,\mu) \, .
\end{equation}
For completeness the explicit result is listed in
Appendix~\ref{app:A}.

\subsection{Computation of the $\epsilon$-scalar jet
  function, $\gamma_{J\epsilon}$ and result for $\bar\gamma_\epsilon$
  in DRED}
\label{sec:epsJet}

In \DRED\ processes with external $\epsilon$-scalars need to be
considered. The discussion of Section~\ref{sec:outlinescet} applies
analogously, and we can determine the anomalous dimension of
$\epsilon$-scalars from an equation like \Eqn{eq:andim2},
\begin{align}\label{eq:andimEps}
\gamma^{\mDRED}_{\gtilde} \equiv \bar\gamma_{\epsilon} 
= \bar\gamma_{J_{\epsilon}}
  -\frac{\gamma^{\mDRED}_{W_{\epsilon}}}{2}\, .
\end{align}
As mentioned in Section~\ref{sec:soft} the soft function is the same
as for external gluons, hence $\gamma^{\mDRED}_{W_{\epsilon}} =
\bar\gamma_W$, from \Eqn{eq:gammaW}. For $\bar\gamma_{J_{\epsilon}}$
an $\epsilon$-scalar jet function is needed. Such an object can be
defined and computed in close analogy to the calculation of the gluon
jet function, with the difference that now the time-ordered product of
two fields $\tilde{\mathcal{A}}_{\mu} =
\gtilde_{\mu\nu}\mathcal{A}^{\nu}$ has to be considered. In light-cone
gauge these fields reduce to the $\epsilon$-scalar field
$\tilde{A}_\mu$. Starting from the propagator
$\bar{\mathcal{J}}_\epsilon(p^{2}) \equiv
\mathcal{J}^{\mDRED}_\epsilon(p^{2})$ given by
\begin{align}
&
\delta^{ab}g_{s}^{2}
\left(-\tilde{g}_{\mu\nu}\right)
\bar{\mathcal{J}}_{\epsilon}(p^{2})
=  \int d^{4}x e^{ipx} \langle 0\vert 
T\{\tilde{\mathcal{A}}_{\mu}^{a}(x)
\tilde{\mathcal{A}}_{\nu}^{b}(0)\}\vert0\rangle
\end{align} 
the $\epsilon$-scalar jet function is obtained as $\bar{J}_\epsilon(p^{2})=
\text{Im} [i\,\bar{\mathcal{J}}_\epsilon(p^{2})]/\pi$.

\begin{figure}[t]
\begin{center}
\includegraphics [width=0.6\textwidth]{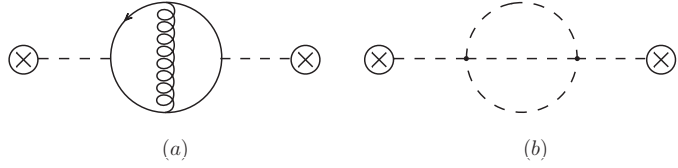}
\caption{ Sample two-loop diagrams contributing to the
  $\epsilon$-scalar jet function both.  Diagram (a) is proportional to
  $\alpha_s\alpha_e$ whereas diagram~(b) is
  $\sim\alpha^2_{4\epsilon}$}
\label{scalarjet1}
\end{center}
\end{figure}

Two examples of diagrams contributing (in light-cone gauge) at
two-loop order are shown in Figure~\ref{scalarjet1}. A new feature is the
appearance of the quartic coupling $\alpha_{4\epsilon}$. We do not
need to distinguish the three different $\alpha_{4\epsilon}$ since the
quartic coupling only appears
at the two-loop level and hence the associated renormalization constants and
$\beta$ functions do not appear.  The only
non-vanishing diagram $\sim\alpha^2_{4\epsilon}$ is depicted in
Figure~\ref{scalarjet1}~b.

Performing a computation analogous to previous cases,  the bare
two-loop $\epsilon$-scalar jet function in Laplace space is found to be
\begin{align}\label{eq:jetebare}
\nonumber
\bar{\jmath}_{\epsilon\,\text{bare}}(Q^2)
&=1+
 a_s\, C_{A}\Big{(}
 \frac{4}{\epsilon^{2}}+\frac{4}{\epsilon}+8-\frac{2\pi^{2}}{3}
 +\epsilon \big(16-\frac{2\pi^{2}}{3}-8\zeta_{3}\big)\Big{)}
\\
& +a_{e}\, N_FT_{R}\Big{(}-\frac{2}{\epsilon}-4
  +\epsilon \big(-8+\frac{\pi^{2}}{3}\big)\Big{)}
\nonumber \\
&+
 a_{s}^{2}\, \Big{(}
   C_{A}^{2}\, \bar{\jmath}^{\, \epsilon;\,AA}_{200}
 + C_{A}N_FT_{R}\, \bar{\jmath}^{\, \epsilon;\,Af}_{200}\Big{)}
\phantom{\frac{1}{1}}\nonumber \\
&+a_{e}^{2}\,N_F T_{R}\, \Big{(}
   C_{A}\, \bar{\jmath}^{\, \epsilon;\,Af}_{020}
 + C_{F}\, \bar{\jmath}^{\, \epsilon;\,Ff}_{020}
 + N_F T_{R}\,
 \bar{\jmath}^{\, \epsilon;\,ff}_{020}\Big{)}
\phantom{\frac{1}{1}}\nonumber \\
&+a_{4\epsilon}^{2}\,
   C_{A}^{2}\, \bar{\jmath}^{\, \epsilon;\,AA}_{002}
\phantom{\frac{1}{1}}\nonumber \\
&+a_{s}a_{e}\,N_F T_{R}\, \Big{(}
   C_{A}\, \bar{\jmath}^{\, \epsilon;\,Af}_{110}
 + C_{F}\, \bar{\jmath}^{\, \epsilon;\,Ff}_{110}\Big{)}
+{\cal O}(a^3)\,.\phantom{\frac{1}{1}}
\end{align}
Due to the presence of $\alpha_{4\epsilon}$, the various coefficients
have now three labels, with the last one indicating the power of
$\alpha_{4\epsilon}$. The explicit NNLO expressions are given in
Appendix~\ref{app:A}. 

Once more, the UV divergences of the bare jet function are absorbed by
a renormalization factor $Z^\mDRED_{\epsilon}(Q^2,\mu)$, which has a
structure similar to \Eqn{eq:lnZfullRS} or \Eqns{eq:lnz}{eq:logZ}. In
fact, it can be written as \Eqn{eq:lnZfullDRED},
\begin{align}\label{eq:lnZJeps}
\ln Z^\mDRED_{\epsilon} &= 
 \left(\frac{\vec{\alpha}}{4\pi}\right)\cdot
\left(\frac{\vec{\Gamma}^{\prime\, \mDRED}_{1}}{4\epsilon^2}
      +\frac{\vec{\mathbf{\Gamma}}^{\mDRED}_{1}}{2\epsilon}\right)\\
 &+
 \sum_{\Sigma=2}  
\left(\frac{\alpha_s}{4\pi}\right)^m
 \left(\frac{\alpha_e }{4\pi}\right)^n
 \left(\frac{\alpha_{4\epsilon}}{4 \pi}\right)^{k}\,
 \nonumber \\
& \quad \Bigg( 
-\frac{3\vec{\beta}^{\mDRED}_{mnk} \cdot \vec{\Gamma}^{\prime\, \mDRED}_{1}}{16\epsilon^3}
- \frac{\vec{\beta}^{\mDRED}_{mnk} \cdot \vec{\mathbf{\Gamma}}^{\mDRED}_{1}}{4\epsilon^2}
+\frac{\Gamma^{\prime\, \mDRED}_{mnk}}{16\epsilon^2}
+\frac{\mathbf{\Gamma}^{\mDRED}_{mnk}}{4\epsilon}
\Bigg)
+ \mathcal{O}(\alpha^3) \, 
\nonumber 
\end{align}
with the identification
\begin{align}
\Gamma^{\prime\,\mDRED}&=-4\,C_A\,\bar\gamma_{\text{cusp}},&
\mathbf{\Gamma}^{\mDRED} &= 2\,C_A\,\bar\gamma_{\text{cusp}}\, L_Q+
2\,\bar\gamma_{J_\epsilon}\, .
\end{align}
We refrain from using the explicit form of \Eqn{eq:logZ} since the
dependence on $\alpha_{4\epsilon}$ leads to a proliferation of similar
terms. The only simplification used is the identification of the
couplings $\alpha_{4\epsilon,i}$, which is possible since the explicit
results show that these couplings appear not at one-loop but only in
the genuine two-loop coefficients.

By comparing with the explicit result for the $\epsilon$-scalar jet
function we determine the renormalization factor using minimal
subtraction and extract from this the anomalous dimension of the
$\epsilon$-scalar jet as 
\begin{align}
\bar\gamma_{J_\epsilon} &=  \left(\frac{\alpha_s}{4\pi}\right)\,
\left(-4\, C_{A} \right) 
+ \left(\frac{\alpha_e}{4\pi}\right)\,
\left(2 N_FT_{R}\right)
\nonumber \\
&+ \left(\frac{\alpha_s}{4\pi}\right)^2\,
\Big[
   C_{A}^{2}\Big{(}-\frac{4603}{108}+\frac{13\pi^{2}}{9}+16\zeta_{3} 
      + N_\epsilon\frac{337}{108}+ N_\epsilon\frac{\pi^{2}}{18} \Big{)}
 + C_{A}N_FT_{R}\Big{(}\frac{338}{27}+\frac{4\pi^{2}}{9}\Big{)}
\Big] \nonumber \\
&+ \left(\frac{\alpha_s}{4\pi}\right)
\left(\frac{\alpha_e}{4\pi}\right)  
\Big[ 10\,C_{F}N_FT_{R}-\frac{4\pi^{2}}{3}C_{A}N_FT_{R} 
\Big] \nonumber \\
&+ \left(\frac{\alpha_e}{4\pi}\right)^2\, 
\Big[ N_FT_{R}\big{(}2\, C_{A}-4\, C_{F}-N_{\epsilon}(C_{A}+C_{F})\big{)}
\Big] \nonumber \\
&+ \left(\frac{\alpha_{4\epsilon}}{4\pi}\right)^2\, 
\Big[C_{A}^{2}\frac{3}{4}(-1+N_{\epsilon}) \Big]
+{\cal O}(\alpha^3)\,.
\label{eq:yJeps}
\end{align} 
Combining this result as prescribed by \Eqn{eq:andimEps} with the soft
anomalous dimension, which has only $\alpha_s^2$ contributions, we
find the $\epsilon$-scalar anomalous dimension
\begin{align}
\bar\gamma_{\epsilon} &=  \left(\frac{\alpha_s}{4\pi}\right)\,
\left(-4\, C_{A} \right) 
+ \left(\frac{\alpha_e}{4\pi}\right)\,
\left(2 N_FT_{R}\right)
\nonumber \\
&+ \left(\frac{\alpha_s}{4\pi}\right)^2\,
\Big[
   C_{A}^{2}\Big{(}-\frac{2987}{108}+\frac{5\pi^{2}}{6}+2\zeta_{3} 
      + N_\epsilon\frac{233}{108}+ N_\epsilon\frac{\pi^{2}}{12} \Big{)}
 + C_{A}N_FT_{R}\Big{(}\frac{226}{27}+\frac{2\pi^{2}}{3}\Big{)}
\Big] \nonumber \\
&+ \left(\frac{\alpha_s}{4\pi}\right)
\left(\frac{\alpha_e}{4\pi}\right)  
\Big[ 10\,C_{F}N_FT_{R}-\frac{4\pi^{2}}{3}C_{A}N_FT_{R} 
\Big] \nonumber \\
&+ \left(\frac{\alpha_e}{4\pi}\right)^2\, 
\Big[ N_FT_{R}\big{(}2\, C_{A}-4\, C_{F}-N_{\epsilon}(C_{A}+C_{F})\big{)}
\Big] \nonumber \\
&+ \left(\frac{\alpha_{4\epsilon}}{4\pi}\right)^2\, 
\Big[C_{A}^{2}\frac{3}{4}(-1+N_{\epsilon}) \Big]
+{\cal O}(\alpha^3)\,.
\label{eq:yeps}
\end{align} 
As discussed in Section~\ref{sec:dred}, $\bar\gamma_{\epsilon}$ is
needed to relate two-loop matrix elements computed in \DRED\ to those
computed in other schemes such as \FDH. With this new result all
anomalous dimensions are known at the two-loop level in all four
schemes.

%% file: 04_Alternative.tex
\section{Alternative determination of $\bar\gamma_{\epsilon}$ from the
  $\mathbf{\epsilon}$-scalar form factor}
\label{sec:alt}

Apart from the new approach of extracting the IR anomalous dimension
of the $\epsilon$-scalar, $\bar\gamma_{\epsilon}$ defined in
\Eqn{eq:Defgammaeps} from the $\epsilon$-scalar jet and soft
functions, it is also possible to obtain this quantity in the more
traditional way, by comparing the generic infrared factorization
formula with a specific amplitude for a process containing external
$\epsilon$-scalars. This procedure is analogous to the determination
of $\bar\gamma_{q}$ and $\bar\gamma_g$ in
Ref.~\cite{Gnendiger:2014nxa}. We now describe the determination of
$\bar\gamma_{\epsilon}$ via a process with two external
$\epsilon$-scalars, the $\epsilon$-scalar form factor, which has been
calculated recently in Ref.~\cite{Broggio:2015ata} up to the two-loop
level.

According to Eq.~\eqref{eq:lnZfullDRED} the one-loop infrared
divergences in the \DRED\ scheme are described by
\begin{align}
\notag
\text{ln}\,\bar{\mathbf{Z}}^{\text{1L}}=&
 \Big(\frac{\alphas}{4\pi}\Big)
   \Bigg[-\frac{\bar\Gamma_{100}}{2 \epsilon^2}
     +\frac{\bar\gamma^{\epsilon}_{100}}{\epsilon}\Bigg]
+\Big(\frac{\alphae}{4\pi}\Big)
   \Bigg[-\frac{\bar\Gamma_{010}}{2 \epsilon^2}
     +\frac{\bar\gamma^{\epsilon}_{010}}{\epsilon}\Bigg]
\\ &
+\Big(\frac{\alphafour}{4\pi}\Big)
   \Bigg[-\frac{\bar\Gamma_{001}}{2 \epsilon^2}
     +\frac{\bar\gamma^{\epsilon}_{001}}{\epsilon}\Bigg].
\label{eq:dredFac1l}
\end{align}
Here the relations $\bar{\Gamma}{'}_{ijk}=-2\,\bar\Gamma_{ijk} = -2
C_A \bar{\gamma}^{\text{cusp}}_{ijk}$ and
$\bar{\mathbf\Gamma}_{ijk}=2\,\bar\gamma^{\epsilon}_{ijk}$ have been
used. The notation with three indices for a common
$\alpha_{4\epsilon}$ coupling and for dropping the superscript
``cusp'' has been explained in Section~\ref{sec:outlinescet}.
\Eqn{eq:dredFac1l} can now be compared with the corresponding IR
divergent one-loop result of the UV renormalized $\epsilon$-scalar
form factor given in Ref.~\cite{Broggio:2015ata}, where
$T_R=\frac{1}{2}$ and $\mu^2 = -s_{12}$ has been used:
\begin{align}
 \bar{F}^{\text{1L}}_{\epsilon} &
 =\Big(\frac{\alphas}{4\pi}\Big)
    \Bigg[-\frac{2}{\epsilon ^2}-\frac{4}{\epsilon }\Bigg]C_A
  +\Big(\frac{\alphae}{4\pi}\Big)\frac{N_F}{\epsilon}
  +\mathcal{O}(\epsilon^0).
  \label{eq:epsFF1l}
\end{align}
The $\frac{1}{\epsilon^2}$-pole of this one-loop form factor confirms
the previous finding that the one-loop cusp anomalous dimension is a
process-independent quantity that has only one non-vanishing component
$\bar\Gamma_{100}=4\,C_A=\bar\gamma^{\text{cusp}}_{100}\,C_A$.  On the
other hand, the $\frac{1}{\epsilon}$-poles in Eq.~\eqref{eq:epsFF1l}
are directly correlated with the components of the anomalous dimension
$\bar\gamma_{\epsilon}$.  The values obtained here agree with the
results from the previous section.

The appropriate two-loop prediction for
$\text{ln}\,\bar{\mathbf{Z}}^{\text{2L}}$ could be given in a
completely general form, as in \Eqns{eq:lnZfullDRED}{eq:lnZJeps}, in
which it would allow to read off once again even the one-loop $\beta$
functions. Here, however, we give the prediction in a more specific
form, where we already use the knowledge that several one-loop
coefficients are zero.  Considering only non-vanishing components of
one-loop anomalous dimensions and $\beta$ functions yields for the
infrared divergence structure at the two-loop level:
\begin{align}
 \notag
 \text{ln}\,\bar{\mathbf{Z}}^{\text{2L}}=&
 \Big(\frac{\alphas}{4\pi}\Big)^2
   \Bigg[\frac{3\,\betaMS{200}\,\bar\Gamma_{100}}{8 \epsilon^3}
   -\frac{\betaMS{200}\,\bar\gamma^{\epsilon}_{100}}{2 \epsilon^2}
   -\frac{\bar\Gamma_{200}}{8 \epsilon^2}
   +\frac{\bar\gamma^{\epsilon}_{200}}{2 \epsilon}\Bigg]
 \\ \notag
 &+\Big(\frac{\alphas}{4\pi}\Big) \Big(\frac{\alphae}{4\pi}\Big)
   \Bigg[
   -\frac{\betaeMS{110}\,\bar\gamma^{\epsilon}_{010}}{2\epsilon^2}
   -\frac{\bar\Gamma_{110}}{8\epsilon^2}
   +\frac{\bar\gamma^{\epsilon}_{110}}{2\epsilon}\Bigg]
 \\ \notag
 &+\Big(\frac{\alphae}{4\pi}\Big)^2\,
   \Bigg[
   -\frac{\betaeMS{020}\,\bar\gamma^{\epsilon}_{010}}{2\epsilon^2}
   -\frac{\bar\Gamma_{020}}{8\epsilon^2}
   +\frac{\bar\gamma^{\epsilon}_{020}}{2\epsilon}\Bigg]
 +\Big(\frac{\alphas}{4\pi}\Big) \Big(\frac{\alphafour}{4\pi}\Big)\Bigg[
   -\frac{\bar\Gamma_{101}}{8\epsilon^2}
   +\frac{\bar\gamma^{\epsilon}_{101}}{2\epsilon}\Bigg]
 \\
 &+\Big(\frac{\alphae}{4\pi}\Big) \Big(\frac{\alphafour}{4\pi}\Big)\Bigg[
   -\frac{\bar\Gamma_{011}}{8\epsilon^2}
   +\frac{\bar\gamma^{\epsilon}_{011}}{2\epsilon}\Bigg]
 +\Big(\frac{\alphafour}{4\pi}\Big)^2\Bigg[
   -\frac{\bar\Gamma_{002}}{8\epsilon^2}
   +\frac{\bar\gamma^{\epsilon}_{002}}{2\epsilon}\Bigg] \, .
\end{align}
Thanks to the simple colour and momentum structure of the form factor,
this has to correspond directly to the divergence structure of the
combination
$\bar{F}^{\text{2L}}_{\epsilon}-\frac{1}{2}(\bar{F}^{\text{1L}}_{\epsilon})^2$,
see Ref.~\cite{Gnendiger:2014nxa}.  Inserting the results for the form
factor of Ref.~\cite{Broggio:2015ata} yields
\begin{align}
\notag
\bar{F}^{\text{2L}}_{\epsilon} &
-\frac{1}{2}\Big(\bar{F}^{\text{1L}}_{\epsilon}\Big)^2
\\* \notag
&=\Big(\frac{\alphas}{4\pi}\Big)^2 \Bigg\{
   C_A^2 \Bigg[\frac{\frac{11}{2}-\frac{\Neps}{4}}{\epsilon ^3}+\frac{\frac{65}{18}+\frac{\pi ^2}{6}-\frac{\Neps}{9}}{\epsilon ^2}
     +\frac{-\frac{2987}{216}+\frac{5 \pi ^2}{12}+\zeta (3)+\Neps\Big(\frac{233}{216}+\frac{\pi ^2}{24}\Big)}{\epsilon }\Bigg]
\\ \notag & \quad\quad\quad\quad\quad
+C_A N_F \Bigg[-\frac{1}{\epsilon ^3}-\frac{7}{9 \epsilon ^2}+\frac{\frac{113}{54}+\frac{\pi ^2}{6}}{\epsilon }\Bigg]\Bigg\}
\\ \notag &
+\Big(\frac{\alphas}{4\pi}\Big) \Big(\frac{\alphae}{4\pi}\Big)
   \Bigg\{C_F N_F \Bigg[-\frac{3}{\epsilon ^2}+\frac{5}{2 \epsilon }\Bigg]-C_A N_F\frac{\pi ^2}{3 \epsilon }\Bigg\}
\\ \notag &
+\Big(\frac{\alphae}{4\pi}\Big)^2 \Bigg\{
   C_A N_F \Bigg[\frac{-1+\frac{\Neps}{2}}{\epsilon ^2}+\frac{\frac{1}{2}-\frac{\Neps}{4}}{\epsilon }\Bigg]
     +C_F N_F \Bigg[\frac{2-\frac{\Neps}{2}}{\epsilon ^2}+\frac{-1-\frac{\Neps}{4}}{\epsilon }\Bigg]+N_F^2\frac{1}{2 \epsilon ^2}\Bigg\}
\\&
+\Big(\frac{\alphafour}{4\pi}\Big)^2 C_A^2\,(1-\Neps)\frac{-3}{8\epsilon}
+\mathcal{O}(\epsilon^0).\phantom{\Bigg\}}
\end{align}
Again, the $\frac{1}{\epsilon}$-poles allow to read off the components
of the anomalous dimension of the $\epsilon$-scalar
$\bar{\gamma}_\epsilon$.  The values found here agree with the results
from the previous section, see \Eqn{eq:yeps}.  Since the remaining
divergence structure is governed by one-loop anomalous dimensions, the
process-independent components of the cusp anomalous dimension and
previously known $\beta$ coefficients, this is further evidence for
the validity of the results obtained in Section~\ref{sec:epsJet}.
With this result, and the results of the previous sections and
Ref.~\cite{Gnendiger:2014nxa}, all two-loop anomalous dimensions
$\gamma_i$ in all \RS\ have been determined both in the SCET approach
and from form factors.

%% file: 05_Examples.tex
\section{Cross check with explicit processes }
\label{sec:examples}

The results of the previous sections allow us to predict the
differences between UV renormalized virtual two-loop amplitudes
squared, as defined in \Eqn{def:M}, computed in different
regularization schemes. In this section we will make these transition
rules more explicit and will check them with explicit examples.

The following discussions will also shed more light on the role of the
various couplings $\alpha_s$, $\alpha_e$ and
$\alpha_{4\epsilon,i}$. In the practical computation of the genuine
two-loop diagrams it is no problem to set these couplings equal from
the beginning. In the process of UV renormalization, i.e.\ in
lower-order diagrams with counterterm insertions, the bare couplings
and the associated renormalization constants appear. It is unavoidable
to keep these distinct, regardless whether \FDH\ or \DRED\ is
used. Once renormalization has been performed, it is possible to set
the renormalized couplings equal and to identify $N_\epsilon$ and
$2\epsilon$.  Likewise, the derivation of the IR subtraction formulas
and the transition rules requires the couplings to be treated
independently, but in the end the transition rules can be easily
written down for the special case of equal couplings.

We will consider the transition rules \FDH\ $\leftrightarrow$ \HV, as
well as \FDH\ $\leftrightarrow$ \DRED. To make connection to the
scheme that is used most often, \CDR, we remind the reader of the
discussion in Section~\ref{sec:chv}. The only difference in the
squared matrix element between \HV\ and \CDR\ is due to the use of
different metric tensors for the polarization sum of external
gluons. All anomalous dimensions are the same in the two schemes.

\subsection{Transition between   FDH and  HV}

Since external gluons are treated in the same way in \FDH\ and \HV, we
can actually relate directly virtual amplitudes and do not need to
work with squared amplitudes. The finite remainders of the scattering
amplitudes are scheme independent. More precisely
\begin{align}\label{scheme1}
\vert \mathcal{A}_{\mathrm{fin}}(\{p\},\mu)\rangle &=
\lim_{\epsilon\to 0}\mathbf{Z}^{-1}(\epsilon,\{p\},\mu)
\vert \mathcal{A}(\epsilon,\{p\})\rangle \nonumber \\ 
&=\lim_{(N)_\epsilon\to 0}\bar{\mathbf{Z}}^{-1}(\epsilon,N_\epsilon,\{p\},\mu)
\vert \bar{\mathcal{A}}(\epsilon,N_\epsilon,\{p\})\rangle \, ,
\end{align}
where $\vert\mathcal{A}\rangle = \vert\mathcal{A}^{\mHV}\rangle$ and
$\mathbf{Z} = \mathbf{Z}^{\mHV}$ denote quantities in the \HV\ scheme
and $\vert\bar{\mathcal{A}}\rangle = \vert\mathcal{A}^{\mFDH}\rangle$
and $\bar{\mathbf{Z}} = \mathbf{Z}^{\mFDH}$ are the corresponding
quantities in the \FDH\ scheme. Suppressing the arguments of the
amplitudes, setting $N_\epsilon = 2\epsilon$ and writing
$\mathbf{Z}^{-1} = 1 + \delta\mathbf{Z}$ in both schemes, we can
rewrite this equation as
\begin{align}
\vert \mathcal{A}\rangle &+ \delta\mathbf{Z}\vert \mathcal{A}\rangle
=\vert \bar{\mathcal{A}}\rangle + \delta\bar{\mathbf{Z}}\vert \bar{\mathcal{A}}\rangle
+\mathcal{O}(\epsilon) \, .
\label{hvfdhv1}
\end{align}
If the expansion coefficients $\delta\textbf{Z}$ are known to
$\mathcal{O}(\alpha^n)$ and the amplitudes $|\mathcal{A}\rangle$ are
known to $\mathcal{O}(\alpha^{n-1})$, this equation allows to obtain a
relation between the $\mathcal{O}(\alpha^n)$ amplitudes
computed in \HV\ and \FDH, up to $\mathcal{O}(\epsilon)$ terms.  
We now give the explicit  results up to the two-loop level.

The tree-level amplitudes in the two schemes are the same
$\vert\bar{\mathcal{A}}_0\rangle = \vert\mathcal{A}_0\rangle$. At
one-loop we can relate the $\mathcal{O}({\alpha_s})$ and
$\mathcal{O}({\alpha_e})$ corrections in the \FDH\ scheme, denoted by
$\vert\bar{\mathcal{A}}_{10}\rangle$ and
$\vert\bar{\mathcal{A}}_{01}\rangle$ respectively, to
$\vert\mathcal{A}_1\rangle$, the $\mathcal{O}({\alpha_s})$ corrections
in the \HV\ scheme
\begin{subequations}
\begin{align}
\vert \bar{\mathcal{A}}_{01} \rangle &= 
-\delta\bar{\mathbf{Z}}_{01} \vert \mathcal{A}_{0}\rangle
 +\mathcal{O}(\epsilon) \, ,\phantom{\frac{1}{1}}
\label{eq:ZA01}\\*
\vert \bar{\mathcal{A}}_{10}\rangle - \vert \mathcal{A}_{1}\rangle &=
 (\delta\mathbf{Z}_{1} - \delta\bar{\mathbf{Z}}_{10}) \vert 
\mathcal{A}_{0}\rangle +\mathcal{O}(\epsilon)\,.\phantom{\frac{1}{1}}
\end{align}
\end{subequations}
In the above equation we have also introduced the expansion
coefficients $\delta\mathbf{Z}_m$ and $\delta\bar{\mathbf{Z}}_{mn}$ of
$\mathbf{Z}^{-1} = 1 + \delta\mathbf{Z}$ in the \HV\ and \FDH\ scheme,
respectively. Substituting in the last equations the explicit
expressions of these expansion coefficients, the explicit form of the
differences for a process with $\# q$ external massless quarks and $\#
g$ external gluons read
\begin{subequations}
\label{app:fdh-hv-1}
\begin{align}
|\bar{\mathcal{A}}_{01}\rangle &= \frac{
\#q \, \bar{\gamma}^{q}_{01} }{2\epsilon} |\mathcal{A}_{0} \rangle
   +\mathcal{O}(\epsilon) = 
\# q\frac{C_{F}}{2} |\mathcal{A}_{0} \rangle
   +\mathcal{O}(\epsilon)\,,\\
|\bar{\mathcal{A}}_{10}\rangle -|\mathcal{A}_{1}\rangle &=\frac{
\# g\, (\bar\gamma^g_{10}-\gamma^g_{10})}{2\epsilon} |\mathcal{A}_{0}\rangle
   +\mathcal{O}(\epsilon) = 
\# g\frac{C_{A}}{6} |\mathcal{A}_{0}\rangle+\mathcal{O}(\epsilon)\, ,
\end{align}
\end{subequations}
which agrees with the results in \cite{Kunszt:1994,Signer:2008va}. In
\Eqn{app:fdh-hv-1} and what follows we use the notation (see footnote
in Section~\ref{sec:outlinescet}) $\gamma_{m0} \equiv \gamma_m^{\mHV}$
for the anomalous dimensions (and the $\beta$-functions) in the
\HV\ scheme. Since in the \HV\ scheme the anomalous dimensions depend
only on $\alpha_s$ but not on $\alpha_e$ the second label is always
zero. Of course, this is not the case in the corresponding quantities
in the \FDH\ scheme, $\bar\gamma_{mn}$. To obtain \Eqn{app:fdh-hv-1}
we have used $\gamma^q_{10} = \bar{\gamma}^{q}_{10}$ and
$\gamma^{\text{cusp}}_{10} = \bar\gamma^{\text{cusp}}_{10}$.

Moving to the two-loop level the corresponding equations are
\begin{subequations}
\begin{align}
\vert\bar{\mathcal{A}}_{02}\rangle &= 
- \delta\bar{\mathbf{Z}}_{01}\vert \bar{\mathcal{A}}_{01} \rangle
  - \delta\bar{\mathbf{Z}}_{02}\vert \mathcal{A}_{0}\rangle 
+\mathcal{O}(\epsilon)\,,\label{scheme6}\phantom{\frac{1}{1}}\\
\vert\bar{\mathcal{A}}_{20}\rangle - \vert\mathcal{A}_{2}\rangle &= 
\delta\textbf{Z}_{1} \vert\mathcal{A}_{1} \rangle
- \delta\bar{\textbf{Z}}_{10} \vert\bar{\mathcal{A}}_{10}\rangle  
+ ( \delta\textbf{Z}_{2} - \delta\bar{\textbf{Z}}_{20}) \vert\mathcal{A}_{0}\rangle
+\mathcal{O}(\epsilon)\,,\label{scheme7}\phantom{\frac{1}{1}}\\
\vert\bar{\mathcal{A}}_{11}\rangle &= 
-\delta\bar{\textbf{Z}}_{01} \vert\bar{\mathcal{A}}_{10}  \rangle
- \delta\bar{\textbf{Z}}_{10} \vert\bar{\mathcal{A}}_{01} \rangle 
- \delta\bar{\textbf{Z}}_{11}\vert\mathcal{A}_{0}\rangle
+\mathcal{O}(\epsilon)\,.\phantom{\frac{1}{1}}
\label{scheme8}
\end{align}
\end{subequations}
The expressions given in (\ref{scheme6}), (\ref{scheme7}) and
(\ref{scheme8}) allow one to move from \FDH\ to \HV\ (and vice versa)
for any process with $\#g$ external gluons and $\#q$ external massless
quarks in QCD up to two-loop order. Exploiting $\gamma^q_{10} =
\bar{\gamma}^{q}_{10}$ and $\gamma^{\text{cusp}}_{10} =
\bar\gamma^{\text{cusp}}_{10}$ we obtain
\begin{subequations}
\begin{align}
|\mathcal{\bar{A}}_{02}\rangle &= 
\bigg{[}\frac{-1}{8\epsilon^{2}}\#q\bar{\gamma}^{q}_{01}
(2\bar{\beta}^{e}_{02}+\#q\bar{\gamma}^{q}_{01})
+\frac{1}{4\epsilon}\#q\bar{\gamma}^{q}_{02}\bigg{]}|\mathcal{A}_{0}\rangle
\nonumber\\
&+\bigg{[}\frac{1}{2\epsilon}\#q\bar{\gamma}^{q}_{01}\bigg{]}
|\mathcal{\bar{A}}_{01}\rangle+\mathcal{O}(\epsilon)
\,,\label{scheme10}\\
|\mathcal{\bar{A}}_{20}\rangle - |\mathcal{A}_{2}\rangle &=
\bigg{[}\frac{-3}{16\epsilon^{3}}\Big{[}(C_{A}\#g + C_{F}\#q)
(\beta^{s}_{20}-\bar{\beta}^{s}_{20})\gamma^{\text{cusp}}_{10}\Big{]}
\nonumber\\
&+\frac{1}{16\epsilon^{2}}\Big{[}(C_{A}\#g+C_{F}\#q) 
(\gamma^{\text{cusp}}_{20}-\bar{\gamma}^{\text{cusp}}_{20})
-2\#g (-2\beta^{s}_{20}\gamma^{g}_{10}
+\#g (\gamma^{g}_{10}-\bar{\gamma}^{g}_{10})^{2}\nonumber\\
&+2\bar{\beta}^{s}_{20} \bar{\gamma}^{g}_{10})+(\beta^{s}_{20}-\bar{\beta}^{s}_{20})
\Big{(}4\#q\gamma^{q}_{10}+2\gamma^{\text{cusp}}_{10}
\sum_{(i,j)}\textbf{T}_{i}\cdot\textbf{T}_{j}\ln(\frac{\mu^{2}}{-s_{ij}})\Big{)}\Big{]}\nonumber\\
&+\frac{1}{8\epsilon}\Big{[}2\#g(\bar{\gamma}^{g}_{20}-\gamma^{g}_{20})+2\#q(
\bar{\gamma}^{q}_{20}-\gamma^{q}_{20})\nonumber\\
&\qquad +(\bar{\gamma}^{\text{cusp}}_{20}-\gamma^{\text{cusp}}_{20})
\sum_{(i,j)}\textbf{T}_{i}\cdot\textbf{T}_{j}\ln(\frac{\mu^{2}}{-s_{ij}})
\Big{]}\bigg{]}|\mathcal{A}_{0}\rangle\nonumber\\
&+\bigg{[}\frac{-1}{4\epsilon^{2}}(C_{A}\#g +C_{F}\#q)\gamma^{\text{cusp}}_{10}
\nonumber\\
&+\frac{1}{4\epsilon}\Big{(}2\#g\bar{\gamma}^{g}_{10}+2\#q\gamma^{q}_{10}+\gamma^{\text{cusp}}_{10}
\sum_{(i,j)}\textbf{T}_{i}\cdot\textbf{T}_{j}\ln(\frac{\mu^{2}}{-s_{ij}})
\Big{)}\bigg{]}|\mathcal{A}^{\text{diff}}_{10}\rangle\nonumber\\
&+\bigg{[}\frac{1}{2\epsilon}\#g(\bar{\gamma}^{g}_{10}-\gamma^{g}_{10})\bigg{]}
|\mathcal{A}^\text{fin}_1\rangle+\mathcal{O}(\epsilon)\,,\\
|\mathcal{\bar{A}}_{11}\rangle &=
\bigg{[}\frac{-1}{4\epsilon^{2}}\#q\Big{(}\bar{\beta}^{e}_{11}
+\#g(\bar{\gamma}^{g}_{10}-\gamma^{g}_{10})\Big{)}\bar{\gamma}^{q}_{01}
+\frac{1}{4\epsilon}(\#g\bar{\gamma}^{g}_{11}
+\#q\bar{\gamma}^{q}_{11})\bigg{]}|\mathcal{A}_{0}\rangle\nonumber\\
&+\bigg{[}-\frac{1}{4\epsilon^{2}}(C_{A}\#g +C_{F}\#q)\gamma^{\text{cusp}}_{10}
\nonumber\\
&+\frac{1}{4\epsilon}\Big{(}2\#g\bar{\gamma}^{g}_{10}
+2\#q\gamma^{q}_{10}+\gamma^{\text{cusp}}_{10}
\sum_{(i,j)}\textbf{T}_{i}\cdot\textbf{T}_{j}\ln(\frac{\mu^{2}}{-s_{ij}})
\Big{)}\bigg{]}|\mathcal{\bar{A}}_{01}\rangle\nonumber\\
&+\bigg{[}\frac{1}{2\epsilon}\#q\bar{\gamma}^{q}_{01}\bigg{]}
|\mathcal{A}^{\text{diff}}_{10}\rangle
+\bigg{[}\frac{1}{2\epsilon}\#q\bar{\gamma}^{q}_{01}\bigg{]}
|\mathcal{A}^\text{fin}_1\rangle+\mathcal{O}(\epsilon)\, ,
\end{align}
\end{subequations}
where we have defined
\begin{subequations}
\begin{align}
\vert\mathcal{A}^{\text{diff}}_{10}\rangle
&=\vert\bar{\mathcal{A}}_{10}\rangle-\vert\mathcal{A}_{1}\rangle \,,
\phantom{\frac{1}{1}}\\
\vert\mathcal{A}^\text{fin}_1\rangle&=
\lim_{\epsilon\to 0}\bigg{[}
\delta\textbf{Z}_{1}\vert\mathcal{A}_{0}\rangle
+\vert\mathcal{A}_{1}\rangle\bigg{]}=
\lim_{\epsilon\to 0}\bigg{[}
\delta\bar{\textbf{Z}}_{10}\vert\mathcal{A}_{0}\rangle
+\vert\bar{\mathcal{A}}_{10}\rangle\bigg{]}\, .
\end{align}
\end{subequations}
$\vert\mathcal{A}^\text{fin}_1\rangle$ is the NLO approximation to
$\vert\mathcal{A}_{\text{fin}}\rangle$ and, thus, a finite and scheme
independent quantity. The one-loop quantities
$\vert\mathcal{A}^{\text{diff}}_{10}\rangle$ and
$\vert\bar{\mathcal{A}}_{01}\rangle$ have to be known up to
$\mathcal{O}(\epsilon^{2})$ terms.

We remark that \Eqn{scheme6} allows to obtain the
$\mathcal{O}(\alpha_e^2)$ contribution of a two-loop amplitude in
\FDH\ up to $\mathcal{O}(\epsilon)$ terms directly from the tree-level
amplitude. This is due to the fact that $\bar{\gamma}^q_{01} \sim
N_\epsilon \sim \epsilon$ and hence the coefficient multiplying
$|\bar{\mathcal{A}}_{01}\rangle$ in \Eqn{scheme10} is
finite. Therefore, we can use \Eqn{eq:ZA01} and with the explicit
expressions of the anomalous dimensions we get
\begin{align}
|\bar{\mathcal{A}}_{02}\rangle &= C_{F}\,\# q\,\bigg{[}
  \frac{2\,C_{F}-C_{A}+N_{F}T_{R}}{2\,\epsilon}
+ \frac{1}{8}\big{(}4\,C_{A}+C_{F}\,(\# q -4)-6\, N_{F}T_{R}\big{)}
  \bigg{]}|\mathcal{A}_{0}\rangle 
  + \mathcal{O}(\epsilon) \,. 
\label{scheme2}
\end{align} 
For a process with no external quarks, $\# q=0$ there are no
$\mathcal{O}(\alpha_e^2)$ terms at NNLO, as can easily be confirmed on
a diagrammatic level.

As mentioned several times, once the UV renormalization has been
carried out, there is no need any longer to distinguish between the
different couplings. After setting $\alpha_{e}=\alpha_{s}$ the full
difference is given by
\begin{align}
|\bar{\mathcal{A}}_{2}\rangle - |\mathcal{A}_{2} \rangle &=
\bigg{[}-\frac{1}{4\epsilon^{2}}C_{A}(\#g \, C_{A}+\#q \, C_{F})
\nonumber\\
&+\frac{1}{36\epsilon}\Big{[}-14\, C_{A}^{2}\#g
-18\,C_{F}\#q(C_{F}-N_{F}T_{R})
+C_{A}(-19\,C_{F}\#q+8\,N_{F}\#g T_{R})\nonumber\\
&+6\,C_{A}\sum_{(i,j)}\textbf{T}_{i}\cdot\textbf{T}_{j}
\log(\frac{\mu^{2}}{-s_{ij}})\Big{]}\nonumber\\
&+\frac{1}{216}\Big{[}C_{A}^{2}\#g(398-3\#g-3\pi^{2})
+C_{A}C_{F}\#q(869-18\#g+9\pi^{2})\nonumber\\
&-9\,C_{F}\Big{(}C_{F}\#q\big{(}3\#q+4(9+\pi^{2})\big{)}
+6\,N_{F}(4\#g+3\#q)T_{R}\Big{)}\nonumber\\
&-96\,C_{A}\sum_{(i,j)}\textbf{T}_{i}\cdot\textbf{T}_{j}
\log(\frac{\mu^{2}}{-s_{ij}})\Big{]}\bigg{]}|\mathcal{A}_{0}\rangle
\nonumber\\
&+\bigg{[}-\frac{1}{\epsilon^{2}}(\#g C_{A}+\#q C_{F})\nonumber\\
&+\frac{1}{6\epsilon}\Big[-11\,C_A\#g-9\,C_F\#q+4\,N_F\#g T_R
+6\sum_{(i,j)}\textbf{T}_{i}\cdot\textbf{T}_{j}
\log(\frac{\mu^{2}}{-s_{ij}})\Big{]}\nonumber\\
&+\frac{1}{6}(\#g C_{A}+3\#q C_{F})\bigg{]}
|\mathcal{A}_{1}^\text{diff}\rangle\nonumber\\
&+\bigg{[}\frac{1}{6}(\#g C_{A}+3\#q
  C_{F})\bigg{]}|\mathcal{A}^\text{fin}_1\rangle\, ,
\label{scheme4}
\end{align}
where we have introduced the notation
\begin{subequations}
\begin{align}
|\bar{\mathcal{A}}_{2}\rangle&=
|\mathcal{\bar{A}}_{20}\rangle+|\mathcal{\bar{A}}_{02}\rangle
+|\mathcal{\bar{A}}_{11}\rangle\, ,\phantom{\frac{1}{1}}\\ 
|\mathcal{A}_{1}^\text{diff}\rangle&=|\mathcal{\bar{A}}_{10}\rangle
+|\mathcal{\bar{A}}_{01}\rangle-|\mathcal{A}_{1}\rangle \, .
\phantom{\frac{1}{1}}
\end{align}
\end{subequations}

\subsection{NNLO $2\to 2$ amplitudes in HV and FDH in massless QCD }
\label{subsec:2To2}

\begin{figure}
\begin{center}
\includegraphics[width=0.9\textwidth]{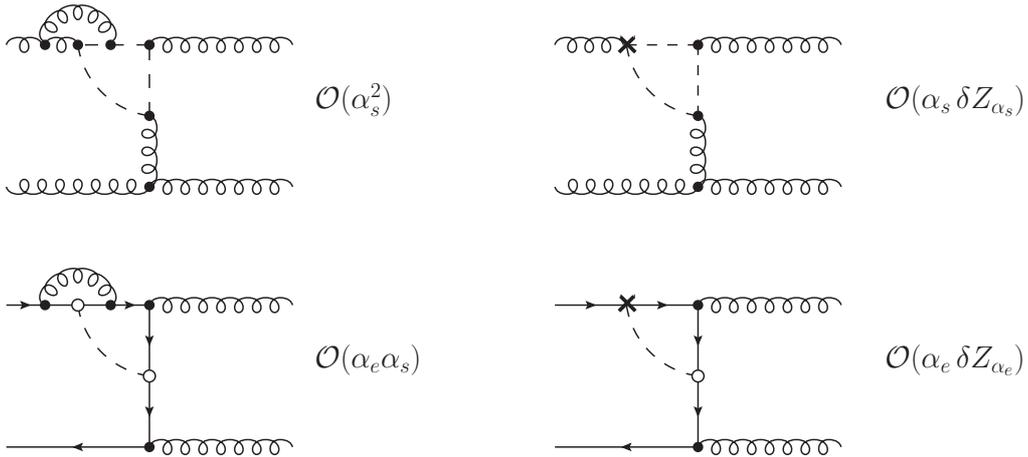}
\caption{\label{fig:2To2} Examples of two-loop (left panel) and
  one-loop counterterm (right panel) diagrams for $gg\to gg$ (top
  panel) and $q\bar{q}\to gg$ (bottom panel). Black vertices denote
  couplings $g_s$ whereas white vertices denote couplings $g_e$, and
  crosses denote counterterm insertions. For $gg\to gg$ at one-loop,
  there are no contributions with couplings $g_e$. The order is given
  relative to the Born term $|\mathcal{A}_0\rangle \sim
  \mathcal{O}(\alpha_s)$.}
\end{center}
\end{figure}

As an example for the transition rules derived in the previous
subsection, we consider the two-loop amplitudes $g g\to g g$ and
$q \bar{q}\to g g $ for massless quarks. Initially the interference of
these two-loop amplitudes with the tree-level amplitudes was
calculated in \CDR~\cite{Glover:2001af, Anastasiou:2001sv}. Later the
helicity amplitudes were computed and explicit results in the \HV\,
and \FDH\, scheme were given~\cite{Bern:2002tk, Bern:2003ck}.
However, for the computation and the UV renormalization procedure in
the \FDH\ scheme, no distinction 
between $\alpha_{s}$ and $\alpha_{e}$ (and $\alpha_{4\epsilon,i}$) was
made. For the process $g g\to g g$ this is of no consequence, but 
for $q \bar{q}\to g g$ this will lead to an incorrect UV
renormalization.  As shown in
Refs.~\cite{Jack:1993ws,Jack:1994bn,Kilgore:2012tb} this leads to
incorrect finite terms which violate unitarity. For our purposes it
also matters because an incorrectly renormalized amplitude cannot be
consistent with the IR structure and transition rules discussed above.

Hence, in order to check the validity of the transition rules we first
need to correct the renormalization of the  $q \bar{q}\to g g$ result
of Ref.~\cite{Bern:2003ck}. Figure~\ref{fig:2To2} shows diagrams which
illustrate the problem. The left panels show genuine two-loop diagrams
to $g g\to g g$
and $q \bar{q}\to g g$. One of them depends on $\alpha_e$, but setting
$\alpha_e=\alpha_s$ in these two-loop diagrams causes no
problem. However, the diagrams have subdivergences, which should be
cancelled by suitable counterterm diagrams, such as the ones in the
right panels. The first of these counterterm diagrams depends on the
one-loop renormalization constant $\delta Z_{\alpha_s}$, but the second one
depends on $\delta Z_{\alpha_e}$, which differs by a divergent amount. If,
as in Ref.~\cite{Bern:2003ck}, this renormalization constant is
effectively replaced by $\delta Z_{\alpha_s}$, the subdivergence is not
properly subtracted, and the final result will not be correct. 

The correct renormalization procedure requires to compute the
lower-order amplitudes for individual couplings. At tree-level, the
amplitudes $|\bar{\mathcal{A}}_{0}\rangle$ for both processes are
proportional to $\alpha_s$
and hence are correctly renormalized by multiplying with
$Z_{\alpha_s}$. At the one-loop level, the amplitudes receive
contributions of $\mathcal{O}(\alpha_s)$  or $\mathcal{O}(\alpha_e)$
relative to tree-level. The latter contribution
$|\bar{\mathcal{A}}_{01}\rangle$ must be renormalized by
multiplication with $Z_{\alpha_s} Z_{\alpha_e}$.

The difference between the two processes $g g\to g g$ and $q
\bar{q}\to g g$ is that for the former process,
$|\bar{\mathcal{A}}_{01}\rangle$ happens to vanish. This is the reason
why for this process the identification $\alpha_s=\alpha_e$ causes no
problem.  In order to restore the correct renormalization for the
latter process, we have computed the $\mathcal{O}(\alpha_s\,
\alpha_e)$ contribution to the one-loop amplitudes. We have then
renormalized this contribution using $Z_{\alpha_s} Z_{\alpha_e}$ and
add the resulting NNLO term to the explicit results of
Ref.~\cite{Bern:2003ck}. We also subtracted the corresponding terms
obtained with the renormalization factor $Z^2_{\alpha_s}$ that had
been applied in Ref.~\cite{Bern:2003ck}.

We have compared the difference between the \FDH\ and \HV\ amplitudes
for both processes with the prediction given by \Eqn{scheme4} and have
found full agreement. This is a further non-trivial confirmation that
our treatment of the scheme dependence is process independent and
applicable at least to NNLO. It is also an independent verification of
the correctness of the anomalous dimensions in \FDH.

\subsection{Transition between  FDH and  DRED}
\label{subsec:fdh-dred}

The transition rules between \DRED\ and \FDH\  can be derived
similarly but are more involved.  To illustrate their structure let us
first consider a process with a single external gluon. The explicit
calculation of the UV renormalized matrix element in \DRED\ yields
$\mathcal{M}^{\mDRED}(g)$ that can be written as
\begin{align}\label{dreduv}
\mathcal{M}^{\mDRED}(g) =
 \mathcal{M}^{\mDRED}(\ghat) 
+ \mathcal{M}^{\mDRED}(\gtilde) = 
2\, \text{Re}\, \langle\mathcal{A}_0^{\ghat}|
\mathcal{A}^{\ghat}\rangle +
2\, \text{Re}\, \langle\mathcal{A}_0^{\epsilon}|
\mathcal{A}^{\epsilon}\rangle \, ,
\end{align}
where we have introduced the shorthand notation $\mathcal{A}^{\ghat}
\equiv \mathcal{A}^{\mDRED}(\ghat)$ and $\mathcal{A}^{\epsilon}
\equiv \mathcal{A}^{\mDRED}(\gtilde)$ etc, and suppressed other
arguments compared to Section \ref{sec:dred}.
We would like to find a relation between  $\mathcal{M}^{\mDRED}(g)$
and the corresponding result in \FDH, 
\begin{align}\label{fdhuv}
\mathcal{M}^{\mFDH}(g) &=
2\, \text{Re}\,
\langle\bar{\mathcal{A}}_0^{g}|
\bar{\mathcal{A}}^{g}\rangle \equiv
2\, \text{Re}\,
\langle\mathcal{A}_0^{\mFDH}(g)|
\mathcal{A}^{\mFDH}(g)\rangle\, .
\end{align}
To do so, we start from the equality of the IR subtracted amplitudes
computed in \DRED\ and \FDH, written with a similar shorthand notation
for the $\textbf{Z}$-factors as
\begin{align}\label{dredfdhsub}
& \langle\mathcal{A}_0^{\ghat}|
\big(\textbf{Z}^{\ghat}\big)^{-1}|
\mathcal{A}^{\ghat}\rangle +
\langle\mathcal{A}_0^{\epsilon}|
\big(\textbf{Z}^{\epsilon}\big)^{-1}|
\mathcal{A}^{\epsilon}\rangle
= \langle\bar{\mathcal{A}}_0^{g}|
\big(\bar{\textbf{Z}}^{g}\big)^{-1}|
\bar{\mathcal{A}}^{g}\rangle + \mathcal{O}(\epsilon)   \, ,
\end{align} 
where we have set $N_\epsilon = 2 \epsilon$.  Writing $\textbf{Z}^{-1}
= 1 + \delta\textbf{Z}$, where $\delta\textbf{Z}$ denote the
perturbatively expanded higher-order terms we obtain an equation
analogous to (\ref{hvfdhv1}),
\begin{align}
\begin{split}
\mathcal{M}^{\mDRED}(g) &+
2\, \text{Re}\, 
\langle\mathcal{A}_0^{\ghat}|
\delta\textbf{Z}^{\ghat}|
\mathcal{A}^{\ghat}\rangle +
2\, \text{Re}\, 
\langle\mathcal{A}_0^{\epsilon}|
\delta\textbf{Z}^{\epsilon}|
\mathcal{A}^{\epsilon}\rangle
\phantom{\frac{1}{1}}\\
=\, \mathcal{M}^{\mFDH}(g) &+
2\, \text{Re}\, 
\langle\bar{\mathcal{A}}_0^{g}|
\delta\bar{\textbf{Z}}^{g}|
\bar{\mathcal{A}}^{g}\rangle + \mathcal{O}(\epsilon) \, .
\phantom{\frac{1}{1}}
\label{fdhdred}
\end{split}
\end{align}
If the expansion coefficients $\delta\textbf{Z}$ are known to
$\mathcal{O}(\alpha^n)$ and the amplitudes $|\mathcal{A}\rangle$ are
known to $\mathcal{O}(\alpha^{n-1})$, \Eqn{fdhdred} allows to obtain a
relation between the $\mathcal{O}(\alpha^n)$ squared matrix element
computed in \DRED\ and \FDH, up to $\mathcal{O}(\epsilon)$ terms.  For
this relation, the knowledge of $\textbf{Z}^{\epsilon} \equiv
\textbf{Z}^{\mDRED}(\gtilde)$ is required, even though
\Eqn{dredfdhsub} is still correct if the second term on the
l.h.s. containing $\textbf{Z}^{\epsilon}$ is dropped.

As a concrete example we consider the process $H\to g\,g$ in \FDH\ and
\DRED\ and work out the transition rules between the two schemes for the
UV renormalized two-loop squared  amplitudes.  For simplicity we also set
$\alpha_{e}=\alpha_{4\epsilon}=\alpha_{s}$.

As we have $\# g=2$ external gluons, in \DRED\ the squared matrix
element is to be written as a sum over $2^{\# g}=4$ terms. However,
in this particular case two of these terms vanish to all orders,
resulting in
\begin{equation}
\mathcal{M}^\mDRED(g,g)
=  \mathcal{M}(\ghat,\ghat) + \mathcal{M}(\gtilde,\gtilde) \, .
\end{equation}
Writing explicitly the equality of the subtracted matrix elements in
\FDH\ and \DRED\ we get
\begin{align}
&\ \langle \bar{\mathcal{A}}_{0}\vert\bigg{(}
1+\delta\bar{\textbf{Z}}_1\Big{(}\frac{\alpha_{s}}{4\pi}\Big{)}
 +\delta\bar{\textbf{Z}}_{2}\Big{(}\frac{\alpha_{s}}{4\pi}\Big{)}^{2}\bigg{)}
\bigg{(}\vert \bar{\mathcal{A}}_{0}\rangle
  +\vert \bar{\mathcal{A}}_{1}\rangle \Big(\frac{\alpha_s}{4\pi}\Big)
  +\vert \bar{\mathcal{A}}_{2}\rangle \Big(\frac{\alpha_s}{4\pi}\Big)^2
  \bigg{)}\nonumber \\
=&\ \langle \mathcal{A}^{\ghat \ghat}_{0}\vert
\bigg{(}1+\delta\textbf{Z}_{1}^{\ghat \ghat}\Big{(}\frac{\alpha_{s}}{4\pi}\Big{)}
+\delta\textbf{Z}_{2}^{\ghat \ghat}\Big{(}\frac{\alpha_{s}}{4\pi}\Big{)}^{2}\bigg{)}
 \bigg{(}\vert \mathcal{A}^{\ghat \ghat}_{0}\rangle
 +\vert \mathcal{A}^{\ghat \ghat}_{1}\rangle \Big(\frac{\alpha_s}{4\pi}\Big)
 +\vert \mathcal{A}^{\ghat \ghat}_{2}\rangle \Big(\frac{\alpha_s}{4\pi}\Big)^2
 \bigg{)} \nonumber \\
+&\ \langle \mathcal{A}^{\epsilon \epsilon}_{0}\vert
\bigg{(}1+\delta\textbf{Z}_{1}^{\epsilon \epsilon}\Big{(}\frac{\alpha_{s}}{4\pi}\Big{)}
+\delta\textbf{Z}_{2}^{\epsilon \epsilon}\Big{(}\frac{\alpha_{s}}{4\pi}\Big{)}^{2}\bigg{)}
 \bigg{(}\vert \mathcal{A}^{\epsilon \epsilon}_{0}\rangle
 +\vert \mathcal{A}^{\epsilon \epsilon}_{1}\rangle \Big(\frac{\alpha_s}{4\pi}\Big)
 +\vert \mathcal{A}^{\epsilon \epsilon}_{2}\rangle \Big(\frac{\alpha_s}{4\pi}\Big)^2
 \bigg{)}  \nonumber \\
+&\ \mathcal{O}(\epsilon) +\mathcal{O}(\alpha^{3}) \, .
\label{fdh-dred}
\end{align}
In \Eqn{fdh-dred} we have introduced a compact notation for the
perturbative coefficients of the amplitudes and $\textbf{Z}^{-1}$ in
\DRED: $\vert \mathcal{A}^{\epsilon \epsilon}_{2}\rangle \equiv \vert
\mathcal{A}_{2}(\gtilde,\gtilde)\rangle$ and
\begin{align}
\textbf{Z}^{-1}(\ghat,\ghat) = 1 +
  \delta\textbf{Z}_{1}^{\ghat\ghat} \Big(\frac{\alpha_s}{4\pi}\Big) +
    \delta\textbf{Z}_{2}^{\ghat\ghat}
      \Big(\frac{\alpha_s}{4\pi}\Big)^2 
+\mathcal{O}(\alpha^{3}) \, ,
\label{Zinvexp}
\end{align}
with analogous expressions for other partonic processes.
Comparing the order $\alpha_{s}$ terms yields
\begin{align}
\label{eq:Dred1transA}
\mathcal{M}^\mDRED_1(g,g)-\mathcal{M}^\mFDH_1(g,g)
&=\mathcal{M}^\mDRED_0(\gtilde,\gtilde)
\frac{(\bar{\gamma}_{010}^{\epsilon}+\bar{\gamma}_{100}^{\epsilon}
 -\bar{\gamma}_{100}^{g})}{\epsilon}
\nonumber \\
&=\mathcal{M}^\mDRED_0(\gtilde,\gtilde)\frac{(2\,N_FT_{R}-C_{A})}{3\epsilon}
 + \mathcal{O}(\epsilon)\, .
\end{align}
This one-loop transition rule is in agreement\footnote{Note that in
  Ref.~\cite{Signer:2008va} a  different convention for the
  $\gamma$'s has been used.} with Ref.~\cite{Signer:2008va}.  To make
this agreement more explicit we write the transition in a more general
way as
\begin{equation}
\label{eq:Dred1transB}
\mathcal{M}^\mDRED_1(g,g)-\mathcal{M}^\mFDH_1(g,g)
=\big(\mathcal{M}^\mDRED_0(g,\gtilde) 
  +\mathcal{M}^\mDRED_0(\gtilde,g) \big) 
  \frac{(2\,N_FT_{R}-C_{A})}{6\epsilon}
 + \mathcal{O}(\epsilon)\, .
\end{equation}
Note that the difference is finite, since the tree-level matrix element
squared on the r.h.s. of \Eqn{eq:Dred1transA} or \Eqn{eq:Dred1transB}
are of $\mathcal{O}(\epsilon)$.

In order to write the scheme difference at NNLO we introduce a similar
short-hand notation for the squared matrix elements as for the
amplitudes, denoting the full tree-level and one-loop contribution for
the $H\to \gtilde\gtilde$ process by
$\mathcal{M}^{\epsilon\epsilon}_{0} \equiv
\mathcal{M}_0(\gtilde,\gtilde)$ and
$\mathcal{M}^{\epsilon\epsilon}_{1} \equiv
\mathcal{M}_1(\gtilde,\gtilde)$, respectively.  The difference can
then be written as
\begin{align}
&\mathcal{M}^\mDRED_2(g,g)-\mathcal{M}^\mFDH_2(g,g) =
\frac{1}{2\epsilon^{3}}\,
C_{A}\,\mathcal{M}^{\epsilon\epsilon}_{0}\bar{\gamma}_{100}^{\textrm{cusp}}
\big{(}\bar{\gamma}_{010}^{\epsilon}+\bar{\gamma}_{100}^{\epsilon}
-\bar{\gamma}_{100}^{g}\big{)}
\nonumber\\
&-\frac{1}{2\epsilon^{2}} \Big{[}
\mathcal{M}^{\epsilon\epsilon}_{0}\big{(}
    \bar{\beta}_{020}^{e}\bar{\gamma}_{010}^{\epsilon}
   +\bar{\beta}_{110}^{e}\bar{\gamma}_{010}^{\epsilon}
   +\bar{\beta}_{200} (\bar{\gamma}_{100}^{\epsilon} - \bar{\gamma}_{100}^{g}) 
   +(\bar{\gamma}_{010}^{\epsilon}+\bar{\gamma}_{100}^{\epsilon})^{2}
   -(\bar{\gamma}_{100}^{g})^{2}\big)
\nonumber\\
&\qquad
+C_{A}\mathcal{M}^{\text{diff}}_{1}\bar{\gamma}_{100}^{\textrm{cusp}}
-C_{A}\mathcal{M}_{0}^{\epsilon\epsilon}\bar{\gamma}_{100}^{\textrm{cusp}}
\bar{\gamma}_{010}^{\epsilon}\ln(-\frac{\mu^{2}}{s}) \Big{]}
\nonumber\\
&+\frac{1}{2\epsilon}\Big{[}
2\,\mathcal{M}^{\epsilon\epsilon}_{1}(\bar{\gamma}_{010}^{\epsilon}
+\bar{\gamma}_{100}^{\epsilon}-\bar{\gamma}_{100}^{g})
+\mathcal{M}^{\epsilon\epsilon}_{0}(\bar{\gamma}_{002}^{\epsilon}
+\bar{\gamma}_{020}^{\epsilon}+\bar{\gamma}_{110}^{\epsilon}
+\bar{\gamma}_{200}^{\epsilon}-\bar{\gamma}_{110}^{g}
-\bar{\gamma}_{200}^{g})
\nonumber\\
&\qquad+2\,\mathcal{M}^{\text{diff}}_1\bar{\gamma}_{100}^{g}
-C_{A}\mathcal{M}^{\text{diff}}_{1}\bar{\gamma}_{100}^{\textrm{cusp}}
\ln(-\frac{\mu^{2}}{s})\Big{]}+
\mathcal{O}(\epsilon)
\label{diff:dred-fdh}\, ,
\end{align}
where we have introduced the one-loop difference 
\begin{align}
\mathcal{M}^{\text{diff}}_{1} \equiv
\mathcal{M}^\mDRED_1(g,g)-\mathcal{M}^\mFDH_1(g,g) \, .
\end{align}
Note that the squared matrix elements
$\mathcal{M}^{\epsilon\epsilon}_{0}$ and
$\mathcal{M}^{\epsilon\epsilon}_{1}$ are of $\mathcal{O}(\epsilon)$
and $\mathcal{M}^{\text{diff}}_{1}$ needs to be known up to
$\mathcal{O}(\epsilon^2)$. Using the explicit results for the
anomalous dimensions \Eqn{diff:dred-fdh} translates into
\begin{align}\label{scheme9}
\mathcal{M}_{2}^{\mDRED}-\mathcal{M}_{2}^{\mFDH}&=
-\frac{2}{3\epsilon^{3}}C_A \mathcal{M}^{\epsilon\epsilon}_{0}\,(C_A-2\,N_FT_R)
+\frac{1}{\epsilon^{2}}\Big{[}
-\frac{2}{3}(3\, C_{A}\,\mathcal{M}^{\text{diff}}_1
+2\, C_{A}^{2}\mathcal{M}^{\epsilon\epsilon}_{0}\nonumber \\
&-5\,C_{A}\,N_FT_{R}\mathcal{M}^{\epsilon\epsilon}_{0}
+3\,C_{F}\, N_FT_{R}\mathcal{M}^{\epsilon\epsilon}_{0})
+4\,C_{A}\, N_FT_{R} \ln(-\frac{\mu^{2}}{s})\mathcal{M}^{\epsilon\epsilon}_{0}
\Big{]}\nonumber \\
&+\frac{1}{18\epsilon}\Big{[}C_{A}(
-66\, \mathcal{M}^{\text{diff}}_1-6\,\mathcal{M}^{\epsilon\epsilon}_{1}
+C_{A}\mathcal{M}^{\epsilon\epsilon}_{0}(-37+2\pi^{2}))\nonumber \\
&+2\,N_F\,T_{R}(12\,\mathcal{M}^{\text{diff}}_1
-9\,C_{F}\mathcal{M}^{\epsilon\epsilon}_{0}
+6\,\mathcal{M}^{\epsilon\epsilon}_{1}
-2\,C_{A}\mathcal{M}^{\epsilon\epsilon}_{0}(-11+\pi^{2}))
\phantom{\frac{1}{1}}\nonumber \\
&-36\,C_{A}\mathcal{M}^{\text{diff}}_1 \ln(-\frac{\mu^{2}}{s})\Big{]}
+\frac{1}{3}C_{A}(\mathcal{M}^{\text{diff}}_1-\mathcal{M}^{\epsilon\epsilon}_{1})
+\mathcal{O}(\epsilon)\, .
\end{align}
We have checked our prediction \Eqn{scheme9} with the explicit
calculation of the gluon form factor in \DRED\ and
\FDH~\cite{Broggio:2015ata} and we have obtained full agreement. This
was of course to be expected, as we have verified in
Section~\ref{sec:alt} that the extraction of $\bar{\gamma}^\epsilon$
from the form factor for $H\to \gtilde\gtilde$ is in agreement with
its determination in SCET.

%% file: 06_Conclusion.tex
\section{Concluding remarks }
\label{sec:conclusion}

With the results presented in this paper we complete the understanding
of the scheme dependence of IR divergent NNLO virtual amplitudes with
massless particles. In particular, we have presented the
generalization of this dependence to \DRED, where we have to consider
amplitudes with external $\epsilon$-scalars and, hence, need the
corresponding anomalous dimension $\bar\gamma_\epsilon$. Furthermore,
we have presented a SCET approach to the scheme dependence and derived
all anomalous dimensions again in this approach. In this way \FDH\ and
\DRED\ are shown to be perfectly consistent IR regularization schemes
(at least) up to NNLO, as long as the UV renormalization is done
consistently.  Concretely, this means that the various couplings
$\alpha_s$, $\alpha_e$ and $\alpha_{4\epsilon, i}$ have to be
distinguished. This is also the case in \FDH, where at NNLO the only
concrete modification appears due to the UV renormalization of the NLO
virtual amplitudes.  Our results and definitions of \FDH\ are
perfectly consistent with the results and definitions proposed in
\cite{Kilgore:2011ta,Kilgore:2012tb}.

Obviously, the virtual amplitudes are not the only ingredients needed
for a calculation of a physical quantity. At NNLO, also double-real
and real-virtual corrections are to be considered. Furthermore, if
there are initial state hadrons, a counterterm for the initial-state
collinear singularities is required. All these additional
contributions are also regularization-scheme dedendent and only once
all parts are combined to a physical cross section, the
regularization-scheme dependence cancels.

In virtually all NNLO calculations of cross sections completed so far,
\CDR\ has been used. The results presented in this paper allow for
using any of the other regularization schemes for the calculation of
the virtual corrections. Using a scheme different from \CDR\ often
facilitates the use of efficient calculational techniques for loop
amplitudes. The results can then be translated to obtain the virtual
corrections in \CDR\ and can be combined with the additional parts
mentioned above, obtained again in \CDR.

Of course, it is not imperative to treat the additional contributions
(i.e. the contributions other than the NNLO virtual corrections) in
\CDR. Also for these terms other schemes might offer advantages. In
fact, a modification of a subtraction scheme at NNLO to the
\HV\ scheme has been presented recently~\cite{Czakon:2014oma},
resulting in a reduction of the algebraic complexity.

The question of the scheme (in)dependence of a full cross section at
NNLO becomes particularly transparent if the calculation is performed
in a SCET inspired way. Following ideas of the slicing
method~\cite{Giele:1991vf} and the $q_T$-subtraction
method~\cite{Catani:2007vq}, the cross section is split into two
regions, a 'hard' region and a 'soft' region. In the hard region not
all radiation in addition to the final state under consideration is
soft (or collinear). At least one of the emitted gluons is hard. Here
we are effectively dealing with a NLO calculation of a process for a
final state with an additional parton and the scheme independence of
cross sections at NLO is well established~\cite{Signer:2008va}. In the
soft region all additional radiation is soft (or collinear) and a true
NNLO calculation is required. For this part a SCET approach is
used. This idea has first been applied to the decay of a top
quark~\cite{Gao:2012ja} $t\to W\, b\, X$ where the invariant mass of
the jet $b+X$ has been used for the split. Recently, the N-jettiness
event-shape variable has been used to obtain a similar setup for
differential NNLO calculations of Higgs plus
jet~\cite{Boughezal:2015dva}, $W$ plus jet~\cite{Boughezal:2015aha}
and Drell-Yan production~\cite{Gaunt:2015pea}.

In the soft region, the cross section factorizes into a product of
hard-, soft- and jet functions (and beam functions if there are
initial-state hadrons). The corresponding bare functions are all IR
divergent and scheme dependent. However, we have shown that the
properly IR subtracted soft function $s_{\text{fin}}$, \Eqn{eq:Sfin},
and jet functions $\jmath_{q\,\text{fin}}$ and
$\jmath_{g\,\text{fin}}$, \Eqns{eq:Jfin}{eq:Jgfin}, are not only
finite but also scheme independent, at least up to NNLO. The same
holds true for the hard function \cite{Kelley:2010fn,Broggio:2014hoa}
that is closely related to $\mathcal{M}_{\mathrm{fin}}$,
\Eqn{eq:fin}. Hence the cross section in the soft limit can be
expressed in terms of these IR subtracted quantities in a manifestly
scheme-independent way.

The soft function that is required for the processes mentioned above
is not the soft function for Drell-Yan or Higgs production as we have
computed. However, the procedure to perform the IR subtraction (or UV
renormalization in SCET language) consistent with the regularization
scheme used in the computation of the bare soft function is exactly
the same.

Since the soft, hard and jet functions are separately scheme
independent, it is possible to use different schemes in the
computation of the various parts contributing to the cross
section. For example, the calculation of the virtual corrections
(i.e. the hard function) in \FDH, where the helicity and unitarity
methods are applicable, can easily be combined with the soft or jet
function computed in \CDR. We are convinced that this flexibility will
be very beneficial for further developments of fully differential NNLO
calculations.

%% file: 10_Appendix.tex
\section{Explicit expressions for the soft and jet functions}
\label{app:A}

In this appendix we give the explicit results for several quantities
as a perturbative expansion. We use the conventions specified in
Section~\ref{sec:outlinescet}. For most results it will be 
sufficient to expand a quantity $X$ in $\alpha_s$ and $\alpha_e$ and
write, instead of \Eqn{eq:generalex},
\begin{align} \label{eq:generalexApp}
X^{\mRS} = \sum^{\infty}_{m,n}
\left(\frac{\alpha_s}{4 \pi}\right)^{m}
\left(\frac{\alpha_e}{4 \pi}\right)^{n}\,
X^{\mRS}_{mn} \, .
\end{align} 
As in \Eqn{barconvention} we will use the short-hand notation $X_{mn}\equiv
X_{mn}^{\mHV}=X_{mn}^{\mCDR}$ and $\bar{X}_{mn}\equiv
X_{mn}^{\mFDH}=X_{mn}^{\mDRED}$. The explicit results for
scheme-dependent quantities will be given in the \FDH/\DRED\ scheme
but we can obtain the corresponding coefficients in the
\HV/\CDR\ scheme as $X_{mn} = \lim_{N_\epsilon\to 0} \bar{X}_{mn}$.

\subsection{Soft functions}
\label{app:S}

It is convenient to solve the RGEs for the soft functions in
\Eqn{rge:softG} order by order in $\alpha_s$. By using the
expansion coefficients of the anomalous dimensions in
\Eqn{def:Cusp}  one obtains the following 
scheme independent result
\begin{align}\label{eq:sfinres}
		s_{\text{fin}}(\kappa,\mu) = 
1 &+ \left(\frac{\alpha_s}{4 \pi}\right)\Big[2\Gamma_{10} L^2_\kappa 
+ 2\gamma^{W}_{10} L_\kappa + c^W_{1}\Big] \nonumber \\
& + \left(\frac{\alpha_s}{4 \pi}\right)^2 \Big[2
  \left(\Gamma_{10}\right)^2 L^4_\kappa 
-\frac{4\Gamma_{10}}{3}\left(\beta^{s}_{20}
-3\gamma^{W}_{10} \right) L^3_\kappa \nonumber \\
& + 2 \left(\Gamma_{20}+ \left(\gamma^{W}_{10}\right)^2 
-\beta^{s}_{20} \gamma^{W}_{10} 
+ \Gamma_{10} c^W_{1}\right) L^2_\kappa \nonumber \\
&+ 2\left(\gamma^{W}_{20} + \gamma^{W}_{10} c^W_1 
-\beta^{s}_{20} c^W_1\right) L_\kappa +c^W_{2}\Big] \, ,
\end{align}
where $\Gamma_{\text{cusp}} = C_R \, \gamma_{\text{cusp}}$ and
\begin{subequations}
\begin{align}
\gamma^{W}_{10} & = 0\, ,\\
\gamma^{W}_{20} & =C_R \bigg[
C_A \Big{(}-\frac{808}{27} + \frac{11}{9}\pi^2 +28 \zeta_3\Big{)} 
+ N_F \Big{(}\frac{112}{27} - \frac{2}{9}\pi^2\Big{)} \bigg]
\, ,
\end{align}
\end{subequations}
and the one and two-loop non-logarithmic coefficients have the expressions
\begin{subequations}
\begin{align}
c^W_{1} &= C_R \frac{\pi^2}{3}\, , \\
c^W_{2} &= C_R \Bigg[C_A\left(-\frac{22 \zeta_3}{9}+\frac{2428}{81}
+\frac{67 \pi^2}{54}-\frac{\pi^4}{3}\right)+C_R \frac{\pi^4}{18}
+N_F\left(\frac{4 \zeta_3}{9}-\frac{5\pi^2}{27}
-\frac{328}{81}\right)\Bigg] \, .
\end{align} 
\end{subequations}
The result in \Eqn{eq:sfinres} is in agreement with previous
calculations in \cite{Belitsky:1998tc,Becher:2007ty}.

\subsection{Quark jet function}

Here we list the explicit two-loop coefficients entering \Eqn{eq:Qbare}:
\begin{subequations}
\begin{align}
\bar{\jmath}^{\, q;F}_{20} &=
\frac{8}{\epsilon^{4}}+\frac{12}{\epsilon^{3}}
+\Big{(}\frac{65}{2}-\frac{8\pi^{2}}{3}\Big{)}\frac{1}{\epsilon^{2}}
+\Big{(}\frac{311}{4}-5\pi^{2}-20\,\zeta_{3}\Big{)}\frac{1}{\epsilon}\nonumber\\*
&+\frac{1437}{8}-\frac{57\pi^{2}}{4}+\frac{5\pi^{4}}{18}-54\zeta_{3}\, ,
\\
\bar{\jmath}^{\, q;A}_{20}&=
\frac{11}{3\epsilon^{3}}
+\Big{(}\frac{233}{18}-\frac{\pi^{2}}{3}\Big{)}\frac{1}{\epsilon^{2}}
+\Big{(}\frac{4541}{108}-\frac{11\pi^{2}}{6}
-20\zeta_{3}\Big{)}\frac{1}{\epsilon}
\nonumber\\*
&+\frac{86393}{648}-\frac{221\pi^{2}}{36}-\frac{37\pi^{4}}{180}
-\frac{142}{3}\zeta_{3}
\nonumber\\*
&+\frac{N_{\epsilon}}{2}\Big{(}
   -\frac{1}{3\epsilon^3}-\frac{25}{18\epsilon^2}
  + \big( \frac{\pi^2}{6} - \frac{523}{108 }\big)\frac{1}{\epsilon}
  -\frac{10219}{648} + \frac{25\pi^2}{36}  
   + \frac{8 \zeta_{3}}{3}\Big{)}\, ,
\\
\bar{\jmath}^{\, q;f}_{20}&=
-\frac{4}{3\epsilon^{3}} -\frac{38}{9\epsilon^{2}}
+\Big{(}-\frac{373}{27}+\frac{2\pi^{2}}{3}\Big{)}\frac{1}{\epsilon}
-\frac{7081}{162}+\frac{19\pi^{2}}{9}+\frac{32}{3}\zeta_{3}\, ,
\\
\bar{\jmath}^{\, q;F}_{02} &=
\frac{N_{\epsilon}^{2}}{4}\Big{(}- \frac{1}{2 \epsilon^2} 
- \frac{7}{4 \epsilon}-\frac{33}{8} + \frac{\pi^2}{4}\Big{)}
+\frac{N_{\epsilon}}{2}\Big{(}\frac{2}{\epsilon^2} 
+ \frac{8}{\epsilon}+24 - \pi^2\Big{)}\, ,
\\
\bar{\jmath}^{\, q;A}_{02}&=
\frac{N_{\epsilon}^{2}}{4}\Big{(}\frac{1}{\epsilon^2} + \frac{4}{\epsilon}
+ 12 - \frac{\pi^2}{2}\Big{)}
 +\frac{N_{\epsilon}}{2}\Big{(}-\frac{1}{\epsilon^2} -\frac{4}{\epsilon}-12 
  + \frac{\pi^2}{2}\Big{)}\, ,
\\
\bar{\jmath}^{\, q;f}_{02} &=
\frac{N_{\epsilon}}{2}\Big{(}\frac{1}{\epsilon^2} 
+ \frac{11}{2 \epsilon}+\frac{89}{4} - \frac{\pi^2}{2}\Big{)}\, ,
\\
\bar{\jmath}^{\, q;F}_{11} &=
 \frac{N_{\epsilon}}{2}\Big{(}-\frac{4}{\epsilon^{3}}-\frac{14}{\epsilon^{2}}
+\big(\frac{5\pi^{2}}{3}-39\big)\frac{1}{\epsilon}
-\frac{201}{2}+6\pi^{2}+18\zeta_{3}\Big{)}\, ,
\\
\bar{\jmath}^{\, q;A}_{11}&=
\frac{N_{\epsilon}}{2}\Big{(}-\frac{11}{2 \epsilon}-\frac{129}{4} 
+ \frac{\pi^2}{3} + 6 \zeta_{3}\Big{)}\, .
\end{align}
\end{subequations}
After renormalization and setting $\epsilon\to 0$ we obtain a finite
and scheme independent quark-jet function. The terms containing
$\alpha_e$ cancel and we are left with only $\alpha_s$ dependent
terms. In Laplace space the quark-jet function reads
\begin{align}
\label{qjetL}
\jmath_{q\,\text{fin}}(Q^2,\mu) &=1+
\frac{\alpha_{s}}{4\pi} \bigg{[}
    \Gamma_{10}\frac{L_Q^{2}}{2}
   +\gamma_{10}^{J_{q}}L_Q+c_{1}^{J_{q}}  \bigg{]}
\nonumber\\
&+\Big{(}\frac{\alpha_{s}}{4\pi}\Big{)}^{2} \bigg{[}
\left(\Gamma_{10}\right)^{2}\frac{L_Q^{4}}{8}
 +\Big(-\beta^s_{20}+3\gamma_{10}^{J_{q}}\Big)
        \Gamma_{10}\frac{L_Q^{3}}{6}
\nonumber\\*
& \qquad\qquad\quad +\Big(\Gamma_{20}+\big(\gamma^{J_{q}}_{10}\big)^2
   -\beta^s_{20}\gamma^{J_{q}}_{10}+c_{1}^{J_{q}}\Gamma_{10}\Big) 
     \frac{L_Q^{2}}{2}
\nonumber\\*
&\qquad\qquad\quad
  +\Big(\gamma_{20}^{J_{q}}+\gamma_{10}^{J_{q}}c_{1}^{J_{q}}
    -\beta^s_{20}c_{1}^{J_{q}}\Big)L_Q+c_{2}^{J_{q}}\bigg{]}\, ,
    \phantom{\frac{L_Q^{3}}{6}}
\end{align} 
where here $\Gamma_{\text{cusp}} = C_F \, \gamma_{\text{cusp}}$ and 
\begin{subequations}
\begin{align}
c_{1}^{J_{q}}&=C_{F}\Big{(}7-\frac{2\pi^{2}}{3}\Big{)}\, ,\\
c_{2}^{J_{q}}&=C^2_{F}\Big{(}\frac{205}{8}
   -\frac{97\pi^2}{12}+\frac{61\pi^{4}}{90}-6\zeta_{3}\Big{)}
   +C_{F}C_{A}\Big{(}\frac{53129}{648}-\frac{155\pi^{2}}{36}
    -\frac{37\pi^{4}}{180}-18\zeta_{3}\Big{)}\nonumber\\
&+C_{F}T_{R}N_F\Big{(}\frac{13\pi^{2}}{9}-\frac{4057}{162}\Big{)}
\end{align}
\end{subequations}
and is in agreement with previous results~\cite{Becher:2006qw}.

\subsection{Gluon jet function}

Here we list the explicit two-loop coefficients entering \Eqn{eq:jetgbare}:
\begin{subequations}
\begin{align}
 \bar{\jmath}^{\, g;\, AA}_{20}&=
\frac{8}{\epsilon^{4}}+\frac{55}{3\epsilon^{3}}
+\frac{1}{\epsilon^{2}}\Big{(}-3\pi^{2}+\frac{152}{3}\Big{)}
+\frac{1}{\epsilon}\Big(-40\zeta_3-\frac{143\pi^2}{18}+\frac{3638}{27}\Big)
\nonumber\\
&+\frac{13\pi^{4}}{180}-\frac{352\zeta_{3}}{3}-\frac{617\pi^{2}}{27}+\frac{57415}{162}
\nonumber\\
&+\frac{N_{\epsilon}}{2}\Big{[}-\frac{5}{3\epsilon^{3}}-\frac{62}{9\epsilon^{2}}
 +\frac{1}{\epsilon}\Big{(}\frac{13\pi^{2}}{18}-\frac{214}{9}\Big{)}
+\frac{85\pi^{2}}{27}-\frac{12371}{162}+\frac{32}{3}\zeta_{3}\Big{]}
\nonumber\\
&+\frac{N_{\epsilon}^{2}}{4}\Big{[}\frac{1}{9\epsilon^{2}}
+\frac{16}{27\epsilon}+\frac{56}{27}-\frac{\pi^{2}}{18}\Big{]}\, ,
\\
\bar{\jmath}^{\, g;\, Af}_{20}&=
-\frac{20}{3\epsilon^{3}}-\frac{188}{9\epsilon^{2}}
+\frac{1}{\epsilon}\Big{(}\frac{26\pi^{2}}{9}-\frac{536}{9}\Big{)}
+\frac{80\zeta_{3}}{3}+\frac{262\pi^{2}}{27}-\frac{12880}{81}
\label{eq:GJETcoeff}
\nonumber\\
&+\frac{N_{\epsilon}}{2}\Big{(}\frac{8}{9\epsilon^{2}}
+\frac{104}{27\epsilon}+\frac{320}{27}-\frac{4\pi^{2}}{9}\Big{)}\, ,
\\
\bar{\jmath}^{\, g;\, Ff}_{20}&=
-\frac{2}{\epsilon}-\frac{55}{3}+16\zeta_{3}\, ,
\\
\bar{\jmath}^{\, g;\, ff}_{20}&=
\frac{16}{9\epsilon^{2}}+\frac{160}{27\epsilon}+16-\frac{8\pi^{2}}{9}\, ,
\\
\bar{\jmath}^{\, g;\, Af}_{11}&=
 3\frac{N_{\epsilon}}{2}\, ,
\\
\bar{\jmath}^{\, g;\, Ff}_{11}&=
\frac{N_{\epsilon}}{2}\Big{(}\frac{2}{\epsilon}+11\Big{)}\, .
\end{align}
\end{subequations}
After renormalization and setting $\epsilon\to 0$ we obtain a finite
and scheme independent gluon jet function. The structure in Laplace
space is the same as for the quark jet function, \Eqn{qjetL},
\begin{align}
\label{gjetL}
\jmath_{g\,\text{fin}}(Q^2,\mu) &=1+
\frac{\alpha_{s}}{4\pi} \bigg{[}
    \Gamma_{10}\frac{L_Q^{2}}{2}
   +\gamma_{10}^{J_g}L_Q+c_{1}^{J_g}  \bigg{]}
\nonumber\\
&+\Big{(}\frac{\alpha_{s}}{4\pi}\Big{)}^{2} \bigg{[}
\left(\Gamma_{10}\right)^{2}\frac{L_Q^{4}}{8}
 +\Big(-\beta^s_{20}+3\gamma_{10}^{J_g}\Big)
        \Gamma_{10}\frac{L_Q^{3}}{6}
\nonumber\\
& \qquad\quad\quad +\Big(\Gamma_{20}+\big(\gamma^{J_g}_{10}\big)^2
   -\beta^s_{20}\gamma^{J_g}_{10}+c_{1}^{J_{g}}\Gamma_{10}\Big) 
     \frac{L_Q^{2}}{2}
\nonumber \\
& \qquad\quad\quad
  +\Big(\gamma_{20}^{J_g}+\gamma_{10}^{J_g}c_{1}^{J_g}
    -\beta^s_{20}c_{1}^{J_g}\Big)L_Q+c_{2}^{J_g}\bigg{]}\, ,
\phantom{\frac{L_Q^{2}}{2}}
\end{align} 
where here $\Gamma_{\text{cusp}} = C_A \, \gamma_{\text{cusp}}$.
The coefficients are given by
\begin{subequations}
\begin{align}
c_{1}^{J_{g}}&=
 C_{A}\Big{(}\frac{67}{9}-\frac{2\pi^{2}}{3}\Big{)}-\frac{20}{9}N_FT_{R}\, ,\\
c_{2}^{J_{g}}&=
 C_{A}^{2}\Big{(}\frac{20215}{162}-\frac{362\pi^{2}}{27}-\frac{88\zeta_{3}}{3}
    +\frac{17\pi^{4}}{36}\Big{)}\nonumber\\
&+C_{A}N_FT_{R}\Big{(}-\frac{1520}{27}+\frac{134\pi^{2}}{27}
   -\frac{16\zeta_{3}}{3}\Big{)}\nonumber\\
&+C_{F}N_FT_{R}\Big{(}-\frac{55}{3}+16\zeta_{3}\Big{)}
   +N_F^{2}T_{R}^{2}\Big{(}\frac{400}{81}-\frac{8\pi^{2}}{27}\Big{)}\, .
\end{align}
\end{subequations}
and agree with Ref.~\cite{Becher:2010pd}.

\subsection{$\epsilon$-scalar jet function}

The results in this subsection depend on $\alpha_{4\epsilon}$ as well
as $\alpha_s$ and $\alpha_e$. We start by listing the explicit
two-loop coefficients entering \Eqn{eq:jetebare}.
\begin{subequations}
\begin{align} 
\bar{\jmath}^{\, \epsilon;\,AA}_{200}
&=\frac{8}{\epsilon^{4}}
 +\frac{1}{\epsilon^{3}}\Big{(}\frac{59}{3}-\frac{N_{\epsilon}}{6}\Big{)}
 +\frac{1}{\epsilon^{2}}\Big{(}\frac{493}{9}-3\pi^{2}-\frac{7\,N_{\epsilon}}{9}\Big{)}
\nonumber \\
& +\frac{1}{\epsilon}\Big{(}\frac{31675}{216}-\frac{17\pi^{2}}{2}
          -40\zeta_{3}+N_{\epsilon}(\frac{\pi^{2}}{12}-\frac{625}{216})\Big{)}
\nonumber \\
&  +\frac{502189}{1296}-\frac{445\pi^{2}}{18}
+\frac{13\pi^{4}}{180}-\frac{376}{3}\zeta_{3}
 +N_{\epsilon}\big{(}-\frac{12787}{1296}+\frac{7\pi^{2}}{18}
 +\frac{4}{3}\zeta_{3}\big{)}\, ,
\\
\bar{\jmath}^{\, \epsilon;\,Af}_{200}
&=-\frac{4}{3\epsilon^{3}}-\frac{44}{9\epsilon^{2}}
  +\frac{1}{\epsilon}\Big{(}\frac{2\pi^{2}}{3}-\frac{457}{27}\Big{)}
  -\frac{9037}{162}+\frac{22\pi^{2}}{9}+\frac{32}{3}\zeta_{3}\, ,
\\
\bar{\jmath}^{\, \epsilon;\,Af}_{020}
&= \frac{1}{\epsilon^{2}}\big(-2+N_{\epsilon}\big)
  +\frac{1}{\epsilon}\big(-9+N_{\epsilon}\frac{9}{2}\big)
  -\frac{61}{2}+N_{\epsilon}\frac{61}{4}+\pi^{2}-N_{\epsilon}\frac{\pi^{2}}{2}\, ,
\\
\bar{\jmath}^{\, \epsilon;\,Ff}_{020}
&=\frac{1}{\epsilon^{2}}\big(4-N_{\epsilon}\big)
  +\frac{1}{\epsilon}\big(18-7\frac{N_{\epsilon}}{2}\big)+61
  -N_{\epsilon}\frac{33}{4}-2\pi^{2}+N_{\epsilon}\frac{\pi^{2}}{2}\, ,
\label{eJETcoeff} \\
\bar{\jmath}^{\, \epsilon;\,ff}_{020}
&=\frac{4}{\epsilon^{2}}+\frac{16}{\epsilon}+48-2\pi^{2}\, ,
\\
\bar{\jmath}^{\, \epsilon;\,AA}_{002}
&= \frac{3}{8\epsilon}\big(1-N_{\epsilon}\big)
  +\frac{39}{16}-N_{\epsilon}\frac{39}{16}\, ,
\\
\bar{\jmath}^{\, \epsilon;\,Af}_{110}
&=-\frac{8}{\epsilon^{3}}-\frac{24}{\epsilon^{2}}
  +\frac{1}{\epsilon}\big{(}\frac{10\pi^{2}}{3}-64\big{)}
  -156+\frac{32\pi^{2}}{3}+24\zeta_{3}\, ,
\\
\bar{\jmath}^{\, \epsilon;\,Ff}_{110}
&=-\frac{6}{\epsilon^{2}}-\frac{29}{\epsilon}-\frac{227}{2}
+3\pi^{2}+24\zeta_{3}\, .
\end{align}
\end{subequations}
The expression for the renormalized $\epsilon$-scalar jet function in
Laplace space is considerably more complicated than the corresponding
expression for the quark- or gluon-jet function. Contrary to the quark-
and gluon-jet function, there is still a dependence on $\alpha_e$ and
$\alpha_{4\epsilon}$. The finite $\epsilon$-scalar jet function is
given by
\begin{align}
&\jmath_{\epsilon\,\text{fin}}(Q^2,\mu)=1
+\frac{\alpha_{s}}{4\pi}\Big{[}\Gamma_{100}\frac{L_Q^{2}}{2}
+\gamma_{100}^{J_\epsilon}L_Q+c_{100}^{J_\epsilon}\Big{]}
+\frac{\alpha_{e}}{4\pi}\Big{[}\Gamma_{010}\frac{L_Q^{2}}{2}
+\gamma_{010}^{J_\epsilon}L_Q+c_{010}^{J_\epsilon}\Big{]}
\nonumber\\
&+\Big{(}\frac{\alpha_{s}}{4\pi}\Big{)}^{2}\Big{[}
\Gamma_{100}^{2}\frac{L_Q^{4}}{8}
+(-\beta^s_{200}+3\gamma_{100}^{J_\epsilon})\Gamma_{100}\frac{L_Q^{3}}{6}
\nonumber\\
&\quad +(\Gamma_{200}+(\gamma^{J_\epsilon}_{100})^{2}
-\beta^s_{200}\gamma^{J_\epsilon}_{100}+c^{J_\epsilon}_{100}\Gamma_{100})\frac{L_Q^{2}}{2}
+(\gamma^{J_\epsilon}_{200}+\gamma^{J_\epsilon}_{100}c_{100}
-\beta^s_{200}c^{J_\epsilon}_{100})L_Q
+c_{200}^{J_\epsilon}\Big{]}
\nonumber\\
&+\Big{(}\frac{\alpha_{e}}{4\pi}\Big{)}^{2}\Big{[}
\Gamma_{010}^{2}\frac{L_Q^{4}}{8}
+(-\beta^{e}_{020}+3\gamma^{J_\epsilon}_{010})\Gamma_{010}\frac{L_Q^{3}}{6}
\nonumber\\
&\quad
+(\Gamma_{020}+(\gamma^{J_\epsilon}_{010})^{2}-\beta^{e}_{020}\gamma^{J_\epsilon}_{010}
+c^{J_\epsilon}_{010}\Gamma_{010})\frac{L_Q^{2}}{2}
+(\gamma^{J_\epsilon}_{020}+\gamma^{J_\epsilon}_{010}c_{010}
-\beta^{e}_{020}c^{J_\epsilon}_{010})L_Q
+c^{J_\epsilon}_{020}\Big{]}
\nonumber\\
&+\Big{(}\frac{\alpha_{4\epsilon}}{4\pi}\Big{)}^{2}
\Big{[}\Gamma_{001}^{2}\frac{L_Q^{4}}{8}
+(-\beta^{4\epsilon}_{002}+3\gamma^{J_\epsilon}_{001})\Gamma_{001}\frac{L_Q^{3}}{6}
\nonumber\\
&\quad
+(\Gamma_{002}+(\gamma^{J_\epsilon}_{001})^{2}
-\beta^{4\epsilon}_{002}\gamma^{J_\epsilon}_{001}
+c^{J_\epsilon}_{001}\Gamma_{001})\frac{L_Q^{2}}{2}
+(\gamma^{J_\epsilon}_{002}+\gamma^{J_\epsilon}_{001}c^{J_\epsilon}_{001}
-\beta^{4\epsilon}_{002}c^{J_\epsilon}_{001})L_Q
+c^{J_\epsilon}_{002}\Big{]}
\nonumber\\
&+\Big{(}\frac{\alpha_{s}}{4\pi}\Big{)}\Big{(}\frac{\alpha_{e}}{4\pi}\Big{)}
\Big{[}\Gamma_{010}\Gamma_{100}\frac{L_Q^{4}}{4}
+(-(\beta^{e}_{110}\Gamma_{010}+\beta_{110}^{s}\Gamma_{100})
+3(\Gamma_{010}\gamma^{J_\epsilon}_{100}
+\Gamma_{100}\gamma^{J_\epsilon}_{010}))\frac{L_Q^{3}}{6}
\nonumber\\
&\quad
+(\Gamma_{110}+2\gamma^{J_\epsilon}_{010}\gamma^{J_\epsilon}_{100}
-(\beta^{e}_{110}\gamma^{J_\epsilon}_{010}+\beta_{110}^{s}\gamma^{J_\epsilon}_{100})
+c^{J_\epsilon}_{100}\Gamma_{010}+c^{J_\epsilon}_{010}\Gamma_{100})\frac{L_Q^{2}}{2}
\nonumber\\
&\quad
+(\gamma^{J_\epsilon}_{110}+\gamma^{J_\epsilon}_{100}c^{J_\epsilon}_{010}
+\gamma^{J_\epsilon}_{010}c^{J_\epsilon}_{100}-(\beta^{e}_{110}c^{J_\epsilon}_{010}
+\beta_{110}^{s}c^{J_\epsilon}_{100}))L_Q
+c^{J_\epsilon}_{110}\Big{]}\, ,
\phantom{\frac{L_Q^{2}}{2}}
\label{eq:jepsfin}
\end{align} 
where we have kept all terms of $\mathcal{O}(\alpha_s^2)$,
$\mathcal{O}(\alpha_e^2)$, $\mathcal{O}(\alpha_{4\epsilon}^2)$ and
$\mathcal{O}(\alpha_s\, \alpha_e)$, that appear in the structure of
the equation, even if they are zero. The limit $\Neps\to 0$ has been
taken and as usual we indicate this in the notation by dropping the
bar, e.g. $\beta^{e} = \lim_{\Neps\to 0}\bar\beta^{e}$. The
coefficients of the anomalous dimension of the $\epsilon$-scalar jet
can be read off \Eqn{eq:yJeps}. In particular
$\gamma^{J_\epsilon}_{001}=0$. The coefficients of the cusp anomalous
dimensions can be read off \Eqn{app:cusp} and only $\Gamma_{100}$ and
$\Gamma_{200}$ are non-vanishing.

The non-logarithmic terms of \Eqn{eq:jepsfin} read
\begin{subequations}
\begin{align}
c^{J_\epsilon}_{100}&=8\,C_{A}-\frac{2\pi^{2}}{3}C_{A}\, ,\phantom{\frac{1}{1}}\\
c^{J_\epsilon}_{010}&=-4N_FT_{R}\, ,\phantom{\frac{1}{1}}\\
c^{J_\epsilon}_{001}&=0\, ,\phantom{\frac{1}{1}}\\
c^{J_\epsilon}_{200}&=\Big{[}\frac{177325}{1296}-\frac{257\pi^{2}}{18}
+\frac{17\pi^{4}}{36}-32\zeta_{3}\Big{]}C_{A}^{2}+\Big{[}\frac{14}{9}\pi^{2}
-\frac{5581}{162}\Big{]}C_{A}N_FT_{R}\, ,\\
c^{J_\epsilon}_{020}&=\Big{[}\frac{\pi^{2}}{3}-\frac{29}{2}\Big{]}C_{A}N_FT_{R}
+\Big{[}29-\frac{2\pi^{2}}{3}\Big{]}C_{F}N_FT_{R}
+\Big{[}16-\frac{2\pi^{2}}{3}\Big{]}N_F^{2}T_{R}^{2}\, ,\\
c^{J_\epsilon}_{002}&=\frac{39}{16}C_{A}^{2}\, ,\\
c^{J_\epsilon}_{110}&=\Big{[}\frac{16\pi^{2}}{3}-28
-8\zeta_{3}\Big{]}C_{A}N_FT_{R}+\Big{[}\pi^{2}
-\frac{131}{2}+24\zeta_{3}\Big{]}C_{F}N_FT_{R}\, .
\end{align}
\end{subequations}

%% file: 11_AppendixB.tex
\section{Anomalous dimensions}
\label{app:B}

In this appendix we collect all results for the anomalous dimensions
relevant for this work without distinguishing the various
$\alpha_{4\epsilon,i}$.


We give the explicit results with $T_R=1/2$ in the \FDH/\DRED\ scheme,
see \Eqns{eq:Defgammabar}{eq:Defgammaeps} for definitions and
relations. The \CDR/\HV\ results are obtained by setting
$N_{\epsilon}=0$. Of course, $\bar\gamma_\epsilon$ is only meaningful
for \DRED.

\begin{subequations}
\begin{align}
\bar{\gamma}_{q}&=\Big{(}\frac{\alpha_{s}}{4\pi}\Big{)}(-3\,C_{F})+\Big{(}\frac{\alpha_{e}}{4\pi}\Big{)}N_\epsilon \frac{C_F}{2}\nonumber\\
&+\Big{(}\frac{\alpha_{s}}{4\pi}\Big{)}^{2}\Big{[} C_{A} C_{F}\Big{(}-\frac{961}{54}-\frac{11}{6}\pi^2+26\zeta_3\Big{)}+C_{F}^{2}\Big{(}-\frac{3}{2}+2\pi^2-24\zeta_3\Big{)}\nonumber \\
& \qquad\qquad +C_{F} N_F\Big{(}\frac{65}{27}+\frac{\pi^2}{3}\Big{)}+  N_\epsilon \Big(\frac{167}{108}+\frac{\pi^2}{12}\Big)C_A C_F\Big{]}\nonumber\\
   &+\Big{(}\frac{\alpha_{s}}{4\pi}\Big{)}\Big{(}\frac{\alpha_{e}}{4\pi}\Big{)} N_\epsilon \Big[\frac{11}{2}C_A C_F-\Big(2+\frac{\pi^2}{3}\Big)C_F^2\Big] \nonumber \\
 &+ \Big{(}\frac{\alpha_{e}}{4\pi}\Big{)}^{2}\Big{[} - N_\epsilon
  \frac{3}{4} C_F N_F - N_\epsilon^2\frac{C_F^2}{8}  \Big{]}  + \mathcal{O}(\alpha^3) \, ,
\\[18pt]
\bar{\gamma}_{g}&=\Big{(}\frac{\alpha_{s}}{4\pi}\Big{)}\Big{[} -\frac{11}{3}C_A+\frac{2}{3}N_F + N_\epsilon \frac{C_A}{6} \Big{]}\nonumber\\
   &+\Big{(}\frac{\alpha_{s}}{4\pi}\Big{)}^{2}\Big{[}C_A^2\Big{(}-\frac{692}{27}+\frac{11}{18}\pi^2+2\zeta_3\Big{)}+C_{A} N_F\Big{(}\frac{128}{27}-\frac{\pi^2}{9}\Big{)}\nonumber\\
   &\qquad\qquad +2C_{F} N_F + N_\epsilon \Big(\frac{98}{27}-\frac{\pi^2}{36}\Big)C_A^2\Big{]}\nonumber\\
   &+\Big{(}\frac{\alpha_{s}}{4\pi}\Big{)}\Big{(}\frac{\alpha_{e}}{4\pi}\Big{)}(-N_\epsilon
C_F N_F)   + \mathcal{O}(\alpha^3) \, ,
\\[18pt]
\bar\gamma_{\epsilon} &=  \left(\frac{\alpha_s}{4\pi}\right)\,
\left(-4\, C_{A} \right)
+ \left(\frac{\alpha_e}{4\pi}\right)\,
\left( N_F\right) \nonumber \\
&+ \left(\frac{\alpha_s}{4\pi}\right)^2\,
\Big[
   C_{A}^{2}\Big{(}-\frac{2987}{108}+\frac{5\pi^{2}}{6}+2\zeta_{3}
      + N_\epsilon\frac{233}{108}+ N_\epsilon\frac{\pi^{2}}{12} \Big{)}
 + C_{A}N_F\Big{(}\frac{113}{27}+\frac{\pi^{2}}{3}\Big{)}
\Big] \nonumber \\
&+ \left(\frac{\alpha_s}{4\pi}\right)
\left(\frac{\alpha_e}{4\pi}\right) 
\Big[ 5\,C_{F}N_F-\frac{2\pi^{2}}{3}C_{A}N_F
\Big] \nonumber \\
&+ \left(\frac{\alpha_e}{4\pi}\right)^2\,
\Big[ N_F\big{(}\, C_{A}-2\, C_{F}-\frac{N_{\epsilon}}{2}(C_{A}+C_{F})\big{)}
\Big] \nonumber \\
&+ \left(\frac{\alpha_{4\epsilon}}{4\pi}\right)^2\,
\Big[C_{A}^{2}\frac{3}{4}(-1+N_{\epsilon}) \Big]  + \mathcal{O}(\alpha^3) \, ,
\\[18pt]
\bar\gamma_{\text{cusp}} &=  \left(\frac{\alpha_s}{4\pi}\right) (\,4\,)
\nonumber \\
&+ \Big{(}\frac{\alpha_s}{4\pi}\Big{)}^2\,
\Big[C_A\Big{(}\frac{268}{9} - \frac{4}{3}\pi^2\Big{)} 
- \frac{40}{9}N_F-N_\epsilon\frac{16}{9} C_A \Big]  + \mathcal{O}(\alpha^3) \, ,
\label{app:cusp}
\end{align}
\end{subequations}
where $\mathcal{O}(\alpha^3)$ stands for a generic coupling
$\alpha\in\{\alpha_s, \alpha_e, \alpha_{4\epsilon,i}\}$.

For the $\beta$ functions we have
\begin{subequations}
\begin{align}
\bar{\beta}^{s}&=-\Big{(}\frac{\alpha_s}{4\pi}\Big{)}^2
\Big{[} \frac{11}{3}C_A-\frac{2}{3}N_F 
+ N_{\epsilon}\Big{(}-\frac{C_A}{6}\Big{)} \Big{]}+ \mathcal{O}(\alpha^3) \, ,
\\[18pt]
\bar{\beta}^{e}&=-\left(\frac{\alpha_s}{4\pi}\right)
\left(\frac{\alpha_e}{4\pi}\right)(  6\,C_F)\nonumber\\
&-\Big{(}\frac{\alpha_e}{4\pi}\Big{)}^2\Big{[}  
-4\,C_F+2\,C_A-N_F+N_{\epsilon}\Big{(}C_F-C_A\Big{)} \Big{]}+ \mathcal{O}(\alpha^3) \, .
\end{align} 
\end{subequations}
A more complete list of coefficients for the $\beta$ functions can be
found in Ref.~\cite{Kilgore:2012tb}.